%% file: PIP76_4MAR2015.tex
\documentclass[longauth,traditabstract]{aa}  
\def\bfm{} % turn off mark-up
\newcommand{\commentproof}[1]{{\bf \color{green}#1}}
\renewcommand{\commentproof}[1]{} % turn off mark-up

\def\bfc{}
\newcommand{\hproof}[1]{}
\newcommand{\hpeter}[1]{}
\newcommand{\hreply}[1]{}

\usepackage{graphicx}
\usepackage{url}
\usepackage{color}
\usepackage{txfonts}
\usepackage{aas_macros}
\usepackage{natbib}
\usepackage[mathscr]{euscript}

\usepackage{tikz}
\usetikzlibrary{calc,fadings,decorations.pathreplacing,positioning}
\usepackage{verbatim}
\newcommand{\replyc}[1]{{\bf \color{green} [#1]}}

\input Planck

\include{polar_definitions}

\include{PIP76_Bootstrap_Results_DX9_22sep2014}

\include{PIP76_PpBootstrap_Results_DX9_22sep2014}

\include{PIP76_Fit_Results_DX9_22sep2014}

\def\fwhm{5}

\def\vers{DX9}
\def\leak{LEAKPlanckDustModel_SKYCOEFF}

\def\CIBCMBrmvd{_FULL_CIBCMBrmvd_rmvzodi}
\def\dir{./}
\def\day{22sep2014}

\def\psimaxPIP75{20}

\def\dRvNotMeasured{0.4}

\def\nsamp{five}
\def\nstarsSNRDISMORTHO{226}
\def\dof{410}

\def\resr{4.2}
\def\ress{0.2} % statistical error
\def\ressplanckBPM{0.2} % systematic Planck BPM
\def\resscombined{0.3} % combination of all systematic 

\def\resrp{5.4}
\def\ressp{0.2} % statistical error
\def\resspplanckBPM{0.2} % systematic Planck BPM
\def\resspcombined{0.3} % systematic error
\def\resri{1.2}
\def\ressi{0.1} % statistical error

\def\PpDF{2.2}
\def\factorPpDF{2.5}
\def\factorIsAvDL{1.9}

\def\avgTdust{18.9}
\def\avgbeta{1.62}
\def\avgTausAv{2.5\times10^{-5}}

\def\sys{{\rm syst.}}
\def\stat{{\rm stat.}}

\def\ge{\geqslant}
\def\le{\leqslant}

\newcommand{\hi}{{\sc Hi}}

\newcommand{\healpix}{{\tt HEALPix}}

\def\ie{i.e.,}
\def\eg{e.g.,}
\def\SNR{\rm S/N}
\def\nside{N_{\rm side}}
\def\NH{N_{\rm H}}

\def\Trans{{\rm T}}

\def\L{(Left)}

\def\LL{\emph{Left:}}
\def\MM{\emph{Middle:}}
\def\RR{\emph{Right:}}

\newlength{\thsize}
\newlength{\hhsize}
\newlength{\qhsize}
\def\dashed{(dashed)}
\def\dotted{(dotted)}

\def\fmcolor{red}
\def\vacolor{orange}
\def\wecolor{blue}
\def\sacolor{mauve}
\def\krcolor{black}

\def\Vband{$V$}
\def\Bband{$B$}
\def\Kband{$K$}
\def\vis{V}
\def\Rv{R_\vis}
\def\Av{A_\vis}
\def\ebv{E({B-V})}

\def\dAv{\sigma_{A_\vis}}
\def\pv{p_\vis}
\def\qv{q_\vis}
\def\uv{u_\vis}
\def\dpv{\sigma_{\pv}}
\def\dqv{\sigma_{\qv}}
\def\duv{\sigma_{\uv}}
\def\tauv{\tau_\vis }
\def\pst{\pv/\tauv}
\def\qst{\qv/\tauv}

\def\dqst{\sigma_{\qv/\tauv}}
\def\dust{\sigma_{\uv/\tauv}}
\def\lmax{\lambda_{\rm max}}

\def\Tdust{T_{\rm dust}}

\def\sub{{\rm S}} 
\def\isub{I_\sub}
\def\psub{P_\sub}
\def\qsub{Q_\sub}
\def\usub{U_\sub}
\def\dpsub{\sigma_{\psub}}
\def\dqsub{\sigma_{\qsub}}
\def\dusub{\sigma_{\usub}}
\def\ebvsub{\ebv_\sub}
\def\PsI{\psub/\isub}
\def\QsI{\qsub/\isub}
\def\UsI{\usub/\isub}

\def\ratio{R_{\sub/\vis}}

\def\NHratio{\ebv_\sub/\ebv}
\def\NHr{R_{\tau_\sub}}
\def\ratiop{R_{P/p}}
\def\ratioI{\isub/\Av}
\def\unitPp{MJy\,sr$^{-1}$}

\def\polang{{\rm \psi}}
\def\Gex{\polang_\vis}
\def\Gem{\polang_\sub}
\def\dGex{\sigma_{\polang_\vis}}
\def\dGem{\sigma_{\polang_\sub}}
\def\DiffG{\polang_{\sub/\vis}}

\def\ICO{W_{\rm CO}}
\def\kkms{\rm K\,km\,s^{-1}}

\def\ebvmin{0.15}
\def\ebvsubmax{0.8}
\def\avsubmax{2.5}
\def\icomax{2}

\def\SNRpmin{3}
\def\SNRAvmin{three}
\def\NHrmax{1.6}
\def\glatlim{6}

\def\Sect{Sect.}

\def\optical{optical}
\def\optical{visible}

\begin{document}

\input PIP_76_Proj_7_9_Martin_authors_and_institutes

\title{\Planck\ intermediate results. XXI.  \\
Comparison of polarized thermal emission from Galactic dust at 353\GHz\ with 
%optical interstellar polarization}
interstellar polarization in the \optical}

\date{Received 28 April 2014 / {\bfm Accepted 3 December 2014} }

\abstract
{The \Planck\ survey provides unprecedented full-sky coverage of the
submillimetre 
\commentproof{[The \Planck\ collaboration has adopted British spelling uniformly.] }polarized emission from Galactic dust.  In addition to
the information on the direction of the Galactic magnetic field, this
also brings new constraints on the properties of dust.
The dust grains that emit the radiation seen by \Planck\ in the
submillimetre also extinguish and polarize starlight in the \optical.
Comparison of the polarization of the emission and of the interstellar
polarization on selected lines of sight probed by stars provides
unique new diagnostics of the emission and light scattering properties
of dust, and therefore of the important dust model parameters,
composition, size, and shape.
Using ancillary catalogues of interstellar polarization and extinction
of starlight, we obtain the degree of polarization, $\pv$, and the
optical depth in the \Vband\ band to the star, $\tauv$.  Toward these
stars we measure the submillimetre polarized intensity, $\psub$, and
total intensity, $\isub$, in the \Planck\ 353\GHz\ channel.  We
compare the column density measure in the \optical, $\ebv$, with that
inferred from the \Planck\ product map of the submillimetre dust
optical depth and compare the polarization direction (position angle)
in the \optical\ with that in the submillimetre.  {\bfc For those
lines of sight through the diffuse interstellar medium with comparable
values of the estimated column density and polarization directions
close to orthogonal, we correlate properties in the submillimetre and
\optical\ to find two ratios, 
\commentproof{[colon to comma] }$\ratio = (\PsI) / (\pst)$ and $\ratiop
= \psub/\pv$, the latter focusing directly on the polarization
properties of the aligned grain population alone.
}
We find $\ratio = \resr$, with statistical and systematic
uncertainties $\ress$ and $\resscombined$, respectively, and $\ratiop
= \resrp \, {\rm MJy\,sr^{-1}}$, with uncertainties $\ressp$ and
$\resspcombined \, {\rm MJy\,sr^{-1}}$, respectively.
\commentproof{[No change in first part of previous sentence, which has parallel construction.] }Our estimate of $\ratio$ is compatible with predictions based on a
range of polarizing dust models that have been developed for the
diffuse interstellar medium.
{\bfm This estimate provides} new empirical validation
of many of the common underlying assumptions of the models, but 
{\bfm is not yet}
%not yet being 
very discriminating among them.
However, our estimate of $\ratiop$ is not compatible with predictions,
which are too low by a factor of about 2.5.  {\bfc This more
discriminating 
\commentproof{[No change.  This means ``Serving to distinguish; distinctive: e.g., a discriminating characteristic." and ``Able to recognize small differences or draw fine distinctions."] }diagnostic, $\ratiop$, indicates that changes to the
optical properties in the models of the aligned grain population are
required.}
These new diagnostics, {\bfc together with the spectral dependence in
the submillimetre from \Planck, } will be important for constraining
and understanding the full complexity of the grain models, and for
%further interpretation of 
{\bfm interpreting}
the \Planck\ thermal dust polarization and
refinement of the separation of this contamination of the cosmic
microwave background.}


\authorrunning{\Planck\ Collaboration} 

\titlerunning{Comparison of \Planck\ 353\GHz\ thermal dust polarization with interstellar polarization in the \optical}

\maketitle

%%%%%%%%%%%%%%%%%%%%%%%%%%%%%%%%%%%%%%%%%%%%%%%%%%%%%%%

\section{Introduction}\label{intro}

\Planck\footnote{\Planck\ (\url{http://www.esa.int/Planck}) is a
project of the European Space Agency (ESA) with instruments provided
by two scientific consortia funded by ESA member states (in particular
the lead countries France and Italy), with contributions from NASA
(USA) and telescope reflectors provided by a collaboration between ESA
and a scientific consortium led and funded by Denmark.}
has the capability of measuring the linear polarization of the cosmic
microwave background (CMB), a valuable probe for precision cosmology
\citep{planck2013-p01,planck2013-p11,planck2014-XXX}.
One of the diffuse foregrounds contaminating the CMB signal is thermal
emission by diffuse interstellar dust.
Because interstellar polarization of starlight is commonly seen in the
\optical\ 
\commentproof{[removed comma] }from differential extinction by aspherical dust particles
that are aligned with respect to the Galactic magnetic field
\citep{Ha49,Hi49,DG51}, it was predicted that the thermal emission
from these grains would be polarized \citep{S66} and indeed this is
the case {\bfc (\citealp{H99, B04,K07,VD08,BICEP11,planck2014-XIX} }
and references therein).  In this paper 
%using 
{\bfm we use}
the new all-sky
perspective of \Planck\ 
%we 
{\bfm to}
derive the ratio of the diffuse dust
polarization in emission in the submillimetre to interstellar
polarization measured in the \optical, 
{\bfm both}
to provide a quantitative
validation of this prediction and to examine the implications for
grain models.
\commentproof{[Not a list of three; inserted ``both."]}

The CMB fades toward higher frequencies, whereas the thermal dust
emission increases, and so dust becomes the dominant signal
in the submillimetre \citep{planck2013-p06}.
The HFI instrument \citep{lamarre2010} on \Planck\ has multifrequency
polarization sensitivity in the ``dust channels" 
\commentproof{[First of six examples where the \Planck\ style is to define a special phrase using quotations the first time it is introduced.] }covering the spectral range where this transition occurs and up to 353\GHz\
\citep{ade2010,planck2013-p01}.
Understanding both the frequency dependence and spatial fluctuations
of the polarized intensity from thermal dust will be important in
refining the separation of this contamination of the CMB.  With its
sensitive all-sky coverage, \Planck\ is providing the most
comprehensive empirical data both for this analysis and for
complementary Galactic science.  Aspects of dust polarization related
to the Galactic magnetic field are explored in two \Planck\ papers
\citep{planck2014-XIX,planck2014-XX}.  \cite{planck2014-XXII}
describes the spectral dependence of dust polarized emission in the
diffuse ISM. 

{\bfc The observed polarization fraction of aligned aspherical grains
(\ie\ the ratio of polarized to total emission in the submillimetre or
the ratio of differential to total extinction in the \optical) is
affected by many different factors that are hard to disentangle:
the degree of asphericity and the shape, whether elongation or
flattening;
the degree of alignment, with respect to magnetic field lines, of dust
grain populations of different composition and size;
the 3D-orientation of the magnetic field along the line of sight; and
the dust chemical composition and corresponding optical properties at
the wavelengths of observation.
\cite{A12book} provides a thorough review of observational insights
into theories of dust alignment; as we discuss below, our study is not
intended to address alignment mechanisms or alignment efficiency.}

On high column density lines of sight in the Galactic plane
and in dense molecular clouds, even when polarization data in the
\optical\ are available, the interpretation of the \optical\ and
submillimetre polarization is further complicated by
\commentproof{[remove colon] }beam dilution; 
distortions in the magnetic field topology;
changes in the degree of alignment along the line of sight; 
grain evolution; and 
ranges of grain temperature and optical depth that affect which grains
dominate the polarized emission in various parts of the submillimetre
spectrum \citep[\eg][]{H99,VA07,VD08,BICEP11,VM12}.
Such complex regions are deliberately not considered here.  We limit
our analysis to lines of sight through the diffuse ISM, for which more
homogeneous properties might be expected, and for which the most
comprehensive observational constraints on dust models are already
available and exploited (\eg\ \citealp{DL07,MC11,J13,SVB14}).

In the diffuse ISM, the spectral dependence of the degree of
polarization of starlight, the interstellar polarization curve
$p(\lambda)$, has a peak in the \optical\ close to the \Vband\ band,
and falls off toward both the infrared and the UV
\citep{SMF75,WM92,MW92,MCW99}.  By contrast, the interstellar
extinction curve $\tau(\lambda)$ decreases with the wavelength from
the UV to the infrared.  From this combination it is inferred
that small grains are either spherical or not aligned \citep{KM95} and
that the polarization is dominated by ``large" grains (around
0.1\,$\mu$m in size, \eg\ \citealp{DL07}) that are in thermal
equilibrium with the interstellar radiation field and radiating in the
submillimetre.
Polarization of absorption bands shows that silicate grains are
aspherical and aligned \citep{DB74}, but that aliphatic carbon grains,
responsible for the 3.4 $\mu m$ band, are not \citep{A99,C06}.
However, the question of whether the large aromatic carbon grains used
in dust models \citep{DL07,MC11} are aligned or not is not directly
constrained by such observations 
%due to 
{\bfm because of}
the lack of characteristic
bands.

{\bfc Various models of diffuse dust have been developed to reproduce
the polarization and extinction spectral dependences with a
combination of aligned and unaligned grains
\citep[\eg][]{LD85,LG97,V12}.  The most recent are further constrained
by fitting the (pre-\Planck) spectral energy distribution of dust
emission in the infrared and submillimetre \citep{DF09,SVB14}.
Efforts have also been made to predict the polarized thermal emission
quantitatively \citep{M07,DF09,DH13}.  }

This paper is organized as follows.  
In \Sect~\ref{Ratios} we introduce the emission to extinction ratios
in polarization that are analyzed in this paper and describe their
diagnostic importance for dust modeling.
The observational data and uncertainties that are available in the
submillimetre from the \Planck\ maps are presented in \Sect~\ref{emi},
supplemented by Appendix~\ref{AppUncertainty}.
Section~\ref{ext}, supplemented by Appendix~\ref{KR09cat}, describes
the data in the \optical\ from catalogues for many lines of sight to stars.
The criteria for selecting suitable lines of sight are given in
\Sect~\ref{criteria}.
The methodology for evaluation of the polarization ratios and the
results for the diffuse ISM follow in \Sect~\ref{Results} and
{\bfc
the robustness of the results is discussed in Appendix~\ref{rob}.
}
{\bfc
In \Sect~\ref{discussion} we discuss how these
new results from polarization both validate and challenge extant dust
models.
}
We conclude with a short summary in \Sect~\ref{con}.

%%%%%%%%%%%%%%%%%%%%%%%%%%%%%%%%%%%%%%%%%%%%%%%%%%%%%%%

\section{Diagnostic polarization ratios involving dust emission and
  extinction}\label{Ratios}

In this paper we evaluate the ratio of the polarization at 353\GHz\ in
the submillimetre, where the signal-to-noise ratio (\SNR) of \Planck\
data is highest for dust polarized emission, to the interstellar
polarization in the \Vband\ band, near the peak of the polarization
curve.

A first condition necessary for this comparison to be meaningful is
met : the \Vband\ band interstellar polarization is dominated by the
same so-called large grains that produce polarized thermal emission
at 353\GHz.  The usual evidence for this was given in
\Sect~\ref{intro}. This is now bolstered by new direct observations of
the strength and spectral shape of the polarized emission in the
submillimetre \citep{planck2014-XXII}.

\begin{figure}
\includegraphics[width=\hsize]{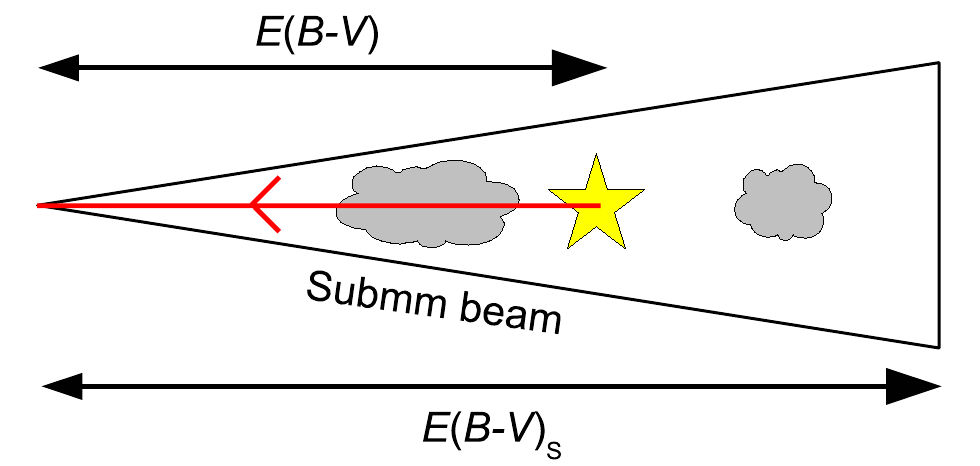}
\caption{Instrumental-beam and line-of-sight components affecting the
comparison of polarized emission with interstellar polarization from
differential extinction of a star.  $\ebv$ is the colour excess to the
star, while $\ebv_{\rm S}$ is the submillimetre optical depth
converted to a colour excess (\Sect~\ref{taus}).}
\label{los}
\end{figure}

A second condition is purely geometrical, as shown schematically in
Fig.~\ref{los}.  In the \optical, interstellar polarization and
extinction arise from dust averaged over the angular diameter of the
star, which is tiny compared to the \Planck\ beam.  Furthermore, these
observations in the \optical\ probe the ISM only up to the distance of
the star, while submillimetre observations probe the whole line of
sight through the Galaxy, thus including a contribution from any
background ISM (see Fig.~\ref{los}).  As will be discussed below, the
effects of these differences can be mitigated and assessed.

As discussed in \Sect~\ref{intro}, current models of interstellar dust
commonly feature multiple grain components and not all components
(even those that are in thermal equilibrium with the interstellar
radiation field and are the major contributors to extinction in the
\optical\ and emission in the submillimetre) might be aspherical and
aligned.  Both the total submillimetre emission, $\isub$, and the
optical depth to the star in the \Vband\ band, $\tauv$, entail the
full complexity from the contributions of aligned and non-aligned
grain populations.  while the polarized emission, $\psub$, and the
degree of polarization toward the star, $\pv$, isolate properties of
the polarizing grains alone.

Because many of the factors driving interstellar polarization, like
grain shape,
alignment efficiency, and 
magnetic field orientation, 
affect $\pv$ and $\psub$ in similar ways \citep{M07}, we are motivated
to examine 
%the following 
{\bfm two}
polarization ratios,
\begin{equation}\label{Eq-ratio}
\ratio = \frac{\PsI}{\pst} %\, ,
\end{equation}
and
\begin{equation}\label{Eq-ratiop}
\ratiop = \psub / \pv\,.
%\frac{\psi}{\pst},
\end{equation}

\commentproof{[New paragraph intended for comments on the first ratio.  The following paragraph is for the second ratio.] }$\ratio$ is the ratio of the polarization fractions at 353 \GHz\ and
in the \Vband\ band.  It is a non-dimensional quantity where both
numerator and denominator are themselves non-dimensional ratios that
apply to the same geometry (common beam and common portion of the line
of sight).  Furthermore, by virtue of the normalization both $\PsI$
and $\pst$, the submillimetre and \optical\ polarization fractions, do
not depend on the column density and become less sensitive to various
factors like the size distribution, grain heating, and opacity that
affect the numerator and denominator in similar ways \citep{M07}.
$\ratio$ is therefore a robust tool for the data analysis.  Being a
mix of aligned and non-aligned grains properties,\footnote{
{\bfm There} are possible contributions by large unaligned grains
to the total thermal emission and extinction.  These contributions
cause a \emph{dilution} of the polarization fractions.
}
$\ratio$ is however complex to interpret.

{\bfc $\ratiop$} characterizes the aligned grains alone, addressing
how efficient they are at producing polarized submillimetre emission
compared to their ability at polarizing starlight in the \Vband\
band. It has the units of polarized intensity, here \unitPp.  It is
easier to interpret, and model, than $\ratio$.  As a drawback, it is
less robust than $\ratio$ from the data analysis point of view.
Although $\psub$ and $\pv$ both depend on column density, $\ratiop$ is
more sensitive to the above geometrical effects and particular
attention must be paid to avoid lines of sight with significant
background emission.  Through $\psub$, $\ratiop$ is also directly
dependent on the submillimetre emissivity of the polarizing grains and
on the intensity of the interstellar radiation field.  Nevertheless,
$\ratiop$ can provide even stronger constraints on the \emph{aligned}
grains than $\ratio$ can.

Despite the overall complexity of the production of dust polarization,
studying polarization ratios like $\ratio$ and $\ratiop$ provide new
insight on some dust properties like optical constants alone.  As a
corollary, the correlation analysis involving data in the
submillimetre and \optical\ for the same line of sight does \emph{not}
provide any information on the dust alignment efficiency.

%%%%%%%%%%%%%%%%%%%%%%%%%%%%%%%%%%%%%%%%%%%%%%%%%%%%%%%

\section{Observations of polarized thermal emission from
  dust}\label{emi}

%==============================================

\subsection{\Planck\ Data}\label{planckdata}

The \Planck\ HFI 353\GHz\ polarization maps that we used for the
Stokes parameters $\qsub$ and $\usub$ \citep{planck2014-XIX} were
those from the full mission with five full-sky surveys.  These have
been generated in exactly the same manner as the data publicly
released in March 2013 and described in \citet{planck2013-p01} and
associated papers.\footnote{
{\bfm However,} the publicly released data include only
temperature (intensity) maps, based on the first two surveys.
}

Intercalibration uncertainties between HFI polarization-sensitive
bolometers and differences in bolometer spectral transmissions
introduce a leakage from intensity $I$ into polarization $Q$ and $U$,
the main source of systematic errors \citep{planck2013-p03}.  The $Q$
and $U$ maps were corrected for this leakage
\citep{planck2014-XIX}.\footnote{
The systematic errors that we quote include uncertainties associated
with residual systematics as estimated by repeating the analysis on
different subsets of the data. We have also checked our data analysis
on the latest version of the maps available to the consortium to
verify that the results are consistent within the uncertainties quoted
in this paper.
}

At 353\GHz\ the dispersion arising from CMB polarization anisotropies
is much lower than the instrumental noise for $\qsub$ and $\usub$
\citep{planck2013-p03} and so has a negligible impact on our analysis
(see Appendix~\ref{RobDataUsed}).  The cosmic infrared background (CIB)
was assumed to be unpolarized \citep{planck2014-XIX}.

For the intensity of thermal emission from Galactic dust, $\isub$, we
begin with the corresponding \Planck\ HFI 353\GHz\ map $I_{353}$ from
the same five-survey internal release, corrected for the CMB dipole.
From a Galactic standpoint, $I_{353}$ contains small amounts of
contamination from the CMB, the CIB, and zodiacal dust emission.
Models for the CMB fluctuations \citep[using {\tt
SMICA};][]{planck2013-p06} and zodiacal emission
\citep{planck2013-pip88} were removed.
For this study of Galactic dust emission we subtract the derived zero
offset from the map, which effectively removes the CIB monopole
\citep{planck2013-p06b}.
The level of the CIB fluctuations (the anisotropies), estimated by
\cite{planck2013-p06b} to be 0.016~MJy\,sr$^{-1}$, introduces this
uncertainty in $\isub$.

To increase the \SNR\ of the \Planck\ HFI measurements on lines of
sight to the target stars (\Sect~\ref{ext}), especially in the diffuse
ISM, the Stokes parameters $\isub$, $\qsub$, and $\usub$, were
smoothed with a Gaussian centred on the star, and the corresponding
noise covariance matrix was calculated (see Appendix~A of
\citealp{planck2014-XIX} for details).  The \Planck\ HFI 353\GHz\ maps
have a native resolution of 5\arcmin\ and a \healpix\footnote{
See \url{http://healpix.jpl.nasa.gov} and
\url{http://healpix.sourceforge.net}}
\citep{GHB05}
grid pixelization corresponding to $\nside = 2048$. Smoothing the
\Planck\ data accentuates the beam difference relative to the stellar
probe (Fig.~\ref{los}).  Therefore, there is a compromise between
achieving higher \SNR\ and maintaining high resolution.  The original
\SNR, and thus any compromise, depends on the region being studied.
However, for simplicity we adopted a common Gaussian smoothing kernel,
with a full width at half maximum (FWHM) of \fwhm\arcmin\ (the
effective beam is then 7\arcmin), and explored the robustness of our
results using different choices (Appendix~\ref{RobDataUsed}).

%==============================================

\subsection{Position angle in polarized emission}\label{pols}

The orientation of the plane of vibration of the electric vector of
the polarized radiation is described by a position angle with respect
to 
\commentproof{[Lower case cardinal directions.] }north, here in the Galactic coordinate system.  North corresponds
to positive $Q$.  In \healpix, the native coordinate system of
\Planck, the position angle increases to the west, whereas in the IAU
convention the position angle increases to the east; this implies
opposite sign conventions for $U$ \citep{planck2014-XIX}.
In this paper, all position angles, whether $\Gem$ in the
submillimetre or $\Gex$ in the \optical, are given in the IAU
coordinate system to follow the use in the ISM literature.

On the other hand, all Stokes parameters, whether \Planck\ $\qsub$,
$\usub$ in the submillimetre or $\qv$, $\uv$ derived in the \optical,
are in the \healpix\ convention.  This accounts for the minus sign
both in Eq.~\ref{Eq-Gem} and in its inverse form, Eq.~\ref{Eq-qv-uv}
below.

Thus from the \Planck\ data we find 
\begin{equation}
\Gem = \frac{1}{2} \, \arctan(-\usub,\qsub) \,\in [-90\deg,90\deg] \, .
\label{Eq-Gem}
\end{equation}  
To recover the correct full range of position angles (either
$[0\deg,180\deg]$, or $[-90\deg,90\deg]$ as used for $\Gem$ here)
attention must be paid to the signs of both $\usub$ and $\qsub$, not
just of their ratio.  This is emphasized explicitly by use of the
two-parameter $\arctan$ function, rather than $\arctan(-\usub/\qsub)$.

%==============================================

\subsection{Column density of the ISM from \Planck}\label{taus}

A standard measure of the column density of dust to a star from data
in the \optical\ is the colour excess $\ebv$.  An estimate of the
column density observed by \Planck\ in the submillimetre is needed to
check for the presence of a significant background beyond the star
(Fig.~\ref{los}).  This independent estimate was based on the \Planck\
map of the dust optical depth $\tau_\sub$ (and its error) at 353\GHz,
at a resolution of 5\arcmin, plus a calibration of $\tau_\sub$ into an
equivalent reddening $\ebvsub$, using quasars: $\ebvsub \simeq 1.49
\times 10^4\, \tau_\sub$ \citep{planck2013-p06b}.

Based on the dispersion of the calibration, and the cross-checks with
ancillary data, our adopted estimate for $\ebvsub$ should be accurate
to about 30\,\% for an individual line of sight.
{\bfm We note} that this uncertainty is not propagated directly into the final
polarization ratios because $\ebvsub$ is not used in those
calculations, but only for selection purposes (\Sect~\ref{Eratio}),
whose robustness is explored in Appendix~\ref{rob}.

%%%%%%%%%%%%%%%%%%%%%%%%%%%%%%%%%%%%%%%%%%%%%%%%%%%%%%%

\section{Observations of polarization and extinction of
  starlight}\label{ext}

Measurements of stellar polarization, here in the \Vband\ band, are
usually reported in terms of the degree of polarization, $\pv$, and
the position angle, $\Gex$, from which we can recover a representation
of the observables in the \healpix\ convention:
\begin{eqnarray}
\qv & = & \phantom{ -} \pv \, \cos{2\,\Gex} \nonumber\,; \\
\uv & = & - \pv \, \sin{2\,\Gex} \, .
\label{Eq-qv-uv}
\end{eqnarray}

We used the \citet{H00} catalogue of polarization in the \optical, a
compilation of several others \citep[\eg][]{Ma71,Ma78}.  This
catalogue provides $\pv$ and its uncertainty $\dpv$, together with
$\Gex\, \in [0\deg,180\deg]$ in the IAU Galactic convention for 9286
stars.  For data with $\SNR > 3$, as will be imposed below, it is
reasonable to use the \cite{SMF75} approximation to the uncertainty
$\dGex$ in $\Gex$:
\begin{equation}
\dGex = 28\pdeg65 \,\dpv/\pv
\label{phierror}
\end{equation}
\citep[\eg][]{NK93}.  The catalogue also provides an estimate of the
distance and colour excess to the star.  However, the colour excess
has too low a precision (0.1 mag) to be used here.

%==============================================

Accurate extinction data are needed both for the selection of stars
(see \Sect~\ref{taus}) and for the calculation of $\ratio$ (but not
$\ratiop$).  We selected stars from various catalogues
\citep{SA85,WE02,WE03,VA04,FM07} sequentially according to the
accuracy of the technique used to derive the colour excess $\ebv$.
In the shorthand notation of Appendix \ref{ExtCat}, in which there is
a description of these catalogues together with our own derivation of
$\ebv$ for remaining stars using the catalogue of \cite{K09}, the
precedence is FM07, VA04, WE23, SA85, and KR09.

By definition, the optical depth is
\begin{equation}
\tauv = \Av / 1.086 \,.
\label{tauv}
\end{equation}
The extinction $\Av$ is found from $\ebv$ through
multiplication by the ratio of total to selective extinction, $\Rv =
\Av/\ebv$, either estimated from the shape of the multifrequency
extinction curve or adopted as 3.1 as for the diffuse ISM
\cite[\eg][]{F04} when such a measure is missing. {\bfm We note} that $\tauv$
is not needed for $\ratiop$.

%%%%%%%%%%%%%%%%%%%%%%%%%%%%%%%%%%%%%%%%%%%%%%%%%%%%%%%

\section{Selection of stars}\label{criteria}

For each sample, we determined the subsample of stars to be used to
calculate the polarization ratios $\ratio$ and $\ratiop$ by applying
four (sets of) selection criteria.  We evaluate the dependence of our
results on these criteria in Appendix~\ref{rob}.

%==============================================

\subsection{\SNR}\label{snrcriterion}

The first criterion was to require a \SNR\ higher than \SNRpmin\ for
$\psub$ and $\pv$, which propagates into an uncertainty in the
position angle of less than $10\deg$ (Eq.~\ref{phierror}).  We also
imposed a \SNR\ higher than \SNRAvmin\ for $\Av$ (and consequently,
$\tauv$), a quantity that might otherwise be poorly
estimated.\footnote{
For the catalogues where $\Rv$ was not measured (SA85 and KR09), the
assumed uncertainty $\delta\Rv=\dRvNotMeasured$ introduced in
\Sect~\ref{ecat} is ignored in the selection process, but not in the
data analysis and fitting.}
In emission, this condition is always met automatically for $\isub$
when it is required for $\psub$.

Lines of sight where the column density is very low are too noisy in
total extinction and in polarized emission to be used with
confidence. We therefore {\bfm also} imposed $\ebv > \ebvmin$ and $\ebvsub
> \ebvmin$.
The latter criterion ensures that any uncertainties in the small
corrections of $I_{353}$ to $\isub$ (\Sect~\ref{planckdata}) are
unimportant.

%==============================================

\subsection{Diffuse ISM}\label{diffusecriterion}
 
Our intent is to characterize dust polarization properties in the
diffuse, largely atomic, ISM.  The column density measure $\ebvsub$
can be used statistically as a selection criterion: the higher
$\ebvsub$, the higher the probability of sampling dense environments.
We found the selection
\begin{equation}
\ebvsub \le \ebvsubmax
\label{Eq-EBV}
\end{equation}
{\bfm ($\Av^\sub \le \avsubmax$)} to be a good compromise between the size of
the selected sample and the exclusion of dense environments (which are
also generally characterized by a lower dust temperature, $\Tdust \le
17$\,K).

As one consistency check, we note that the selected lines of sight
have low $\ICO$ (less than about $\icomax\,\kkms$) as judged from the
\Planck\ ``type~3" CO map \citep{planck2013-p03a} smoothed to
30\arcmin\ resolution.
As another, adopting the average opacity $\sigma_{\rm e}(353) =
\tau_{353}/N_{\rm H}$ found by \citet{planck2013-p06b} over the range
$0.15 < \ebvsub \le \ebvsubmax$ (or equivalently $0.87 < N_{\rm H} \le
4.6$ in units of $10^{21}$\,cm$^{-2}$ using the diffuse ISM conversion
between $\ebv$ and $N_{\rm H}$ from \citealp{Bo78}) together with this
diffuse ISM conversion, we find a consistent calibration between
$\ebvsub$ and $\tau_\sub$.  Furthermore, \citet{planck2013-p06b} show
that over the above range $\ebvsub$ compares favourably to estimates
of colour excess based on stellar colours in the 2MASS data base
\citep{Skrutskie06}.

%==============================================

\subsection{Compatibility between the column densities in the
  submillimetre and the \optical}\label{Eratio}

The selection on $\ebvsub$ helps to remove some lines of sight with
potential emission background beyond the star. This can be
supplemented by comparing the \Planck\ $\ebvsub$ with $\ebv$ for the
star (see Fig.~\ref{los}).
Significant disagreement between the two column density estimates,
whether an effect of different beams or an effect of the medium beyond
the star, would mean that the polarization data cannot be compared
usefully.  The effect of slightly mismatched columns is mitigated
somewhat by the normalization in $\ratio$, which is a ratio of ratios;
it is of heightened concern for $\ratiop$.

We define the column density ratio between the submillimetre and the
\optical:
\begin{equation}
\NHr = \NHratio\,.
\label{Eq-Re}
\end{equation}

\begin{figure}
%\center
\includegraphics[width=\hhsize]{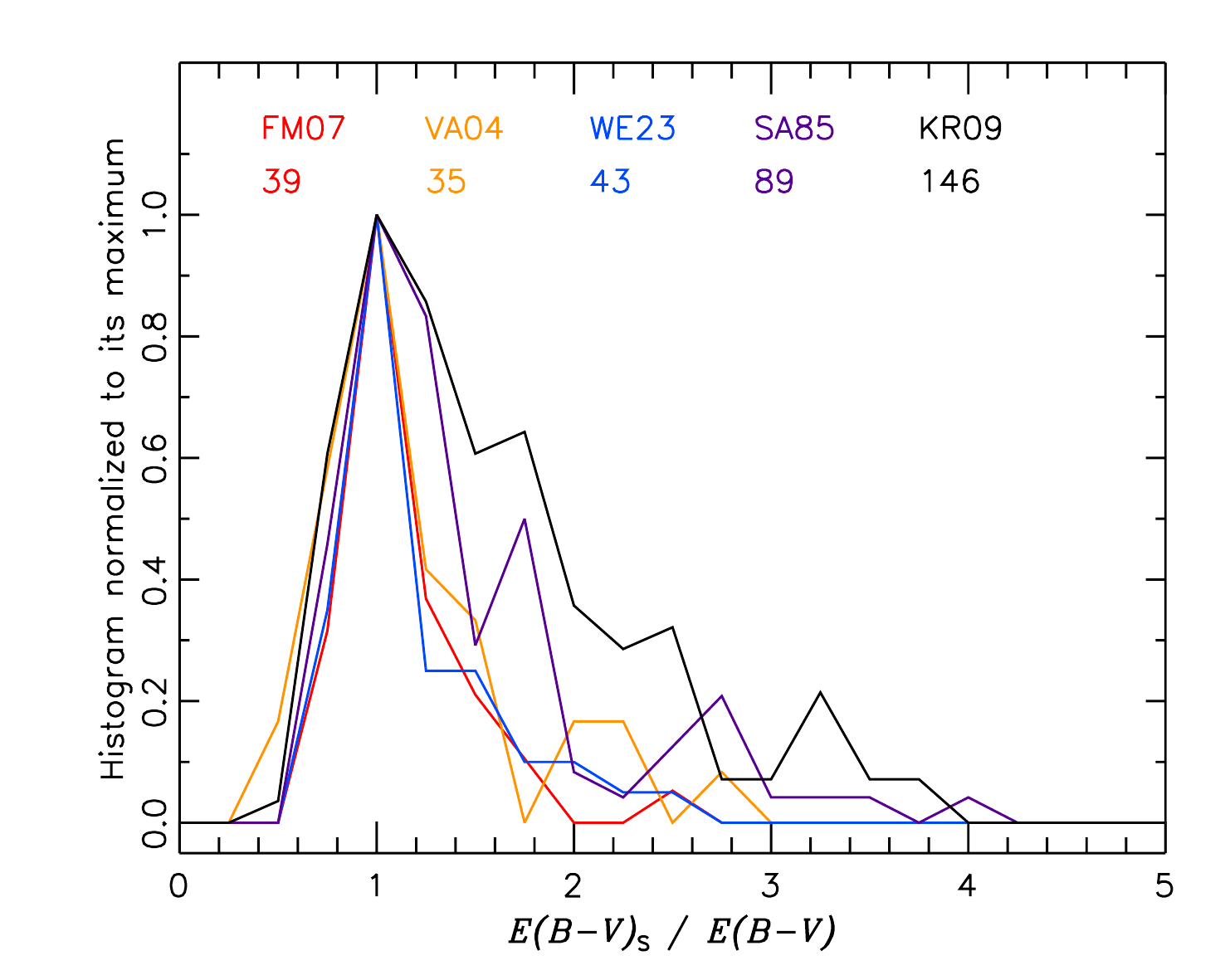}
\caption{Normalized histograms of the column density ratio $\NHr$ for
those lines of sight passing the first two selection criteria for $\SNR$
and diffuse ISM, for the independent FM07 (\fmcolor), VA04 (\vacolor),
WE23 (\wecolor), SA85 (\sacolor), and KR09 (\krcolor) samples. The number
of stars in each sample is indicated.}
\label{Fig-hist-NHratio}
\end{figure}

Figure~\ref{Fig-hist-NHratio} presents such a comparison in the form
of a normalized histogram of $\NHr$ for each sample.  For the FM07,
VA04, and WE23 samples the histograms correspond to what we would
expect for lines of sight with little emission background beyond the star,
namely a peak near $\NHr \simeq 1$, and we take this as a first
indication of the good quality of the $\ebvsub$ and $\ebv$ estimates.
In \Sect~\ref{taus} we estimated that for a given line of sight
$\ebvsub$ might have a 30\,\% uncertainty and from the \SNR\ criterion
on $\tauv$ (\Sect~\ref{snrcriterion}) the uncertainty in $\ebv$ is
less than 33\,\%.  These uncertainties would readily account for the
width of the distribution about this peak.
 
Nevertheless, the SA85 and KR09 samples are not so well peaked,
containing many lines of sight with $\NHr \ge 2$.
This might indicate a significant background beyond the star, which
must in principle arise for some lines of sight. The stars in these
independent samples, absent from the other more accurate samples,
probe regions of the sky not represented by the other samples.
Alternatively, for some lines of sight the dust opacity might be
higher than for the diffuse ISM adopted here to derive $\ebvsub$ 
({\bfm see,} e.g., \citealp{M12,R13}), leading to an overestimation.

Whatever the reason for this disagreement between $\ebvsub$ and
$\ebv$, we need to be wary about including lines of sight with high
$\NHr$ in our analysis.  Therefore, as a third criterion we removed
all lines of {\bfm sight} with $\NHr$ higher than a certain threshold.

To determine this threshold, we made use of the histograms of the
difference in position angles in emission and in extinction:
\begin{equation}\label{deltapsi}
\DiffG \equiv \frac{1}{2} \, \arctan{\left[\left(\usub\,\qv - \qsub\,\uv \right),-\left(\qsub\,\qv+\usub\,\uv\right)\right]}\, . 
\end{equation}
In the ideal case where measurements of emission and extinction probe
the same medium, the polarization directions measured in extinction
and in emission should be orthogonal (e.g., \citealp{M07}). With
Eq.~\ref{deltapsi}, orthogonality corresponds to $\DiffG =
0\deg$.\footnote{
The expression for $\DiffG$ follows from the $\arctan$ addition rule
as for Eq.~7 in \cite{planck2014-XIX}, with a minus sign before each
argument allowing for the rotation by 90\deg\ of the polarization
direction in emission as measured by $\Gem$ and an additional sign
change in the first argument because $\DiffG$, like $\Gem$ and $\Gex$,
follows the IAU convention for angles, increasing from north through
east (Eq.~\ref{Eq-Gem}).}
Because the systematic presence of backgrounds beyond the stars would
induce some deviations from orthogonality, we expect a decline of the
quality of position angle agreement as $\NHr$ increases.

\begin{figure}
\center
\includegraphics[width=\hhsize]{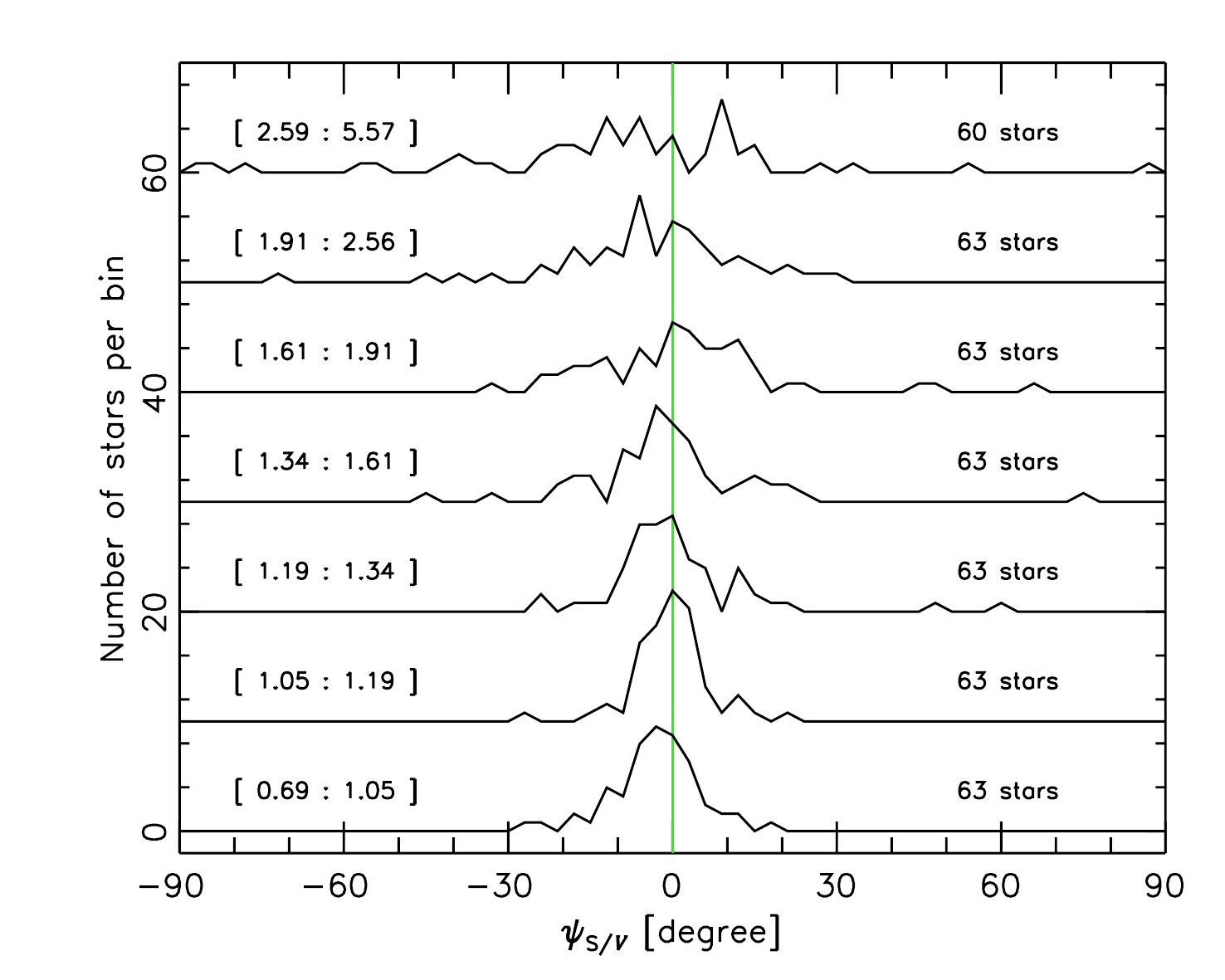}
\caption{Histograms of difference in position angles $\DiffG$ for
successive ranges in column density ratio, $\NHr$,
as indicated on the left, with the corresponding number of stars on the
right.  Only lines of sight of our \nsamp\ samples satisfying the
first (\SNR) and second (diffuse ISM) selection criteria have been
used.  For clarity, the histogram has been shifted upward by 10 units
for each range.
}
\label{Fig-histdiffG-NHratio}
\end{figure}

This hypothesis is tested in Fig.~\ref{Fig-histdiffG-NHratio}, for
lines of sight selected by only the first two criteria ($\SNR$ and
diffuse ISM).  The form of the histogram is observed to depend on the
range considered for the column density ratio, $\NHr$.  When the
column densities agree ($\NHr \simeq 1$), the histogram of $\DiffG$ is
well peaked around zero, as expected. This agreement persists as long
as $\NHr$ is not too large, here below~$\NHrmax$.  Whether we correct
for leakage (\Sect~\ref{planckdata}) or not has no effect on these
conclusions.  As a corollary, the agreement of column densities
appears to be a good indicator of consistency between position angles,
at least statistically.

Based on this discussion of Figs.~\ref{Fig-hist-NHratio} and
\ref{Fig-histdiffG-NHratio}, we defined our third selection criterion
to be
\begin{equation}
%\NHr = \NHratio \le \NHrmax \, .
\NHr \le \NHrmax \, .
\label{NH-criteria}
\end{equation}
An alternative way to select lines of sight with little background
would be to select according Galactic height or sufficient distance
using the \emph{Hipparcos} catalogue.  Although we did not adopt this
as an additional criterion, we tested its impact in
Appendix~\ref{robcriteria}.

%==============================================

\subsection{Consistency of polarization directions
  (orthogonality)}\label{orthocriterion}

\begin{figure}
\includegraphics[width=\hhsize]{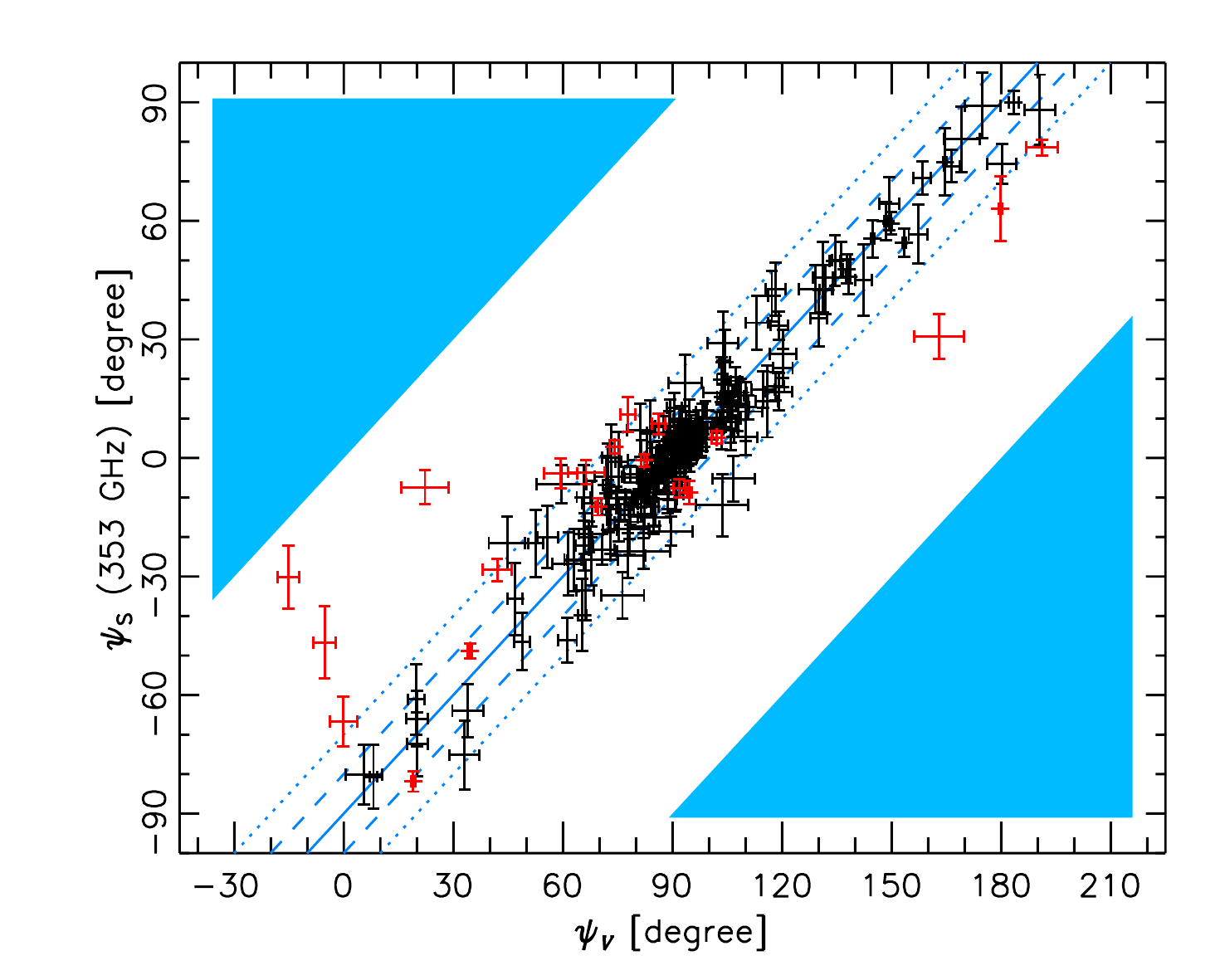}
\caption{Correlation plot of position angles in emission, $\Gem$, and
in extinction, $\Gex$, for the merged sample, for lines of sight
satisfying the first three selection criteria.  Data for lines of
sight failing the fourth angle criterion are marked in red; those
finally selected are in black.  The central diagonal solid line
indicates perfect agreement (orthogonal polarization pseudo-vectors),
and other lines are for offsets of $10\deg$ \dashed\ and $20\deg$
\dotted.  {\bfm We note} that when the arithmetic difference in angles falls
outside the allowed $\pm 90\deg$ range for $\Gem$ (filled zones), the
plotted $\Gex$ is adjusted by $\pm 180\deg$ (Eq.~\ref{deltapsi}).
}
\label{Gem-Gex}
\end{figure}

{\bfc The fourth selection criterion is a check for the consistency
\emph{within the uncertainties} of the polarization directions in the
visible and in the submillimetre.}\footnote{
\bfc Consistency of the orientation of polarization in the \optical\
at $9\pdeg2$ resolution with the direction of the interstellar
magnetic field inferred from K band (23 GHz) polarization measured by
WMAP has been noted by \citet{Pa07}.  This is of interest whether the
K band polarization arises from synchrotron emission or dust.
Likewise, in their BICEP millimetre-wave polarization Galactic plane
survey in the longitude range $260\deg < l < 340\deg$, \cite{BICEP11}
found general consistency with polarization angles in the \optical.
}
We adopt
\begin{equation}
|\DiffG| \le 3\,  \sqrt{\dGem^2+\dGex^2}\, .
\label{angle-criteria}
\end{equation}

Figure~\ref{Gem-Gex} presents a comparison of the position angles for
the sample of $\nstarsSNRDISMORTHO$ stars selected above.  As
anticipated by the histograms of $\DiffG$ in
Fig.~\ref{Fig-histdiffG-NHratio}, some lines of sight (plotted in red)
are rejected by this fourth criterion.
The outliers in Fig.~\ref{Gem-Gex} arise at least in part from
systematic errors attributable to the small leakage of intensity into
polarization that is imperfectly corrected in the March 2013 internal
release of the \Planck\ data (\Sect~\ref{planckdata}).  Differing
beams and paths probed by measurements in the submillimetre and
\optical\ (Fig.~\ref{los}) can also contribute.\footnote{
\citet{planck2014-XIX} have found that the dispersion of the position
angles measured in neighbouring \Planck\ beams is anti-correlated with
$\PsI$ (their Fig.~23).  Dispersions comparable to the angle $\DiffG$
of the outliers in Fig.~\ref{Gem-Gex} occur statistically at low
$\PsI$, below a few percent.  
\commentproof{[The original should be kept.  It is the same phrasing as in ``the cost of living rose by a few percent."  Using the proposed alternatives would be awkward and would not change the meaning.] }We find that $\DiffG$ anti-correlates
with $\PsI$ too, suggesting that the same processes are responsible
for the dispersion of the position angles in both extinction and
emission at low fractional polarization.
}

{\bfm We note} that the final sample (plotted in black) covers a considerable
range in position angle, \ie\ the sample does not only probe
environments where the orientation of the polarization in the
\optical\ is close to parallel to the Galactic plane ($\Gex = 90\deg$,
$\Gem = 0\deg$).  We will see in the following section that this
dynamic range is essential for deriving the polarization ratios
$\ratio$ and $\ratiop$ using correlation analysis.

%==============================================

\subsection{Selected sample of stars}

Combining the four sets of criteria (regarding the \SNR, the diffuse
ISM, the agreement in column densities, and the consistency of
position angles) for each sample we selected those lines of sight that
would be suitable for a comparison of polarization in the diffuse
ISM. Table \ref{Table-Selection} presents the numbers of stars
remaining in our sample after the selection criteria were applied in
sequence.  Starting from 9286 stars, we retain only \nfitQUcorrBPM.
We assess the impact of this systematic reduction in
Appendix~\ref{rob} by relaxing our selection criteria.
Our full sample is spread between the different catalogues, thus
avoiding any strong dependence on any one in particular.

\begin{table}[h] 
\begingroup % this + \endgroup at the end keep table things local
\caption{Evolution of the number of stars remaining after successive
selection criteria are applied.
}
\label{Table-Selection}
\nointerlineskip
\vskip -3mm
\footnotesize % good font size for a table, but can be changed
\setbox\tablebox=\vbox{ %
\newdimen\digitlsmwidth % see \S\,17.12 for the purpose of the next 10 lines
\setbox0=\hbox{\rm 0}
%\digitwidth=\wd0
\catcode`*=\active
\def*{\kern\digitwidth}
\newdimen\signwidth
\setbox0=\hbox{+}
%\signwidth=\w
\catcode`!=\active
\def!{\kern\signwidth}
%%
%\halign{#\hfil\tabskip=01em&\hfil#&\hfil#&\hfil#&\hfil#&\hfil#&\hfil\tabskip=02em#\cr
\halign{#\hfil\tabskip=01em&\hfil#&\hfil#&\hfil#&\hfil#&\hfil#&\hfil\tabskip=02em#\tabskip=0pt\cr
\noalign{\doubleline}
Selection criteria &  FM07 & VA04 & WE23 & SA85 &KR09 & Total \cr %of stars remaining \cr
\noalign{\vskip 5pt\hrule\vskip 3pt}
\cite{H00}             & & & & & & 9286 \cr
$\pv /\dpv >  3$ & & & & & & 5579 \cr
$\psub/\dpsub >  3$ @ 5\arcmin && & & & & 3030 \cr
%\noalign{\vskip 5pt\hrule\vskip 3pt}
%catalogues & FM & VA & WE & SA &KR &  \cr %of stars remaining \cr
$\Av/\dAv >  3 $ & 128 & 245 & 338 & 575 & 980 & 2266 \cr
Indep. samples & 128 & 147 & 207 & 324 & 653 & 1459 \cr
Diffuse ISM & 39 & 35 & 43 & 89 & 146 & 352 \cr
Column density & 34 & 26 & 34 & 58 & 74 & 226 \cr %\nstarsSNRDISMORTHO \cr
Position angles  & 32 & 22 & 31 & 54 & 67 & 206 \cr %\nfitQUcorrBPM \cr
\noalign{\vskip 5pt\hrule\vskip 3pt}
}
}
\endPlancktable %or % for a one-column table; defined in Planck.tex.
\endgroup
\end{table}

\begin{figure}
%?? hard coded to keep label from sticking out
\includegraphics[width=9.0 cm]{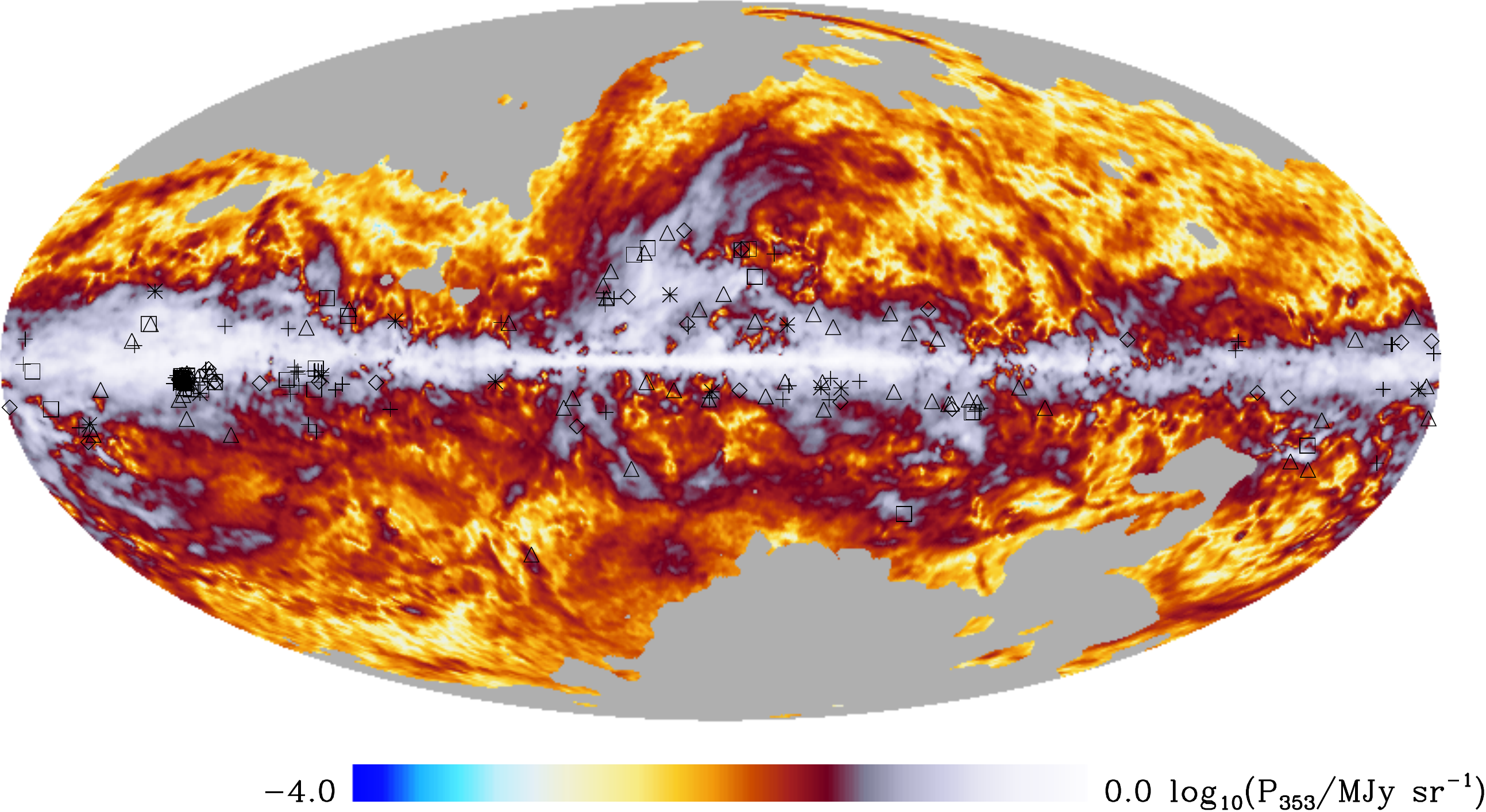}
% larger width for referee mode, but label sticks out --does not matter here
%\includegraphics[width=\hhsize]{\dir glat-glon.pdf}
%
\caption{Galactic lines of sight selected in our \nsamp\ independent
samples: FM07 (squares), VA04 (asterisks), WE23 (diamonds), SA85
(crosses), and KR09 (triangles).  The background image is of debiased
polarized intensity, $\psub$, at a resolution of 1\deg
\citep{planck2014-XIX} in a Mollweide projection centred on the
Galactic centre.  Near Galactic coordinates ($-134\pdeg7$, $-3\pdeg7$)
the selected stars are highly concentrated.  Nevertheless, for $\nside
= 2048$ there is only one pixel containing a pair of stars; there are
two such pixels for $\nside=1024$.}
\label{starmap}
\end{figure}

Figure~\ref{starmap} presents the Galactic coordinates of our selected
stars in the five independent samples. Stars are more concentrated in
some local ISM regions of interest where the polarization fraction in
the submillimetre is known to be high \citep{planck2014-XIX}, in
particular the Auriga-Fan region around $l=135\deg$ and $b=-5\deg$
(accounting for about one third of our sample, see Table
\ref{Table-Geo}), the Aquila Rift around $l=20\deg$ and $b=20\deg$,
and the Ara region at $l=330\deg$ and $b=-5\deg$. The fact that all
extinction catalogues provide data in these regions allows us to study
local variations of $\ratio$ with a limited bias (see
Appendix~\ref{SpatialDep}).

\begin{figure*}
\includegraphics[width=\hhsize]{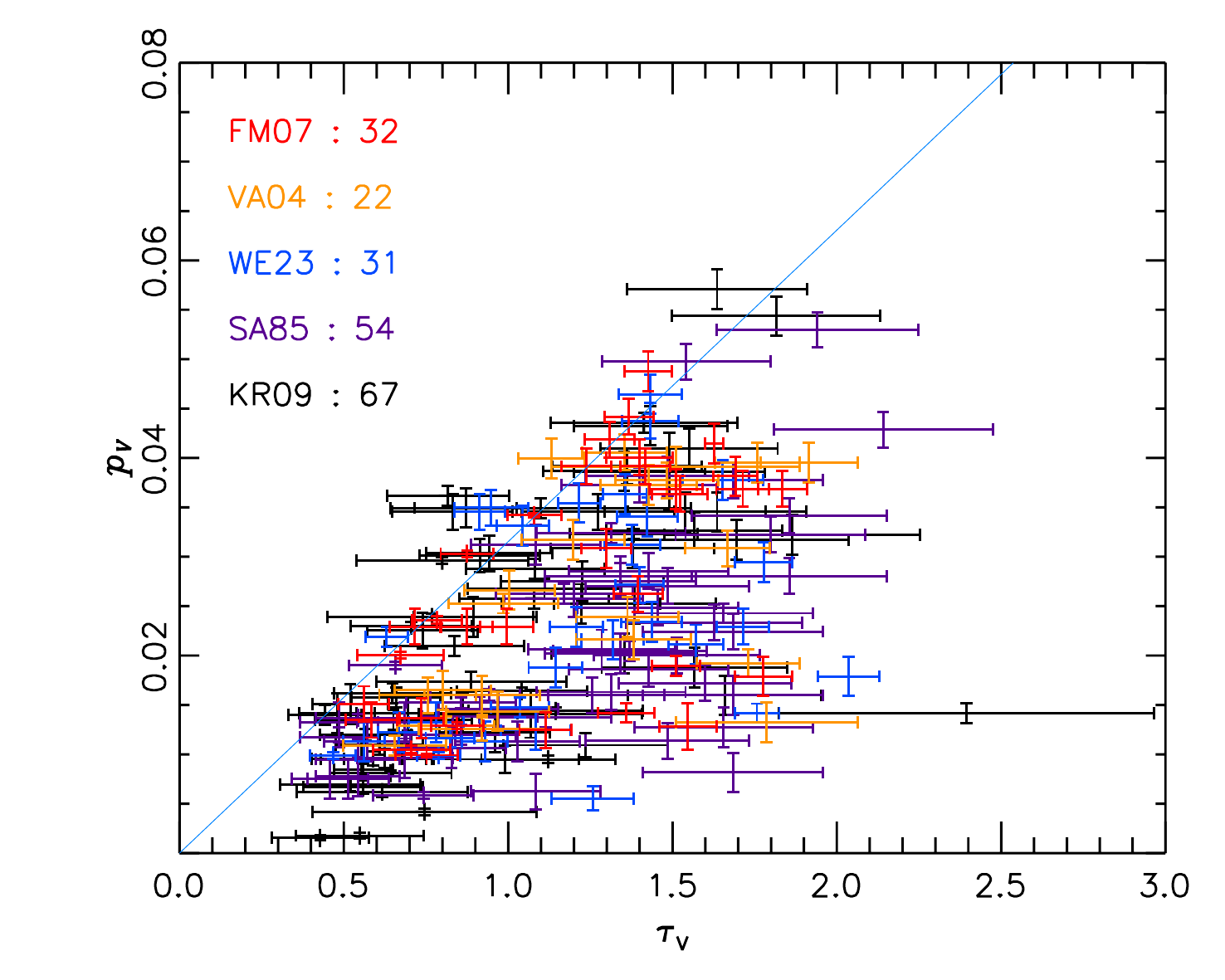}
\includegraphics[width=\hhsize]{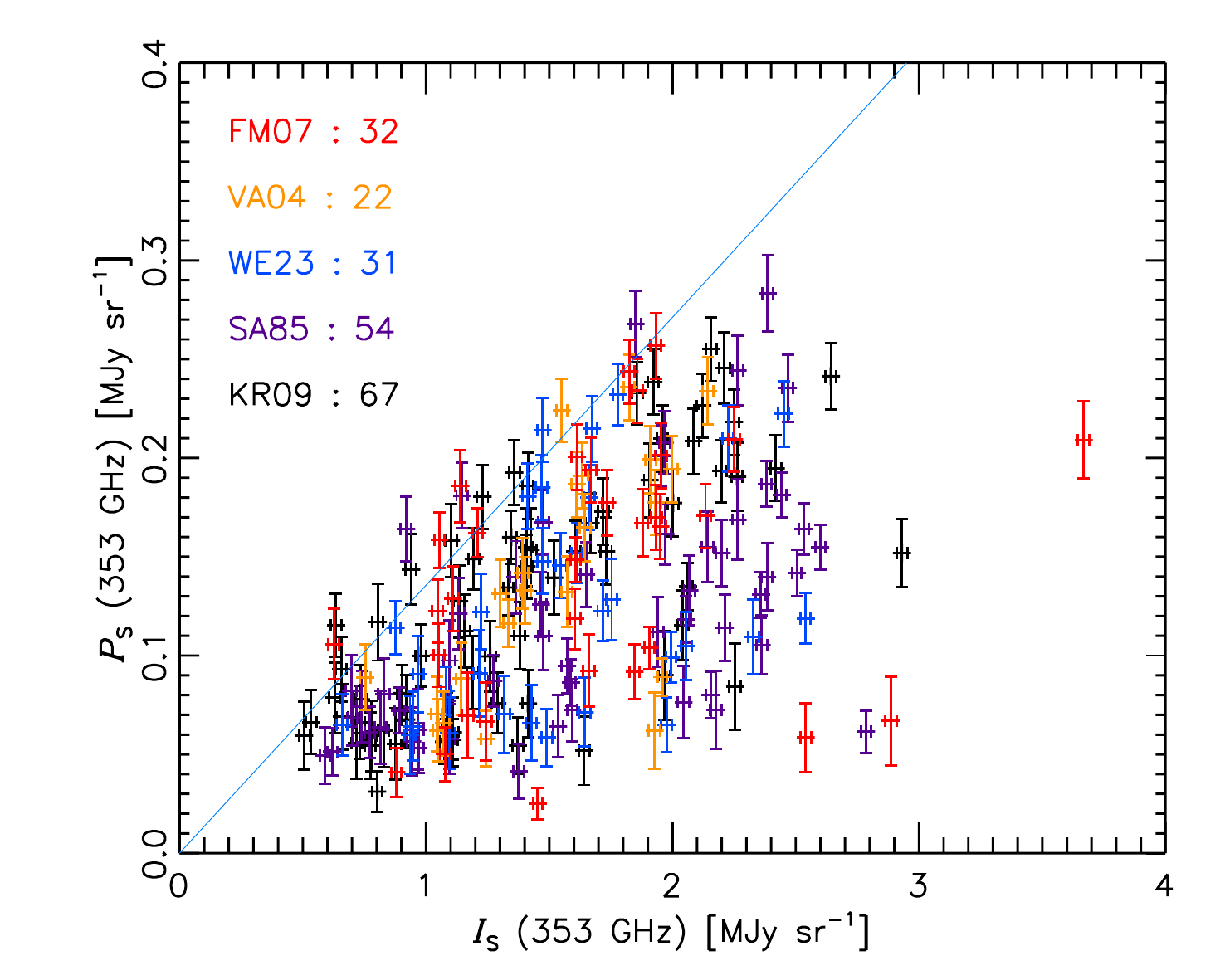}
\caption{Scatterplots of data for the selected lines of sight 
( \LL\ Extinction;  \RR\ Submillimetre emission).
Polarization degree $\pv$ and intensity $\psub$ were debiased
with the Modified Asymptotic method \citep{PLA13}.
{\bfm We note the variable that has the larger error in each plot:} in the \optical, $\tauv$
($x$ axis); in the submillimetre, $\psub$ ($y$ axis).
In the left panel, the line (blue) represents the ``classical" upper
envelope, $\pv=0.0315\,\tauv$ \citep{SMF75}.  This upper envelope has
been transferred to the right panel using the derived value for
$\ratio$.  }
\label{pvstau-pvsi}
\end{figure*}

{\bfc Figure~\ref{pvstau-pvsi} shows scatterplots of the data for the
selected lines of sight.  The similar distributions are discussed in
Appendix~\ref{MaxPolarFraction} in the context of $\ratio$ and its
relationship to the maximum observed polarization fractions indicated
by the envelopes shown.
Contributing factors leading to a polarization fraction below the
upper envelope(s) include dust grains being less aspherical, a lower
grain alignment efficiency, and a suboptimal orientation of the
magnetic field with respect to the line of sight (either systemic from
viewing geometry or through changes of field orientation along the
line sight).
None of these factors is addressed by the selection criteria,
resulting in a rather diverse set of lines of sight as {\bfm shown} in
these scatterplots.
}

%%%%%%%%%%%%%%%%%%%%%%%%%%%%%%%%%%%%%%%%%%%%%%%%%%%%%%%

\section{Estimates of the polarization ratios}\label{Results}

The polarization ratios, $\ratio$ and $\ratiop$, defined in
Eqs.~\ref{Eq-ratio} and \ref{Eq-ratiop}, respectively, can in
principle be obtained by correlating $\psub$ with $\pv$ and $\PsI$
with $\pst$, respectively.  However, both the submillimetre polarized
intensity, $\psub = \sqrt{\qsub^2+\usub^2}$, and the polarization
degree, $\pv$, are derived non-linearly from the original data, the
Stokes parameters.  In the presence of errors, these are biased
estimates of the true values (\citealp{S58,WK74,SS85}; see also
\citealp{Q12,PLA13,planck2014-XIX} and references therein).
The polarization ratio, $\ratiop$, and the polarization fractions
$\PsI$ and $\pv/\tauv$ -- thus also the polarization ratio, $\ratio$
-- would be affected by the same problem.  We revisit this in
Appendix~\ref{CorrP}, but use the original data here.

%==============================================

\subsection{Correlation plots in $Q$ and $U$ for an unbiased
  estimate}\label{CorrelationPlots}

In the ideal case where noise is negligible and the polarization
pseudo-vectors in extinction and emission are orthogonal, from
Eqs.~\ref{Eq-Gem} and \ref{Eq-qv-uv} we
have\footnote{Eq.~\ref{Eq-qv-uv} changes the signs of both $\qv$ and
$\uv$ when the position angle $\Gex$ is changed by $90\deg$.}
$\qsub/\psub=-\qv/\pv$ and $\usub/\psub=-\uv/\pv$, which yields
\begin{equation}
\qsub= - \frac{\psub}{\pv} \, \qv= - \ratiop \,\qv \,,
\end{equation}
and the same for $U$.  Introducing $\isub$ and $\tauv$ in the
denominator on the left and right, respectively, and rearranging
slightly, we obtain similarly
\begin{equation}
\QsI= - \frac{\PsI}{\pst} \, \qst= - \ratio \,\qst \,,
\end{equation}
and the same for $U$. Therefore, the polarization ratios can be
measured not only by correlating $\psub$ with $\pv$ and $\PsI$ with
$\pst$, but also by correlating their projections in $Q$ and $U$.

We correlate in $Q$ and $U$ first separately, and then jointly.  This
approach has several advantages.
First, while $\psub$, $\pv$, $\PsI$ and $\pst$ are biased, their
equivalents in $Q$ and $U$ are not biased.\footnote{
Analysis using $Q$ and $U$ makes it
possible to skip the \SNR\ criteria for $\pv$ and $\psub$, but not for
$\Av$ and $\isub$, as is explored in
Appendix~\ref{robcriteria}.
}

\begin{figure*}
\includegraphics[width=\hhsize]{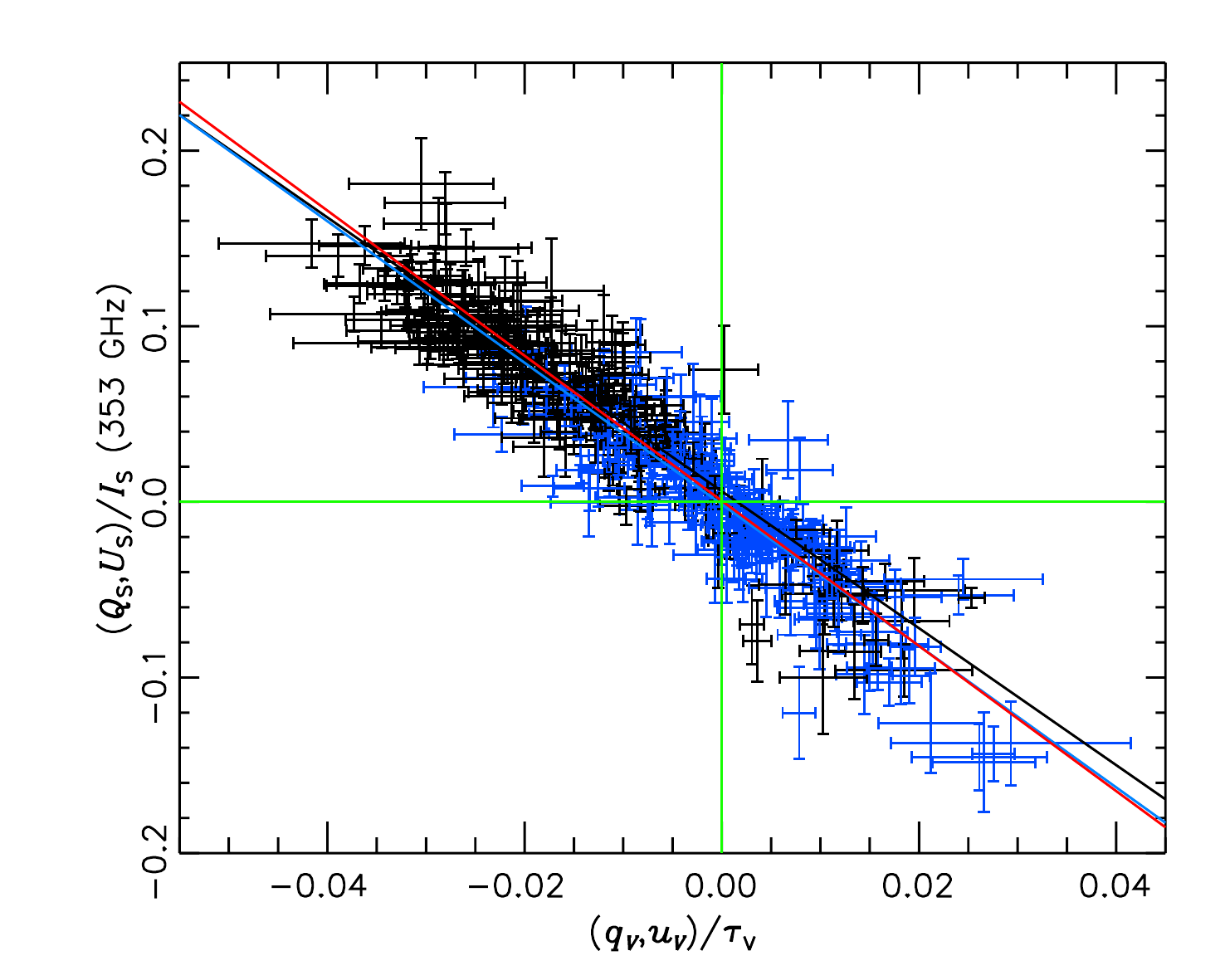}
\includegraphics[width=\hhsize]{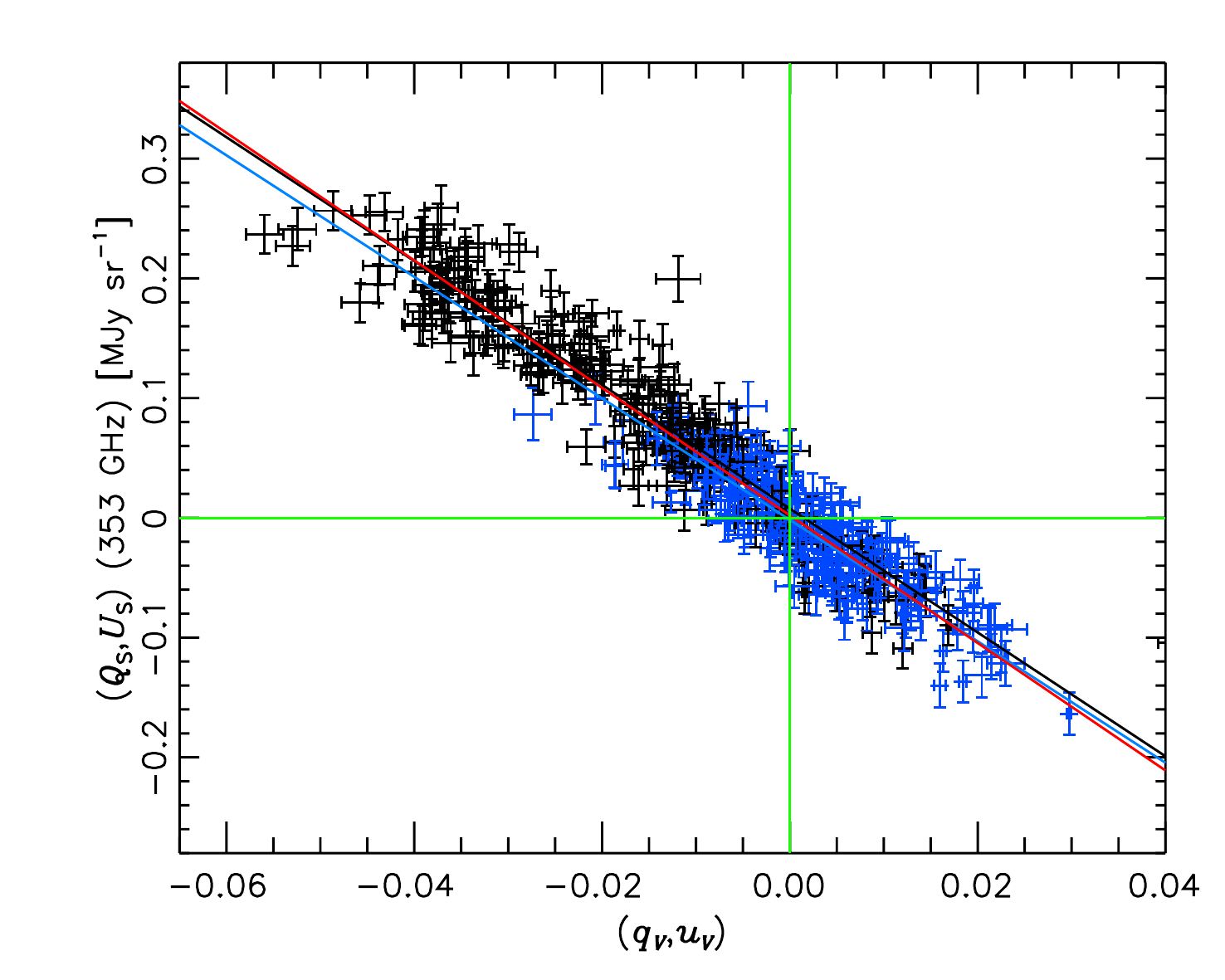}
\caption{\LL\ Correlation of polarization fractions in emission
with those in extinction for the joint fit in $Q$ (black) and $U$
(blue).  Using Eq.~\ref{Eq-QsIUsI} the best linear fit (red line) has
slope and $y$-intercept $\RfitQUcorrBPM\pm\dRfitQUcorrBPM$ and
$\bfitQUcorrBPM\pm\dbfitQUcorrBPM$, respectively.  The Pearson
correlation coefficient is $\PearsonfitQUcorrBPM$ and $\chi^2_{\rm
r}=\chifitQUcorrBPM$.
\RR\ Correlation of polarized intensity in emission (\unitPp) with the
degree of interstellar polarization. Using Eq.~\ref{Eq-QU}, the best
linear fit (red line) has slope and $y$-intercept
$(\RfitPpQUcorrBPM\pm\dRfitPpQUcorrBPM)$\,\unitPp\ and
$(\bfitPpQUcorrBPM\pm\dbfitPpQUcorrBPM)$\,\unitPp, respectively.  The
Pearson correlation coefficient is $\PearsonfitPpQUcorrBPM$ and
$\chi^2_{\rm r}=\chifitPpQUcorrBPM$.  Lines for the independent fits
to $Q$ (black) and $U$ (blue) are also shown.}
\label{qsiusi_qstust}
\end{figure*}

Second, the data in $Q$ and $U$ each present a better dynamic range
than in $P$, because they can be both positive and negative and
because they can vary from line of sight to line of sight if the
position angle $\polang$ changes, even while $P$ and $p$ (or
equivalently $P/I$ or $p/\tau$) remain fairly constant. This allows
for a better definition of the correlation\footnote{
Comparing Figs.~\ref{qsiusi_qstust} and \ref{psi_pst-biased}, the
Pearson correlation coefficients for the $\ratio$ fit are
$\PearsonfitQcorrBPM$ in $Q$ and $\PearsonfitUcorrBPM$ in $U$ as
opposed to $\PearsonfitPcorrBPM$ in $P$, and for the $\ratiop$ fit are
$\PearsonfitPpQcorrBPM$ in $Q$ and $\PearsonfitPpUcorrBPM$ in $U$ as
opposed to $\PearsonfitPpPcorrBPM$ in $P$.
}
and hence a better constraint on the slope, \ie\ the polarization
ratio.

Third, we can obtain two independent estimates of the polarization
ratios from the slope of separate correlations for $Q$ and for $U$.
Under our hypothesis that for the samples of stars being selected the
measured polarization in emission and extinction arises from the same
aligned grains, these two estimates of the polarization ratio ought to
be the same and the intercepts ought to be close to zero. 
This is what we find.
For the $Q$ and $U$ independent correlations analyzed using the scalar
equivalent of Eq.~\ref{vector} the slopes of the $\ratio$ fit
are, respectively,  $\RfitQcorrBPM\pm\dRfitQcorrBPM$ and
$\RfitUcorrBPM\pm\dRfitUcorrBPM$ and the $y$-intercepts are
$\bfitQcorrBPM\pm\dbfitQcorrBPM$ and $\bfitUcorrBPM\pm\dbfitUcorrBPM$.
For the $\ratiop$ fit, the slopes are $(\RfitPpQcorrBPM\pm\dRfitPpQcorrBPM)$ \unitPp\ and
$(\RfitPpUcorrBPM\pm\dRfitPpUcorrBPM)$ \unitPp\ and the $y$-intercepts are
$(\bfitPpQcorrBPM\pm\dbfitPpQcorrBPM)$ \unitPp\ and
$(\bfitPpUcorrBPM\pm\dbfitPpUcorrBPM)$ \unitPp.
In both cases the $y$-intercepts are small compared to the dynamic
range in $Q$ and $U$ (see Fig.~\ref{qsiusi_qstust}).
The uncertainties quoted were derived in the standard way from the
quality of the fit.
As reinforced by our bootstrapping analysis below
(\Sect~\ref{meanratio}), the results of these independent fits are
compatible with our hypothesis that the two correlations are measuring
the same phenomenon, and furthermore reflect the quality of the
selected data.  See also Appendix~\ref{CorrP} for a comparison with
the fits in $P$.

Fourth, given this satisfactory consistency check, measuring the
polarization ratios from the correlation of the joint data $(Q,U)$ is
both motivated and justified.
We compute the linear ($y=ax+b$) best fit to the data by minimizing a
$\chi^2$, which for the joint fit has the form\footnote{
The $x$ and $y$ coordinates can be inverted in the fitting routine
without affecting the measure of the polarization ratio.}
\begin{equation}
\chi^2(a,b) = \sum_i V(a,b)\,M(a,b)^{-1}\,V(a,b)^\Trans \,, \nonumber \\
\label{Eq-chi2}
\end{equation}
with
\begin{eqnarray}
\label{vector}
V(a,b) & = & \left( \qsub/\isub-a\,\qv/\tauv-b, \usub/\isub -a\,\uv/\tauv -b \right) \,,  \nonumber \\
M(a,b) & = & \left( \begin{array}{cc}
\sigQsIQsI +a^2\dqst^2& \sigQsIUsI \\ 
\sigQsIUsI & \sigUsIUsI+a^2\dust^2 \, 
\end{array} \right)\,,
\label{Eq-QsIUsI}
\end{eqnarray}
for the $\ratio$ fit, and
\begin{eqnarray}
V(a,b) & = & \left( \qsub-a\,\qv-b, \usub -a\,\uv -b \right) \,,  \nonumber \\
M(a,b) & = & \left( \begin{array}{cc}
\sigQQ +a^2\dqv^2& \sigQU \\ 
\sigQU & \sigUU+a^2\duv^2 \, 
\end{array} \right)\,,
\label{Eq-QU}
\end{eqnarray}
for the $\ratiop$ fit.
The calculation of the elements $\covar$ of the noise covariance
matrix, $\MatC$, for the \Planck\ data is presented in
Appendix~\ref{AppUncertainty}.  The slopes of the joint
correlation are $\RfitQUcorrBPM\pm\dRfitQUcorrBPM$ and
$(\RfitPpQUcorrBPM\pm\dRfitPpQUcorrBPM)$ \unitPp, respectively.

{\bfc Figure~\ref{qsiusi_qstust} shows that the joint correlations are
  remarkably tight.  } The Pearson correlation coefficients are
$\PearsonfitQUcorrBPM$ for $\ratio$ and $\PearsonfitPpQUcorrBPM$ for
$\ratiop$.  Values of the reduced $\chi^2$ of the fit ($\chi^2_{\rm r}
= \chifitQUcorrBPM$ for $\ratio$ and $\chifitPpQUcorrBPM$ for
$\ratiop$) are higher than expected given the large number of degrees
of freedom ($\dof$), in part because the noise covariance matrix does
not capture the systematic errors in the data, primarily from the
leakage correction (see \Sect~\ref{planckdata}).
{\bfc Nevertheless, the tight correlation in Fig.~\ref{qsiusi_qstust}
is in sharp contrast to the scatter among the underlying observables
in Fig.~\ref{pvstau-pvsi}.  The contributing factors that move points
below the upper envelopes in Fig.~\ref{pvstau-pvsi} move points toward
the origin along the relevant \optical\ (horizontal) and submillimetre
(vertical) axes in Fig.~\ref{qsiusi_qstust}, left, and actually along
the correlation line toward the origin if the changes in the
submillimetre and \optical\ polarization fractions are related by the
same $\ratio$ for all lines of sight, as they evidently are.  Similar
comments apply to Fig.~\ref{qsiusi_qstust}, right.}
As a complement, we show in Appendix~\ref{MaxPolarFraction} that a
statistical analysis of the maximum polarization fractions seen in the
\optical\ and at 353\GHz\ gives a result consistent with $\ratio$.

These good correlations confirm our initial idea that the polarization
ratios can be obtained without limiting the analysis to the case of
optimal alignment (magnetic field in the plane of the sky, perfect
alignment), and that the dependences of polarization on the magnetic
field orientation and on the dust alignment efficiency are similar in
emission and in extinction.
{\bfc We conclude that $\ratio$ and $\ratiop$ are each characterizing
a property of the dust populations that is homogeneous across a
diverse set of lines of sight in the diffuse interstellar medium.
}

%==============================================

\subsection{Mean values and uncertainties for $\ratio$ and $\ratiop$ in the diffuse ISM}\label{meanratio}

Because of the finite sample size and potential sensitivity to the
exact membership in the samples, we used the technique of
bootstrapping \citep{Efron93}, in particular random sampling with
replacement or case resampling, to carry out many instances of the
fit, and then from these solutions we calculated the mean slope, the
mean intercept, and their dispersions.  The number of trials, 500, was
large enough to ensure convergence of the results of resampling. 

For all of the fits -- $\ratio$ and $\ratiop$ and $Q$, $U$, and joint
-- we find the same slopes using bootstrapping as we found in the
direct fits.  The uncertainties are somewhat more conservative, by up
to a factor {\bfm of 4} for $Q$.
In the rest of this paper we report the results from bootstrapping in
terms of $\ratio$ and $\ratiop$ (the negative of the values of the
slopes) and to be conservative the dispersions were rounded to the
upper first decimal (\eg\ 0.12 gives 0.2) to give the statistical
uncertainties quoted below.

{\bfc
In Appendix~\ref{rob} we investigated
} 
the robustness of the polarization ratios $\ratio$ and $\ratiop$ with
respect to the selection criteria defining the sample, the data used,
the region analyzed, and the methodology.  We derived each time the
mean values and uncertainties of the polarization ratios.
{\bfc
We showed that all systematic variations of $\ratio$ and $\ratiop$ are
small, less than twice the statistical uncertainty derived from the fit.
}

On the basis of these results we adopt
\begin{equation}
\ratio=  (\PsI)/(\pst) = \resr\pm\ress\ (\stat)\pm\resscombined\ (\sys)\,.
\end{equation}
This appears to be a homogeneous characterization of diverse lines of
sight in the diffuse ISM.  The statistical uncertainty is
representative of the standard deviation of the histogram of $\ratio$
found by bootstrap analysis in the many robustness tests
(Appendix~\ref{rob}).  The systematic error gathers all potential
contributions, but is dominated by our incomplete knowledge of the
small correction of \Planck\ data for leakage of intensity into
polarization (\Sect~\ref{planckdata}).

For the other polarization ratio we adopt
\begin{equation}
\ratiop=\psub / \pv = \left[\resrp\pm\ressp\ (\stat)\pm\resspcombined\ (\sys)\right]\,{\rm MJy\,sr^{-1}}\,.
\end{equation}
The statistical uncertainty from the bootstrap analyses is again rather
small. Uncertainties in $\ratiop$ are dominated by systematic errors,
namely from uncertainties in the leakage correction and from the
tendency to overestimate $\ratiop$ in the presence of backgrounds
beyond the stars.

The corresponding total intensity per unit magnitude in the \Vband\
band, $\ratioI$, is obtained by the direct calculation
$\ratioI=\ratiop/\ratio / 1.086 = \resri$\,\unitPp, with an uncertainty
$\pm\,\ressi$\,\unitPp.

To help in constraining dust models, we provide the mean SED for the
lines of sight in our sample in the form of a modified blackbody fit
to the data \citep{planck2013-p06b}: $\Tdust=\avgTdust$\,K;
$\beta_{\rm FIR}=\avgbeta$; and a dust opacity at 353\GHz,
$\tau_\sub/\Av=\avgTausAv$\,mag$^{-1}$.

%%%%%%%%%%%%%%%%%%%%%%%%%%%%%%%%%%%%%%%%%%%%%%%%%%%%%%%

\section{Discussion}\label{discussion}

Our measurement of the polarization ratios $\ratio$ and $\ratiop$
provide new constraints on the submillimetre properties of dust. In
fact, in anticipation of results from \Planck, $\ratio$ for the
diffuse ISM has already been predicted in two studies.

%==============================================

\subsection{$\ratio$ from dust models}\label{modelRSV}

\citet{M07} discussed how $\PsI$ at 350\GHz\ can be predicted by
making use of what is already well known from dust models of \optical\
(and infrared and ultraviolet) interstellar polarization and
extinction.  From that study, the polarization ratio of interest here,
$\ratio$, can be recovered by dividing the estimated values of $\PsI$
in the last row of his 
Table~2, first by 100 (to convert from percentages) and then by
0.0267, the value of $\pst$ on which the table was based.  For
example, for aligned silicates in the form of spheroids of axial ratio
$1.4$, the entry is 9.3, so that $\ratio = 3.5$.  Similar values are
obtained for other shapes and axial ratios, because changes in these
factors affect polarization in both the \optical\ and submillimetre.
Sources of uncertainty in this prediction include: whether imperfect
alignment reduces the polarization equally in the \optical\ and
submillimetre (as was assumed); how the extinction in the \optical\,
including contributions by unaligned grains, is modelled, whether by
aligned or by randomly oriented grains ($\pm 25$\,\%); and the amount
by which the {\bfm submillimetre} polarization fraction is diluted by
unpolarized emission from carbonaceous grains in the model (adopting a
dilution closer to that in the \cite{DF09} mixture discussed below
could raise $\ratio $ to about 4.5).

\cite{DF09} also made predictions of $\PsI$ for mixtures of silicate
and carbonaceous grains, where again the dust model parameters (size
distributions, composition, alignment, etc.) were constrained by a
detailed match to the infrared to ultraviolet extinction and
polarization curves, with the normalization $\pst = 0.0326$.
According to their Fig.~8, at 350\GHz\
$\PsI$ is about 13--14\,\% for models in which only the silicate
grains are aspherical and aligned (axial ratio 1.4--1.6),\footnote{
{\bfm This} prediction of 13--14\,\% for $\PsI$ when graphite
grains are spherical is overestimated. As can been seen from Fig. 7 in
\cite{DF09}, the models with spherical graphite grains are about
30\,\% lower in intensity than the models with aligned oblate graphite
grains. This discrepancy alone explains the 30\,\% difference between
the models for $\PsI$. Once corrected, all these models would predict
$\PsI$ of about 10\,\%.
} 
and about 9--10\,\% for models in which both silicate and carbonaceous
grains are aspherical and aligned.
Dividing as above yields $\ratio = 4.1$ and 2.9, respectively.  

These are challenging calculations that encounter similar issues to
those discussed in \citet{M07}.  Given the uncertainties, we conclude
that the predictions of $\ratio$ are in reasonable agreement with what
has now been observed,
{\bfc
providing empirical validation of many of the common basic
assumptions underlying polarizing grain models.
Although validating the basic tenets of the models, the new empirical
results on $\ratio$ do not allow a choice between different models.
}

%==============================================

\subsection{$\ratiop$ and $\ratioI$ from dust models}

The polarization ratio $\ratiop$ is a more direct and easier-to-model
observational constraint. Unlike for $\ratio$, one does not need to
model the extinction and emission of non-aligned or spherical grain
populations.  Model predictions for $\ratiop$ will depend not only on
the size distribution, optical properties, and shape of the aligned
grain population, but also on the radiation field intensity 
{\bfc
through the dependence of $\psub$ on the grain temperature. 
}
Because $\ratiop$ is not a dimensionless quantity like $\ratio$, it
provides a new constraint on grain models, i.e., models that are able
to reproduce $\ratio$ will not automatically satisfy $\ratiop$.

From the \cite{DF09} prediction of $\psub/\NH$ we measure
$\nu_\sub\psub/\NH \simeq 1.0 \times 10^{-11}$ W\,sr$^{-1}$ per H at
353\GHz.  In the \optical\ the same models produce $\pv / \NH \simeq
1.4\times10^{-23}$ cm$^2$ per H (their Fig.~6, taking into account
their Erratum).  These can be combined to give $\ratiop
\simeq\PpDF$\,\unitPp.  Our empirical results are significantly
higher, by a factor {\bfm of} $\factorPpDF$,
{\bfc
and so represent a considerable challenge to these first polarizing
grain models. 
A basic conclusion is that the optical properties of the materials in the
model need to be adjusted so that the grains are more emissive.
}

As this example illustrates, there is great diagnostic
power in focusing directly on the polarization properties alone;
$\ratiop$ describes the aligned grain population and so is important,
along with $\ratio$, for constraining and understanding the full
complexity of grain models.

Complementing the discussion in Sect.~\ref{modelRSV}, from the
\cite{DF09} prediction of $\isub/\NH$ we measure
$\nu_\sub\isub/\NH\simeq\,$ 0.7--1.1$\,\times10^{-10}$ W\,sr$^{-1}$
per H at 353\GHz.  In the \optical\ the same models produce $\tauv /
\NH~\simeq~4.5\times10^{-22}$ cm$^2$ per H (their Fig.~4).
These can be combined to give $\isub/\tauv \simeq\,$0.5--0.8\,\unitPp,
after a colour correction by 10\,\%, taking into account the width of
the 353\GHz\ \Planck\ band \citep{planck2013-p03d}.
Our empirical results are significantly higher than these models, by a
factor {\bfm of} 1.5--2.4, depending on the model.\footnote{
By fitting \Planck\ intensity maps with the \cite{DL07} dust model,
\citet{planck2014-XXIX} derived $\Av$ maps that were found to be
overestimated compared to data by a factor {\bfm of} $\factorIsAvDL$;
alternatively, normalizing the models to the observed $\Av$, the
submillimetre emission is underpredicted by a factor {\bfm of} $\factorIsAvDL$,
like in \cite{DF09}.}
{\bfc
Again, this is in the sense that the grains need to be more emissive,
though the magnitude of the discrepancy is somewhat lower than it is
for polarized emission.
}

%==============================================

%%%%%%%%%%%%%%%%%%%%%%%%%%%%%%%%%%%%%%%%%%%%%%%

\section{Conclusion}\label{con}

Comparison of submillimetre polarization, as seen by \Planck\ at
353\GHz, with interstellar polarization, as measured in the \optical,
has allowed us to provide new constraints relevant to dust models for
the diffuse ISM. After carefully selecting lines of sight in the
diffuse ISM suitable for this comparison, a correlation analysis
showed that the mean polarization ratio, defined as the ratio between
the polarization fractions in the submillimetre and the \optical, is
\begin{equation}
\label{ratiosummary}
\ratio=  (\PsI)/(\pst) = \resr\pm\ress\ (\stat)\pm\resscombined\ (\sys) \, ,
\end{equation}
where the statistical uncertainty is from the bootstrap analysis
(\Sect~\ref{meanratio}) and the systematic error is dominated
(Appendix~\ref{RobDataUsed}) by our incomplete knowledge of the small
correction of \Planck\ data for leakage of intensity into polarization
(\Sect~\ref{planckdata}).

Similarly we found the ratio between the polarized intensity in the
submillimetre and the degree of polarization in the \optical:
\begin{equation}
\label{ratioPsummary}
\ratiop=  \psub / \pv=\left[\resrp\pm\ressp\ (\stat)\pm\resspcombined\ (\sys)\right]\,{\rm MJy\,sr^{-1}}\,.
\end{equation}

This analysis using the new \Planck\ polarization data suggests that
the measured $\ratio$ is compatible with a range of polarizing dust
models, validating the basic assumptions, but not yet very
discriminating among them.
By contrast, the measured $\ratiop$ is higher than model predictions
by a factor of about 2.5.
{\bfc
To rectify this in the dust models, changes will be needed in the
optical properties of the materials making up the large polarizing
grains that are emitting in thermal equilibrium.
}
Thus, the simpler polarization ratio $\ratiop$ turns out to provide a
more stringent constraint on dust models than $\ratio$.

Future dust models are needed that will satisfy the constraints
provided by both $\ratio$ and $\ratiop$, as well as by the spectral
dependencies of polarization in both the \optical\ and submillimetre.  
{\bfc 
How the optical properties of the aligned grain population
(including the silicates) should be revised in the submillimetre
and/or \optical\ needs to be investigated through such detailed modeling.
}
Understanding the polarized intensity from thermal dust will be
important in refining the separation of this contamination of the CMB.

%%%%%%%%%%% ORDER: Main, Acknowledgements, References, Appendix

\begin{acknowledgements}

The development of \Planck\ has been supported by: ESA; CNES and
CNRS/INSU-IN2P3-INP (France); ASI, CNR, and INAF (Italy); NASA and DoE
(USA); STFC and UKSA (UK); CSIC, MICINN, JA and RES (Spain); Tekes,
AoF and CSC (Finland); DLR and MPG (Germany); CSA (Canada); DTU Space
(Denmark); SER/SSO (Switzerland); RCN (Norway); SFI (Ireland);
FCT/MCTES (Portugal); and PRACE (EU). A description of the Planck
Collaboration and a list of its members, including the technical or
scientific activities in which they have been involved, can be found
at
\url{http://www.sciops.esa.int/index.php?project=planck&page=Planck_Collaboration}.
The research leading to these results has received funding from the
European Research Council under the European Union's Seventh Framework
Programme (FP7/2007-2013) / ERC grant agreement no. 267934.
This research has made use of the SIMBAD database and the VizieR
catalogue access tool, operated at CDS, Strasbourg, France, and NASA's
Astrophysics Data System Service.

\end{acknowledgements}

%%%%%%%%%%%%%%%%%%%%%%%%%%%%%%%%%%%%%%%%%%%%%%%

\bibliographystyle{aa}
% in order of precedence, these should be as follows
% biblio_[some date] as generated by Papers
% biblio_pgm some additions not in above, from pgm

% Planck_bib the master Planck bibliography file; grab most recent update!!
% https://scisvn01.esac.esa.int/Planck_Publication_Management/Repositories/BibTeX/

% Planck_bib_4polar, a mock-up of the forthcoming entries in the first
% 4 polarization papers; REMOVE when Planck_bib contains these

% Planck_bib_thermaldustentries NEW or planned Planck papers not yet in the above master file

% bootstrap

{\raggedright
%\bibliography{biblio_7fev,biblio_pgm,Planck_bib,Planck_bib_otherPIPs_9jul,Planck_bib_thermaldustentries,bootstrap,referee}
\bibliography{biblio_7fev,biblio_pgm,Planck_bib,Planck_bib_thermaldustentries,bootstrap,referee}
}

%%%%%%%%%%%%%%%%%%%%%%%%%%%%%%%%%%%%%%%%%%%%%%%

\begin{appendix}

%%%%%%%%%%%%%%%%%%%%%%%%%%%%%%%%%%%%%%%%%%%%%%%%%%%%%%%

%%==============================================

\section{Uncertainties for polarization quantities}\label{AppUncertainty}

We start with the values of the Stokes parameters, $I$, $Q$, and $U$,
and the noise covariance matrix, $\MatC$, at the position of each
star.  By definition, the uncertainties $\dqsub$ of $\qsub$ and
$\dusub$ of $\usub$ are calculated from the variances $\dqsub^2 \equiv
\sigQQ$ and $\dusub^2 \equiv \sigUU$, respectively.

The variance of the polarized intensity
$\psub\equiv\sqrt{\qsub^2+\usub^2}$ (without bias correction) is
\begin{eqnarray}
\dpsub^2 & = &  \left<\left(\frac{\partial P}{\partial Q}\,\delta Q+\frac{\partial P}{\partial U}\,\delta U\right)^2\right> \nonumber \\
& = & \frac{\qsub^2\,\sigQQ+\usub^2\,\sigUU+2\,\qsub\,\usub\sigQU}{\psub^2} \, ,
\end{eqnarray}
from which we derive the $\SNR$ of the polarization intensity,
$P/\sigma_P$.  For the uncertainty in the position angle $\Gem$ we use
the approximate formula in Eq.~\ref{phierror}, which is appropriate
because we always require the \SNR\ to be larger than 3.\footnote{The
uncertainties in $\psub$ and $\Gem$ are used just for the selection
process and for plotting, not for the fitting, which involves the full
noise covariance matrix; see \Sect~\ref{CorrelationPlots}.}

The uncertainty of the projected polarization fraction $\QsI$ has the
following dependence:
\begin{equation}
\delta\left(\frac{Q}{I}\right) = \frac{\partial \left(Q/I\right)}{\partial Q} \delta Q +  \frac{\partial \left(Q/I\right)}{\partial I} \delta I  = \frac{1}{I}\delta Q - \frac{Q}{I^2}\delta I \,,
\end{equation}
and the same holds for $\UsI$.  The covariance of $\QsI$ with $\UsI$
follows:
\begin{eqnarray}
\sigQsIUsI & = & \left< \,\delta\left(\frac{Q}{I}\right)\,\delta\left(\frac{U}{I}\right)\,\right>  \nonumber \\
& =& \left<\, \frac{1}{I^2}\delta Q\delta U +\frac{QU}{I^4}\delta I^2 - \frac{U}{I^3}\delta I \delta Q - \frac{Q}{I^3}\delta I \delta U \,\right>  \nonumber \\
& = & \frac{\isub^2\,\sigQU+\qsub\,\usub\,\sigII-\isub\,\qsub\,\sigIU-\isub\,\usub\,\sigIQ}{\isub^4}\,.
 \end{eqnarray}
The uncertainties $\sigma_{\QsI}$ and $\sigma_{\UsI}$, used for
plotting only, are derived simply from the variances $\sigQsIQsI$ and
$\sigUsIUsI$.

When the data are smoothed, with the method described in Appendix~A of
\cite{planck2014-XIX},\footnote{The coordinates of the central pixel,
$\mathrm{J}$, of the smoothing beam are replaced by those of the
star.} the formulae hold substituting the smoothed Stokes parameters
and the elements of the corresponding covariance matrix, [$\Scovar$].

%%%%%%%%%%%%%%%%%%%%%%%%%%%%%%%%%%%%%%%%%%%%%%%%%%%%%%%

\section{Extinction catalogues}\label{ExtCat}

\subsection{Extinction and colour excess catalogues}\label{ecat}

Using stellar atmosphere models, \citet{FM07} provide well-determined
$\ebv$ and $\Av$ measurements for 328 stars, 14 of which could be
identified in the \citet{H00} polarization catalogue \textit{via}
their catalogue identifiers (HD -- Henry Draper, BD -- Bonner
Durchmusterung, CD -- Cordoba Durchmusterung, or CPD -- Cape
Photographic Durchmusterung identifiers).  This was the basis for what
we refer to as the ``FM07 sample."  Similarly, from the
\cite{VA04} and \cite{WE02,WE03} extinction catalogues (derived with
the more standard technique based on ``unreddened" reference stars),
we generated the VA04 and WE23 samples; {\bfm we} note that we have removed
stars in common with previously-defined samples, with the same order
in priority of the samples.  These three samples all contain
measurements of both $\Av$ and $\ebv$, providing an estimate of $\Rv$,
a useful diagnostic of the diffuse ISM where $\Rv$ is close to 3.1.

The \cite{SA85} catalogue provides measures of $\ebv$ to 1415 stars,
1085 of which were identified in the \cite{H00} catalogue.  Lacking a
measure of $\Rv$, we assumed the standard value for the diffuse ISM,
$\Rv = 3.1$; its uncertainty $\delta\Rv =
\dRvNotMeasured$\footnote{$\delta\Rv$ was set equal to the standard
deviation of $\Rv$ in our final selected samples FM07, VA04, and
WE23.}  adds another uncertainty to our estimate of $\tauv$. Again
removing stars in common with previous samples, we built the SA85
sample.

\subsection{Deriving $\ebv$ from the Kharchenko \& Roeser star catalogue}\label{KR09cat}

\begin{figure}[h]
\includegraphics[width=\hhsize]{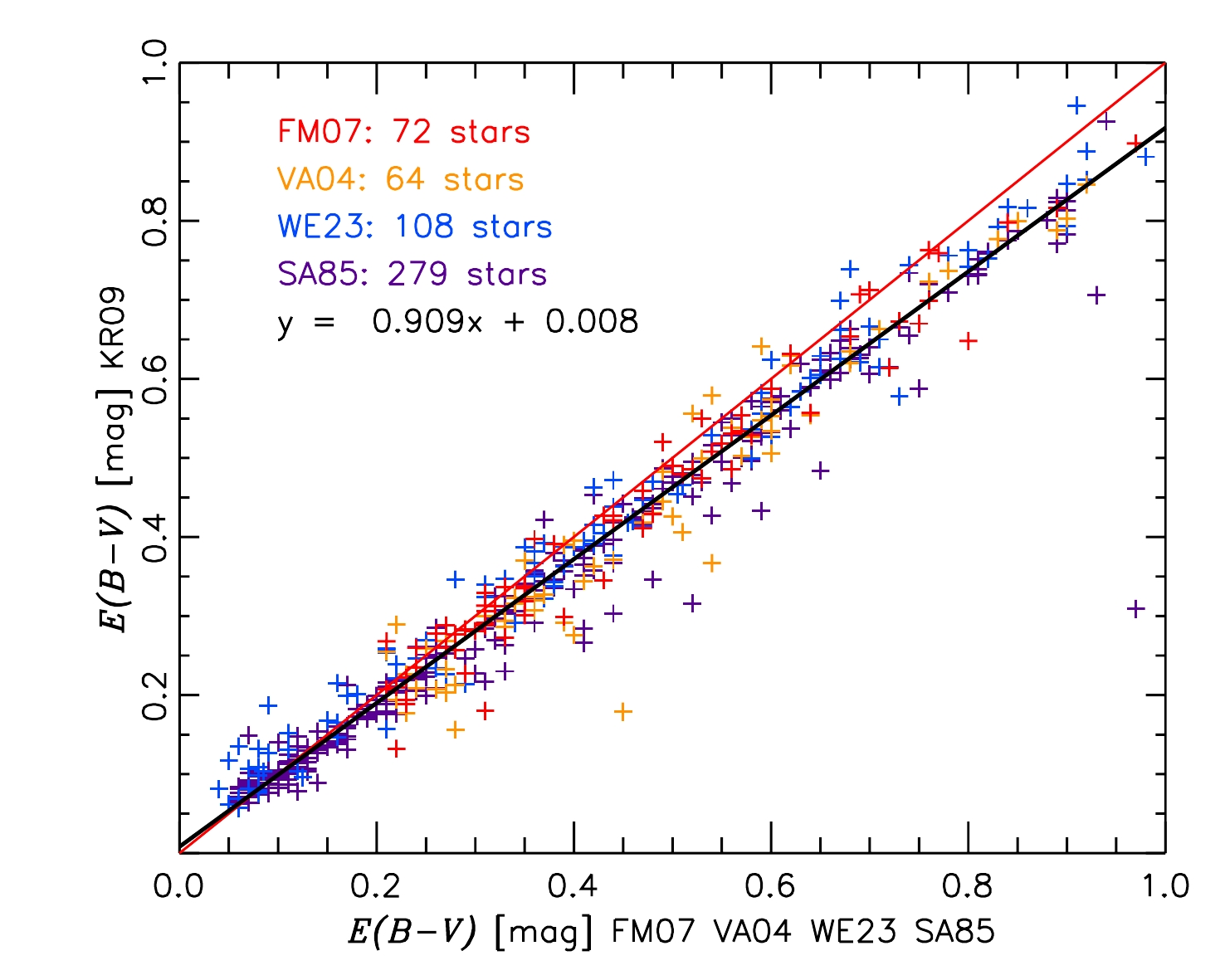}
\caption{Correlation between our derived $\ebv$ from the \cite{K09}
catalogue with that for common stars in the FM07, VA04, and WE23
samples.  Only data with $\SNR$ on $\ebv$ larger than 3 are
presented. The red line is a 1:1 correlation, while the black line is
a fit.}
\label{Fig-compareAv}
\end{figure}

As described below, using the high-quality photometry in the all-sky
catalogue of \cite{K09} we were able to derive $\ebv$ and its
uncertainty for more than 3000 stars present in the \cite{H00}
catalogue.  Stars absent from other samples then form the KR09 sample.
As for the SA85 catalogue, we assumed $\Rv = 3.1\pm\dRvNotMeasured$
for all stars.

The \cite{K09} catalogue (SIMBAD reference code I/280 B) used for the
derivation of $\ebv$ is a compilation of space and ground-based
observational data for more than 2.5 million stars. The catalogued
data include, among others, \Bband\ and \Vband\ magnitudes in the
Johnson system, \Kband\ magnitude, HD number, and the spectral type
and luminosity class of the star. 
TOPCAT\footnote{\url{http://www.starlink.ac.uk/topcat/}} 
\citep{taylor2005} was used to cross-match the
polarization \citep{H00} and extinction \citep{K09} catalogues with
the HD number as the first-order criterion. Where two or more stars
are identified with the same number, the one for which the visual
magnitude is closest to that in the polarization catalogue was
retained. For the rest of the catalogue, a coordinate-match criterion
with a 2\arcsec\ radius was applied, using the visual magnitude to
choose between candidates if necessary.

We have used the intrinsic colors $(B-V)_0$ derived by \cite{F70} for
different spectral classifications.  The colour excess for each of
star was calculated according to $\ebv=(B-V)_{\rm KR09} - (B-V)_0$,
using the \Bband\ and \Vband\ magnitudes and the corresponding
intrinsic colour deduced from the spectral classification in the
\cite{K09} catalogue.  Following \cite{SA85}, Be stars and stars with
spectral types B8 and B9 were removed.

A few hundred stars overlapping the FM07, VA04, WE23, or SA85 samples
allowed us to check the quality of our derivation of $\ebv$.
Figure~\ref{Fig-compareAv} reveals a good correlation with the other
samples, with our KR09 $\ebv$ tending to underestimate the reddening
to the star by about 9\,\%.
This small discrepancy has only a small impact on the derived
$\ratio$ (the KR09 sample represents one third of our sample, see
Table~\ref{Table-Selection}) and does not affect $\ratiop$.

%%%%%%%%%%%%%%%%%%%%%%%%%%%%%%%%%%%%%%%%%%%%%%%%%%%%%%%

\section{Robustness tests}\label{rob}

From the joint fits and bootstrap analysis of uncertainties in
Sects.~\ref{CorrelationPlots} and \ref{meanratio} we found
$\ratio=\RmainQUcorrBPM\pm\dRmainQUcorrBPM$ and
$\ratiop=(\RPpQUcorrBPM\pm\dRPpQUcorrBPM)$\,\unitPp.
In this appendix we investigate the robustness of these polarization
ratios with respect to the selection criteria defining the sample, the
data used, the region analyzed, and the methodology.  We derive each
time their mean values and uncertainties.

As a potential drawback owing to its simplicity, $\ratiop$ involves
systematic dependences on parameters such as the ambient radiation
field, the submillimetre opacity of aligned grains, and the presence
of a background beyond the star. Because $\psub$ and $\pv$ are
proportional to the column density (of polarizing dust) probed with
their respective observations in the submillimetre and \optical, the
correlation presented in Fig.~\ref{qsiusi_qstust} is potentially
biased (an overestimate) if there is systematically a background
beyond the stars selected (Fig.~\ref{los}).  Thanks to the
normalization of $\psub$ by $\isub$ and $\pv$ by $\tauv$, such
dependences are weakened\footnote{
For multiple grain populations the opacity and $\Tdust$ affecting
$\isub$ could be different than for $\psub$ and so these effects might
not cancel completely in the polarization fraction $\PsI$ used in
$\ratio$.
}
in the analysis of $\ratio$.

%==============================================

\subsection{Selection criteria}\label{robcriteria}

We explore the effects of varying the limits of the four selection
criteria presented in \Sect~\ref{criteria}. We do this one criterion
at a time, with the others unchanged.  We also examine other
alternatives for defining the sample.

%++++++++++++++++++++++++++++++++++++++++++++++

\paragraph{\SNR}
The accuracy of the polarization degree in extinction data is not a
limiting factor because the mean $\SNR$ is about 10 for the selected
stars. Asking for a $\SNR$ threshold higher than \SNRpmin\
(\Sect~\ref{snrcriterion}) for $\psub$ and for $\Av$ could bias our
estimates of $\ratio$, which is proportional to these quantities. It
would also exclude many diffuse regions where such a high \SNR\ cannot
be achieved at 5\arcmin\ resolution. Nevertheless, we find no
significant variation of the polarization ratios when imposing $\SNR >
\PpSNRA$ (\nPpSNRAcorrBPM\ stars, $\ratio =
\RSNRAcorrBPM\pm\dRSNRAcorrBPM$,
$\ratiop=(\RPpSNRAcorrBPM\pm\dRPpSNRAcorrBPM)$ \unitPp) or $\SNR >
\SNRC$ (\nPpSNRCcorrBPM\ stars,
$\ratio=\RSNRCcorrBPM\pm\dRSNRCcorrBPM$, $\ratiop
=(\RPpSNRCcorrBPM\pm\dRPpSNRCcorrBPM)$ \unitPp).

%++++++++++++++++++++++++++++++++++++++++++++++

\paragraph{Diffuse ISM}
The $\ebvsub$ criterion (Eq.~\ref{Eq-EBV} in
\Sect~\ref{diffusecriterion}) is responsible for the removal of lines
of sight toward denser environments or toward the Galactic plane.
Ignoring this criterion so that these stars are included gives
$\ratio=\RDISMcorrBPM\pm\dRDISMcorrBPM$ and
$\ratiop=(\RPpDISMcorrBPM\pm\dRPpDISMcorrBPM)$ \unitPp, for
\nPpDISMcorrBPM\ stars.
On the other hand, we can be more strict in our selection by imposing
lower $\ebvsub$.  
{\bfc
A limit $\ebvsub \le \ebvsubmaxC$ rather than our reference criterion
$\ebvsubmax$ has no effect.
\ifdefined\CHECK \replyc{Check:
$\ratio=\RebvsubmaxCcorrBPM\pm\dRebvsubmaxCcorrBPM$ and
$\ratiop=(\RPpebvsubmaxCcorrBPM\pm\dRPpebvsubmaxCcorrBPM)$ \unitPp\
($\nebvsubmaxCcorrBPM$ stars).}\fi 
With even lower column densities ($\ebvsub \le \ebvsubmaxB$ and
$\ebvsubmaxA$), we get
$\ratio=\RebvsubmaxBcorrBPM\pm\dRebvsubmaxBcorrBPM$ and 
$\ratio=\RebvsubmaxAcorrBPM\pm\dRebvsubmaxAcorrBPM$,
for $\nebvsubmaxBcorrBPM$ and $\nebvsubmaxAcorrBPM$ stars,
respectively.  While an increase {\bfm in} $\ratio$ with decreasing column
density could arise through the inverse dependence of $\ratio$ on
$\isub$ (Eq.~\ref{Eq-ratio}), the evidence is not strong.
Changes in $\ratiop$ are not significant:
($\RPpebvsubmaxBcorrBPM\pm\dRPpebvsubmaxBcorrBPM)$ \unitPp\ and
$(\RPpebvsubmaxAcorrBPM\pm\dRPpebvsubmaxAcorrBPM)$ \unitPp,
respectively.
}

%++++++++++++++++++++++++++++++++++++++++++++++

We can restrict our sample to lines of sight where the ratio of the
total to selective extinction, $\Rv$, is close to 3.1, its
characteristic value for the diffuse ISM \cite[\eg][]{F04}.
Specifically, we exclude those lines of sight where $\Rv$ was not
measured (SA85, KR09), and impose $2.6 < \Rv < 3.6$. Our sample is
then reduced to \nRvcorrBPM\ stars and gives
$\ratio=\RRvcorrBPM\pm\dRRvcorrBPM$ and
$\ratiop=(\RPpRvcorrBPM\pm\dRPpRvcorrBPM)$ \unitPp.

A similar selection can be made on the basis of the wavelength
corresponding to the peak of the polarization curve in extinction,
$\lmax$, as taken from \citet{SMF75}. Imposing
$0.5\,\mu$m~$<\lmax<0.6\,\mu$m, we find $\ratio=\RlmaxcorrBPM \pm
\dRlmaxcorrBPM$ and $\ratiop=(\RPplmaxcorrBPM\pm\dRPplmaxcorrBPM)$
\unitPp, for $\nlmaxcorrBPM$ stars.

%++++++++++++++++++++++++++++++++++++++++++++++

\paragraph{Column density ratio}
The polarization ratio $\ratio$ is, by construction, proportional to
$\ebv$ and could therefore anticorrelate with $\NHr$.  However, we do
not find such dependance when varying the upper limit of $\NHr$ from \NHratioA\
($\ratio = \RNHratioAcorrBPM\pm\dRNHratioAcorrBPM$, \nNHratioAcorrBPM\
stars) to \NHratioD\ ($\ratio=\RNHratioDcorrBPM\pm\dRNHratioDcorrBPM$,
\nNHratioDcorrBPM\ stars).  Going beyond the limits where angles
agree, with upper limits of \NHratioE\ and \NHratioF, yields the same
value $\RNHratioEcorrBPM\pm\dRNHratioEcorrBPM$ for the polarization
ratio, for \nNHratioEcorrBPM\ and \nNHratioFcorrBPM\ stars,
respectively.
{\bfc 
We also tested other proxies to estimate the total column density
observed by \Planck.  Replacing the dust optical depth at 353\GHz,
used to derive $\ebvsub$, by the \hi\ 21 cm emission or the dust
radiance \citep[the total power emitted by dust,][]{planck2013-p06b}
did not affect our polarization ratios significantly.
}

%++++++++++++++++++++++++++++++++++++++++++++++

\paragraph{Orthogonality}
If we become more restrictive in our selection based on the difference
between position angles (Eq.~\ref{angle-criteria} in
\Sect~\ref{orthocriterion}), by requiring a $1\sigma$ agreement 
our sample shrinks to \nOrthocorrBPM\ stars.  The quality of the fit is
preserved, as expected: $\ratio=\ROrthocorrBPM\pm\dROrthocorrBPM$ and
$\ratiop=(\RPpOrthocorrBPM\pm\dRPpOrthocorrBPM)$ \unitPp.

%++++++++++++++++++++++++++++++++++++++++++++++

\paragraph{Galactic height}
As mentioned in \Sect~\ref{Eratio}, the Galactic height of the star
can play a role, similar to that of the column density ratio $\NHr$,
in selecting lines of sight with a low probability of background
emission.
Still requiring $\NHr < \NHrmax$ and then selecting on $H> \HeightA$
pc, $H> \HeightB$ pc, and $H > \HeightC$ pc, we find the same
polarization ratios, $\ratio=\RHeightAcorrBPM\pm\dRHeightAcorrBPM$ for
\nHeightAcorrBPM, \nHeightBcorrBPM\ and \nHeightCcorrBPM\ stars,
respectively.
\ifdefined\CHECK \replyc{Check:
  $\RHeightBcorrBPM\pm\dRHeightBcorrBPM$,
  $\RHeightCcorrBPM\pm\dRHeightCcorrBPM$} \fi
Selecting on $H > \Heightalone$ pc without selecting on the column
density ratio, we obtain
$\ratio=\RHeightalonecorrBPM\pm\dRHeightalonecorrBPM$
($\nHeightalonecorrBPM$ stars).  Results for $\ratiop$ are also
similar, with an average
$\ratiop=(\RPpHeightAcorrBPM\pm\dRPpHeightAcorrBPM)$ \unitPp.
\ifdefined\CHECK \replyc{Check: $\ratiop =
  \RPpHeightAcorrBPM\pm\dRPpHeightAcorrBPM,
  \RPpHeightBcorrBPM\pm\dRPpHeightBcorrBPM,
  \RPpHeightCcorrBPM\pm\dRPpHeightCcorrBPM,
  \RPpHeightalonecorrBPM\pm\dRPpHeightalonecorrBPM $} \fi

%++++++++++++++++++++++++++++++++++++++++++++++
%%%%%%%%%%%%%%%%%%%%%%%%%%%%%%%%%%%%%%%%%%%%%%%%%%%%%%%

\begin{figure}
\includegraphics[width=\hhsize]{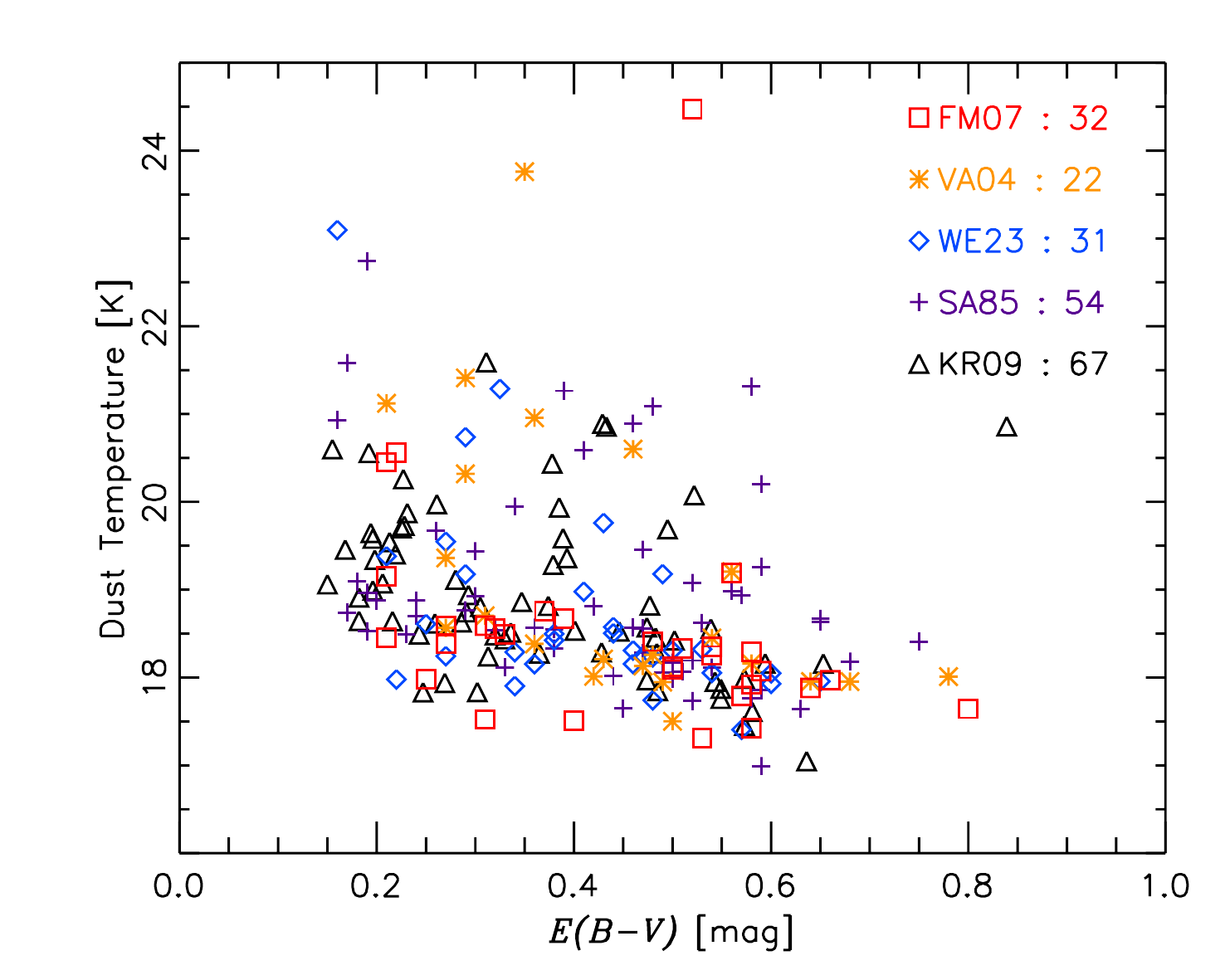}
\caption{\Planck\ line of sight dust temperature $\Tdust$
\citep{planck2013-p06b} and the column density to the star, $\ebv$,
for the independent samples.}
\label{Tebv}
\end{figure}

\begin{figure*}
\includegraphics[width=\thsize]{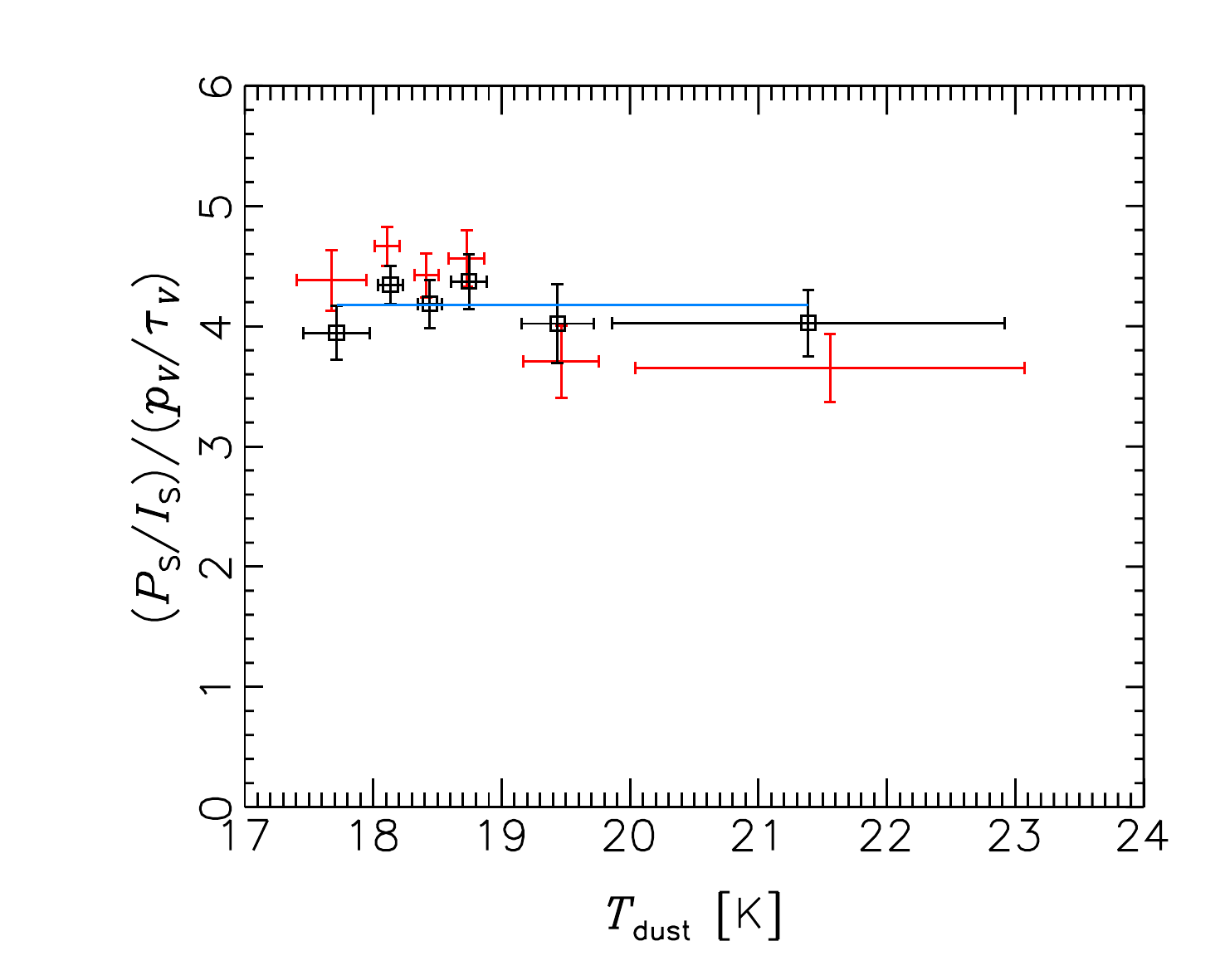}
\includegraphics[width=\thsize]{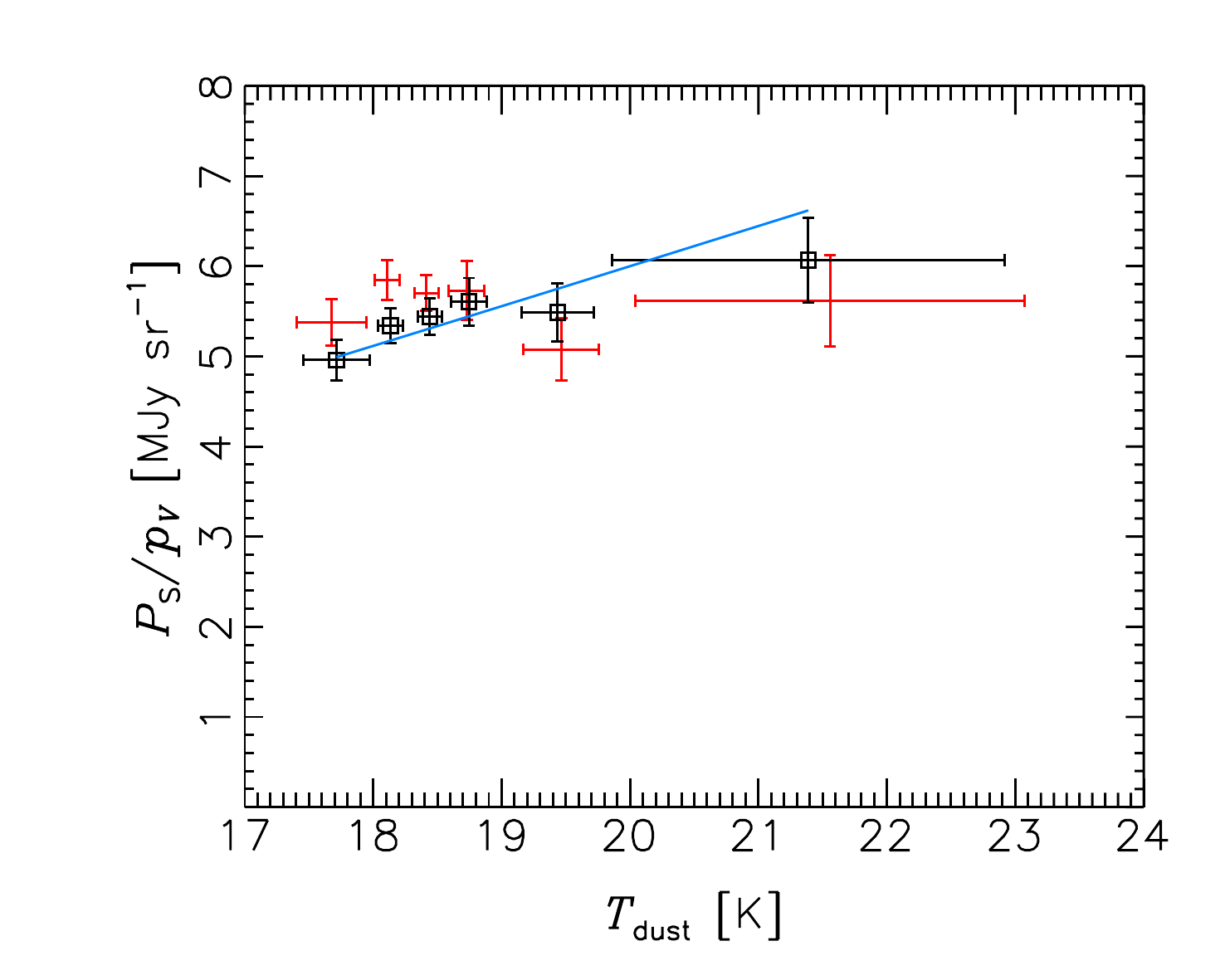}
\includegraphics[width=\thsize]{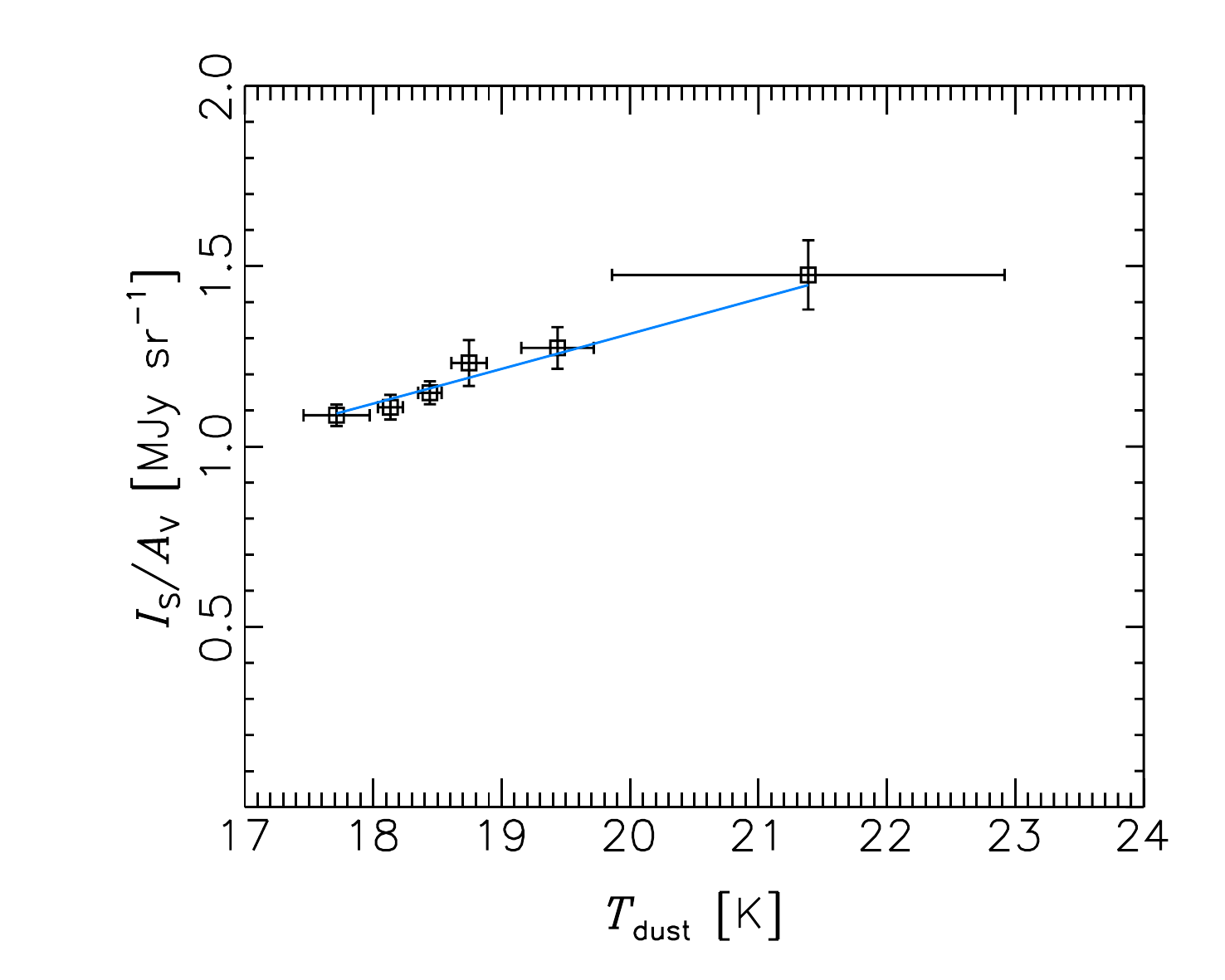}
\caption{\LL\ Mean $\ratio$ as a function of the mean
$\Tdust$, each plotted with the standard deviation, in bins of equal number. 
Two versions of the \Planck\ data have been used: with (black) and
without (red) correction for leakage of intensity into
polarization.
\MM\ The same, but for $\ratiop$.
\RR\ The same, but for $\ratioI$.  
{\bfc The blue curves, motivated in the right panel, show the expected
response of the ratios to an increase in $\Tdust$, according to a
simple model in which the subset of grains that are polarizing had the
same temperature as characterized the total emission (see text).}
}
\label{Fig-ratio-Temp}
\end{figure*}

\paragraph{Thermal dust temperature}
The dust temperature $\Tdust$ from \citet{planck2013-p06b}
characterizes the spectral energy distribution of the combined
emission $\isub$ from all dust components in the column of dust along
the line of sight and has no direct connection to the sample
selection.  Figure \ref{Tebv} shows the distribution in the
$\Tdust$~--~column density (as measured by $\ebv$) plane.  The ranges
of $\ebv$ and $\Tdust$ are considerable for each sample.  In this
plane, there is a band showing a slight anti-correlation of $\Tdust$
and $\ebv$; there are also several lines of sight with $\Tdust \simeq
21$\,K but with a range of $\ebv$. 

We looked for any dependencies of $\ratio$, $\ratiop$, and $\ratioI$
on $\Tdust$, for data with and without the correction for leakage of
intensity into polarization.  In Fig.~\ref{Fig-ratio-Temp} the data
were binned in $\Tdust$, each bin containing the same number of stars.
{\bfc The blue curve in the right panel of Fig.~\ref{Fig-ratio-Temp}
shows the relative change arising from the expected increase in
$\isub$ (but not $\Av$) when $\Tdust$ increases.  It has been fit to
the data in the vertical direction.
A similar trend in $\ratiop$ arising from $\psub$ would be expected in
the middle panel if the subset of grains that are polarizing had the
same temperature as characterized the total emission.  This appears to
be consistent with the corrected data.
Under the same hypothesis, the trend for $\ratio$ would be flat in the
left panel.  This too appears to be consistent with the corrected
data.
}

%==============================================

\subsection{Data used}\label{RobDataUsed}

Here we test the sensitivity of our results to the choice of extinction
catalogues and to the smoothing and original processing of the
\Planck\ data.  The main sources of data uncertainty are, for the
extinction data, the measure of $\Av$ and, for the polarized emission
data, the instrumental systematics related to the correction for the
leakage of intensity to polarization (\Sect~\ref{planckdata}).  

%++++++++++++++++++++++++++++++++++++++++++++++

\paragraph{Extinction catalogues} 
One important source of uncertainty in $\ratio$ (not $\ratiop$) is the
measure of the dust extinction in the \optical, $\Av$.
In Table \ref{Table-perCat} we summarize our results for each
catalogue taken separately, independently of the others (\ie\ we do
not remove common stars).  Although the catalogues neither share all
of the same stars nor have the same extinction data for stars in
common, the estimates obtained for $\ratio$ and $\ratioI$ are
compatible.  $\ratiop$ is independent of $\Av$, therefore of any
extinction samples. Its variations among catalogues helps to constrain
its statistical uncertainty, here less than $\ressp$ \unitPp.

\begin{table}[h] %or % use table for a one-column table
\begingroup % this + \endgroup at the end keep table things local
\caption{$\ratio$, $\ratiop$, and $\ratioI$ with their uncertainties
obtained with our bootstrap method for each sample taken independently
of the others (common stars are \emph{not removed}, unlike in our full
sample; see \Sect~\ref{ecat}).}
\label{Table-perCat}
\nointerlineskip
\vskip -3mm
\footnotesize % good font size for a table, but can be changed
\setbox\tablebox=\vbox{ %
\newdimen\digitlsmwidth % see \S\,17.12 for the purpose of the next 10 lines
\setbox0=\hbox{\rm 0}
%\digitwidth=\wd0
\catcode`*=\active
\def*{\kern\digitwidth}
\newdimen\signwidth
\setbox0=\hbox{+}
%\signwidth=\w
\catcode`!=\active
\def!{\kern\signwidth}
\halign{#\hfil\tabskip=01em&\hfil#\hfil\tabskip=01em&\hfil#\hfil\tabskip=01em&\hfil#\hfil\tabskip=01em&\hfil#\hfil\tabskip=0pt\cr
\noalign{\doubleline}
Sample & No. of stars & $\ratio$ & $\ratiop$ & $\ratioI$ \cr
& & & (\unitPp)  & (\unitPp) \cr
\noalign{\vskip 5pt\hrule\vskip 3pt}
FM07 & \nFMcorrBPM &$\RFMcorrBPM\pm\dRFMcorrBPM$&$\RPpFMcorrBPM\pm\dRPpFMcorrBPM$&$\RIsAvFMcorrBPM\pm\dRIsAvFMcorrBPM$  \cr
VA04 & \nVAcorrBPM &$\RVAcorrBPM\pm\dRVAcorrBPM$&$\RPpVAcorrBPM\pm\dRPpVAcorrBPM$ &$\RIsAvVAcorrBPM\pm\dRIsAvVAcorrBPM$ \cr
WE23 & \nWEcorrBPM &$\RWEcorrBPM\pm\dRWEcorrBPM$&$\RPpWEcorrBPM\pm\dRPpWEcorrBPM$&$\RIsAvWEcorrBPM\pm\dRIsAvWEcorrBPM$ \cr
SA85 & \nSAcorrBPM &$\RSAcorrBPM\pm\dRSAcorrBPM$&$\RPpSAcorrBPM\pm\dRPpSAcorrBPM$&$\RIsAvSAcorrBPM\pm\dRIsAvSAcorrBPM$ \cr
KR09 & \nKRcorrBPM &$\RKRcorrBPM\pm\dRKRcorrBPM$&$\RPpKRcorrBPM\pm\dRPpKRcorrBPM$&$\RIsAvKRcorrBPM\pm\dRIsAvKRcorrBPM$ \cr
\noalign{\vskip 5pt\hrule\vskip 3pt}
}
}
\endPlancktable %or % for a one-column table; defined in Planck.tex.
\endgroup
\end{table}

%++++++++++++++++++++++++++++++++++++++++++++++

\paragraph{Smoothing} 
In order to increase the $\SNR$ of the \Planck\ polarization data and
increase the quality of the correlations in $Q$ and $U$, we chose to
smooth our maps with a 5\arcmin\ FWHM Gaussian, making the effective
map resolution 7\arcmin. Because $Q$ and $U$ are algebraic quantities
derived from a polarization pseudo-vector, the smoothing of
polarization maps statistically tends to diminish the polarization
intensity $\psub$, which would propagate into the polarization ratios
$\ratio$ and $\ratiop$.  Using the raw (not smoothed) data, or data
smoothed with a beam of \SmoothB\arcmin\ and \SmoothD\arcmin\ (keeping
the same sample as was selected using data smoothed with a 5\arcmin\
beam to allow for an unbiased comparison), we find $\ratio=$
$\RSmoothEcorrBPM\pm\dRSmoothEcorrBPM$,
$\RSmoothCcorrBPM\pm\dRSmoothCcorrBPM$,
$\RSmoothDcorrBPM\pm\dRSmoothDcorrBPM$, and
$\ratiop=(\RPpSmoothEcorrBPM\pm\dRPpSmoothEcorrBPM)$ \unitPp,
$(\RPpSmoothCcorrBPM\pm\dRPpSmoothCcorrBPM)$ \unitPp,
$(\RPpSmoothDcorrBPM\pm\dRPpSmoothDcorrBPM)$ \unitPp), respectively.

{\bfm We note} that the mean values of $\ratio$ and $\ratiop$ could still be
underestimated owing to depolarization in the \Planck\ beam, which
does not have a counterpart in the visible measurement (see
Fig.~\ref{los}); however, this effect should be small and within the
uncertainties.

%++++++++++++++++++++++++++++++++++++++++++++++

\paragraph{Zodiacal emission}
Removing zodiacal emission, or not, in deriving $I_\sub$
(\Sect~\ref{planckdata}) does not affect our result, with
$\ratio=\RrmvzodicorrBPM\pm\dRrmvzodicorrBPM$ in both cases.
\ifdefined\CHECK
\replyc{Check keep zodi: $\RkeepzodicorrBPM\pm\dRkeepzodicorrBPM$,
remove zodi : $\RrmvzodicorrBPM\pm\dRrmvzodicorrBPM$}.
\fi

%++++++++++++++++++++++++++++++++++++++++++++++

\paragraph{Contamination by CMB polarization at 353\GHz}
Following the approach in \cite{planck2014-XXII}, we can remove the
CMB patterns in intensity and polarization from the 353\GHz\ maps by
subtracting the 100\GHz\ ($I$,$Q$,$U$) maps in CMB thermodynamic
temperature units.  This method unfortunately adds noise to the
353\GHz\ maps, and is therefore used only as a check. It also has the
drawback of subtracting a fraction of dust emission that is present in
the 100\GHz\ channel. However, dust is then subtracted both in
intensity and in polarization, though perhaps not proportionally to
the polarization fraction at 353 \GHz\ (however, this is a {\bfm second-order}
effect).  With this 100\GHz-subtracted version of $Q$ and $U$,
we obtain $\ratio=\RCMBsubstractedcorrBPM\pm\dRCMBsubstractedcorrBPM$
and $\ratiop=(\RPpCMBsubstractedcorrBPM\pm\dRPpCMBsubstractedcorrBPM)$
\unitPp, for $\nPpCMBsubstractedcorrBPM$ stars.

%++++++++++++++++++++++++++++++++++++++++++++++

\begin{table*} %or % use table for a one-column table
\begingroup % this + \endgroup at the end keep table things local
\caption{Values of $\ratio$ and $\ratiop$ in specific regions.}
\label{Table-Geo}
\nointerlineskip
\vskip -3mm
\footnotesize % good font size for a table, but can be changed
\setbox\tablebox=\vbox{ %
\newdimen\digitlsmwidth % see \S\,17.12 for the purpose of the next 10 lines
\setbox0=\hbox{\rm 0}
%\digitwidth=\wd0
\catcode`*=\active
\def*{\kern\digitwidth}
\newdimen\signwidth
\setbox0=\hbox{+}
%\signwidth=\w
\catcode`!=\active
\def!{\kern\signwidth}
\halign{#\hfil\tabskip=01em&\hfil#\hfil\tabskip=01em&\hfil#\hfil\tabskip=01em&\hfil#\hfil\tabskip=01em&\hfil#\hfil\tabskip=01em&\hfil#\hfil\tabskip=01em&\hfil#\hfil\tabskip=01em&\hfil#\hfil\tabskip=0pt\cr
\noalign{\doubleline}
Region & Longitude  & Latitude & No. stars &$\ratio$ &$\ratiop$ & $\ratioI$& $\langle \Tdust \rangle$\cr
 &  [deg]  &  [deg] 	&  & & [\unitPp] & [\unitPp] & [K] \cr
\noalign{\vskip 5pt\hrule\vskip 3pt}
Fan & $[125\deg:140\deg]$& $[-8\pdeg5:-1\pdeg5]$& \nFancorrBPM&$\RFancorrBPM\pm\dRFancorrBPM$&$\RPpFanQUcorrBPM\pm\dRPpFanQUcorrBPM$ & $\RIsAvFancorrBPM\pm\dRIsAvFancorrBPM$ & 18.3 \cr
Aquila\,Rift & $[320\deg:360\deg]$& $[+10\deg:+35\deg]$& \nAquilacorrBPM&$\RAquilacorrBPM\pm\dRAquilacorrBPM$&$\RPpAquilaQUcorrBPM\pm\dRPpAquilaQUcorrBPM$ & $\RIsAvAquilacorrBPM\pm\dRIsAvAquilacorrBPM$ & 19.4 \cr
Ara&$[340\deg:~~40\deg]$&$[-12\deg:-2\pdeg0]$&\nAriacorrBPM&$\RAriacorrBPM\pm\dRAriacorrBPM$&$\RPpAriaQUcorrBPM\pm\dRPpAriaQUcorrBPM$ & $\RIsAvAriacorrBPM\pm\dRIsAvAriacorrBPM$ & 20.7 \cr
\noalign{\vskip 3pt\hrule\vskip 5pt}
}
}
\endPlancktable 
\endgroup
\end{table*}

%++++++++++++++++++++++++++++++++++++++++++++++

\paragraph{Leakage correction}
A small correction for leakage of intensity into polarization has been
applied to the \Planck\ polarization data used here (see
\Sect~\ref{planckdata}). While this correction is imperfect, the
alternative of ignoring this correction leaves systematic errors in
the data.  For the version of the data \emph{not} corrected for
leakage, we obtain figures similar to Fig.~\ref{qsiusi_qstust}, for
$\nmainQUNOTcorrBPM$ selected stars, with Pearson correlation
coefficients $\PearsonfitQUNOTcorrBPM$ and $\PearsonfitPpQUNOTcorrBPM$
and $\chi^2_{\rm r} = \chifitQUNOTcorrBPM$ and
$\chifitPpQUNOTcorrBPM$, for $\ratio$ and $\ratiop$, respectively.
Running the bootstrap analysis we find
$\ratio=\RmainQUNOTcorrBPM\pm\dRmainQUNOTcorrBPM$ and
$\ratiop=(\RPpmainQUNOTcorrBPM\pm\dRPpmainQUNOTcorrBPM)$ \unitPp,
systematic changes of $+\ressplanckBPM$ and $+\resspplanckBPM$
\unitPp\ compared to our values for data corrected for leakage.  
Therefore, the correction of the March 2013 \Planck\ polarization data
for this leakage is a significant source of systematic uncertainty in
$\ratio$ and $\ratiop$, though perhaps the uncertainty is not as much
as $\ressplanckBPM$ or of the same sign.

%==============================================

%\subsection{Dependence on Galactic latitude or hemisphere}\label{SpatialDep}
\subsection{Region analyzed}\label{SpatialDep}

%++++++++++++++++++++++++++++++++++++++++++++++
The polarization ratios that we derived are an average over the sky.
Here we examine the ratios for spatial subsets of the data.

\paragraph{Galactic hemisphere or latitude}
We find no significant variation of the polarization ratios between
the two hemispheres: $\ratio=\RNorthcorrBPM\pm\dRNorthcorrBPM$ and
$\ratiop=(\RPpNorthcorrBPM\pm\dRPpNorthcorrBPM$) \unitPp\ for the
northern Galactic hemisphere, and
$\ratio=\RSouthcorrBPM\pm\dRSouthcorrBPM$) and
$\ratiop=(\RPpSouthcorrBPM\pm\dRPpSouthcorrBPM$) \unitPp\ for the
southern.

Polarization ratios might depend on the latitude of stars if that were
indicative of different potential backgrounds. Selecting high latitude stars
from both hemispheres ($|b| > \glatlim\deg$, \nHighLatcorrBPM\ stars)
to limit the presence of backgrounds, we find
$\ratio=\RHighLatcorrBPM\pm\dRHighLatcorrBPM$,
$\ratiop=(\RPpHighLatcorrBPM\pm\dRPpHighLatcorrBPM)$ \unitPp, and
$\ratioI=(\RIsAvHighLatcorrBPM\pm\dRIsAvHighLatcorrBPM)$ \unitPp.
\ifdefined\CHECK \replyc{$\RHighLatNOTcorrBPM\pm\dRHighLatNOTcorrBPM$
  and $\RPpHighLatNOTcorrBPM\pm\dRPpHighLatNOTcorrBPM$ \unitPp\
  without correction for leakage.}  \fi 
For low latitude stars ($|b| \le \glatlim\deg$, \nLowLatcorrBPM\
stars), $\ratio=\RLowLatcorrBPM\pm\dRLowLatcorrBPM$,
$\ratiop=(\RPpLowLatcorrBPM\pm\dRPpLowLatcorrBPM)$ \unitPp, and
$\ratioI=(\RIsAvLowLatcorrBPM\pm\dRIsAvLowLatcorrBPM)$ \unitPp, with no
indication of any contamination by backgrounds.
\ifdefined\CHECK \replyc{$\RLowLatNOTcorrBPM\pm\dRLowLatNOTcorrBPM$ and
  $\RPpLowLatNOTcorrBPM\pm\dRPpLowLatNOTcorrBPM$ \unitPp\ without
  correction for leakage.}  \fi

\paragraph{Selected regions on the sky}
Table~\ref{Table-Geo} presents the polarization ratios for three
regions, among them the Fan which contains almost one third of our
selected stars; these results are close to the overall average. If we
select all stars except those from the Fan, we find
$\ratio=\RnoFancorrBPM\pm\dRnoFancorrBPM$ and
$\ratiop=(\RPpnoFanQUcorrBPM\pm\dRPpnoFanQUcorrBPM)$ \unitPp.  
Table~\ref{Table-Geo} also presents the polarization ratios for two
other regions in the local ISM where stars in our sample are more
concentrated (see Fig.~\ref{starmap}): the Aquila Rift, and the Ara
region \citep{planck2014-XIX}.  Taking into account the uncertainties,
we conclude that our total sample is not biased by any particular
region
{\bfc
and there is no evidence for spatial variations.
}

%==============================================

\subsection{Correlation plots in $P$}\label{CorrP}

\begin{figure*}
\includegraphics[width=\hhsize]{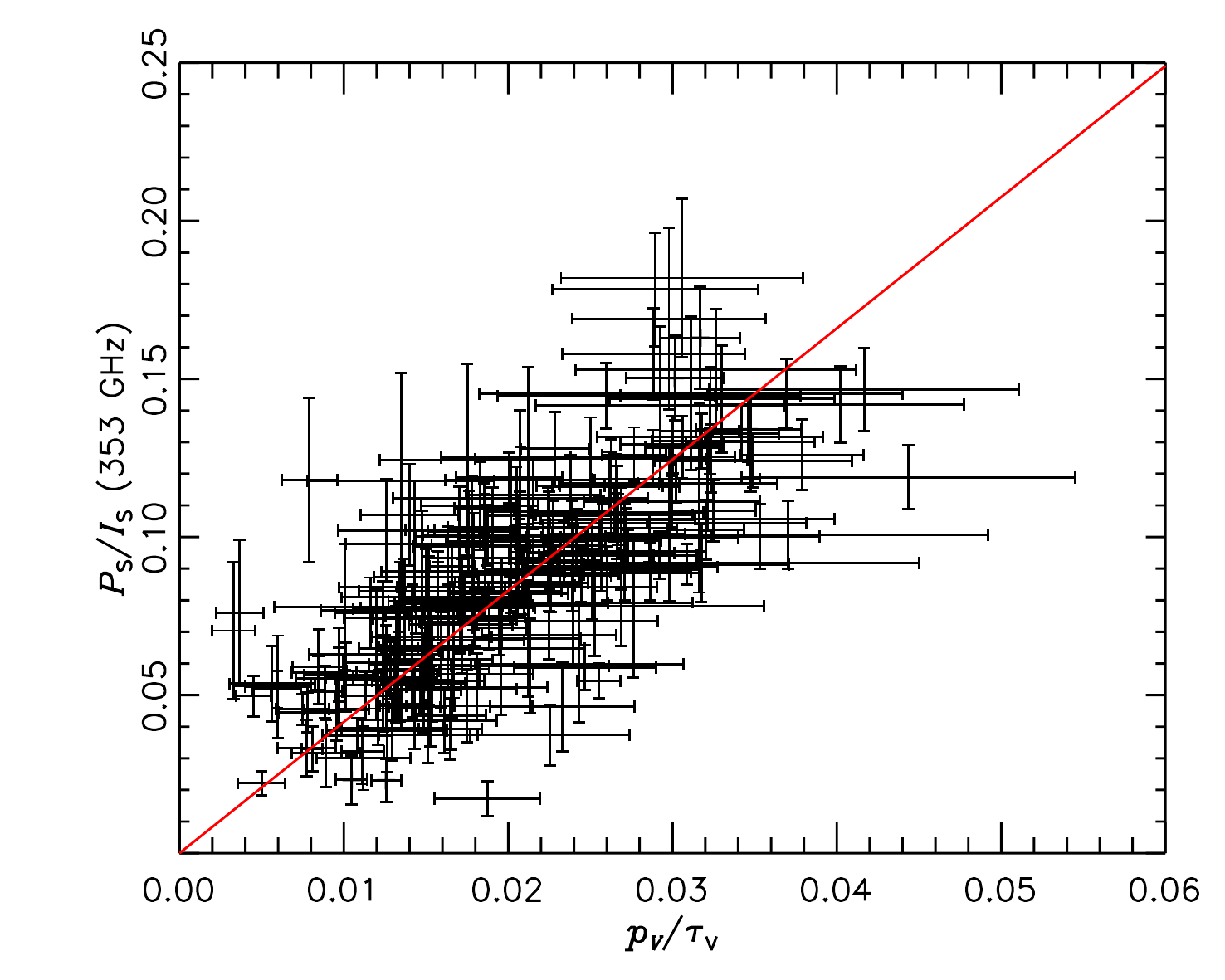}
\includegraphics[width=\hhsize]{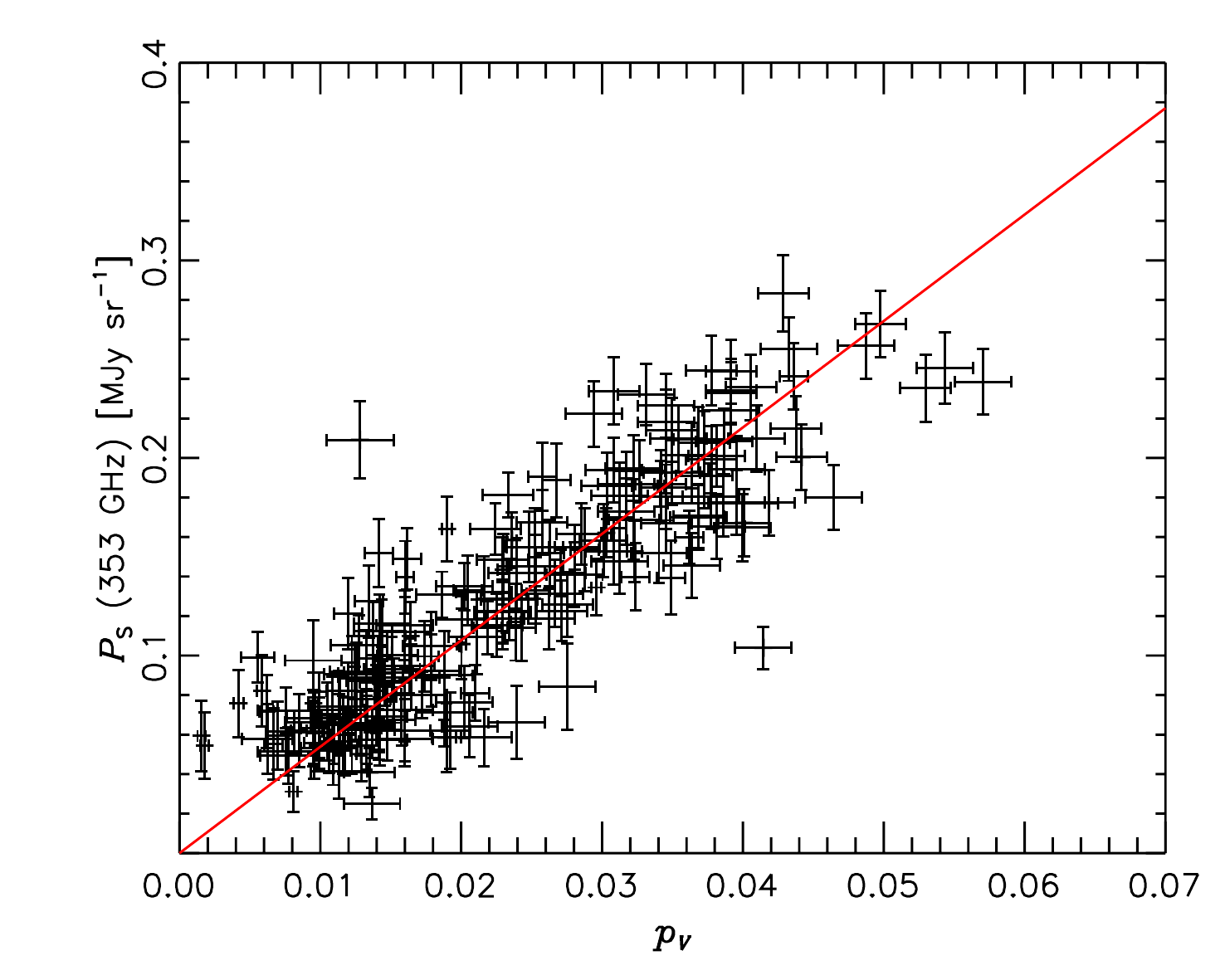}
\caption{\LL\ Correlation of debiased polarization fraction in
emission with that in extinction (Pearson coefficient
\PearsonfitPcorrBPM). \RR\ Correlation of the debiased polarized
emission (in \unitPp) with starlight polarization degree (Pearson
coefficient \PearsonfitPpPcorrBPM).  The range corresponds to one
quadrant in Fig.~\ref{qsiusi_qstust}. The fits are forced to go
through the origin and have slopes $\ratio =
\RfitPcorrBPM\pm\dRfitPcorrBPM$ ($\chi^2_{\rm r}=\chifitPcorrBPM$) and
$\ratiop=(\RfitPpPcorrBPM\pm\dRfitPpPcorrBPM)$ \unitPp\
($\chi^2_{\rm r}=\chifitPpPcorrBPM$), respectively.}
\label{psi_pst-biased}
\end{figure*}

In \Sect~\ref{CorrelationPlots} we derived $\ratio$ and $\ratiop$
using joint correlation plots in $Q$ and $U$ rather than in the biased
quantity $P$.
However, our selection of $\psub$ and $\pv$ with $\SNR > 3$ implies
that the bias should not be too significant and it is possible to
debias $P$ at least statistically (the Modified Asymptotic debiasing
method of \citealp{PLA13} was used; see also references in
\Sect~\ref{CorrelationPlots}).  This is confirmed by the correlation
plot in Fig.~\ref{psi_pst-biased} for debiased polarization fractions
(which are almost identical to those for the original data).  The data
in the submillimetre and \optical\ show a fairly good correlation,
though, compared with Fig.~\ref{qsiusi_qstust}, have a smaller dynamic
range and a smaller Pearson correlation coefficient.  We fit the
slopes, forcing the fit to go through the origin unlike for the fits
in Fig.~\ref{qsiusi_qstust}. Whether with debiased data or not, the
polarization ratios (from bootstrapping) are essentially identical:
$\ratio=\RmainPdebiasedcorrBPM\pm\dRmainPdebiasedcorrBPM$
\ifdefined\CHECK \replyc{Check:
  $\RmainPbiasedcorrBPM\pm\dRmainPbiasedcorrBPM$ }, \fi 
and
$\ratiop=(\RPpmainPdebiasedcorrBPM\pm\dRPpmainPdebiasedcorrBPM)$
\unitPp,
\ifdefined\CHECK \replyc{Check:
  $\RPpmainPbiasedcorrBPM\pm\dRPpmainPbiasedcorrBPM$ }, \fi
and also the same as found in the preferred analysis in
Fig.~\ref{qsiusi_qstust}.

%==============================================

\section{$\ratio$ and its relationship to the maximum observed
  polarization fractions}\label{MaxPolarFraction}

For a given dust model, including the grain shape, the maximum
polarization fraction that can be observed corresponds to the ideal
case of optimal dust alignment: the magnetic field lies in the plane
of the sky, has the same orientation (position angle) along the line
of sight, and the dust alignment efficiency with respect to the field
is perfect.  The maximum $\pst \simeq 3\,\%$ observed in extinction
\citep[corresponding to $\pv \le 9\,\%\,\ebv$,][]{SMF75}, is supposed
to be close to this ideal case
{\bfc
\citep{DF09}.
}
For our selected sample of lines of sight, Fig.~\ref{pvstau-pvsi}
(left panel) shows this classical envelope and the corresponding
envelope ($\PsI= 3\,\% \times\,\ratio = 12.9\,\%$) transferred to
emission (right panel).

We have also investigated the upper envelope that might be derived
independently from the emission data.
At a resolution of $1\deg$ \Planck\ HFI has revealed regions with
$\PsI$ greater than \psimaxPIP75\,\% \citep{planck2014-XIX}, albeit
for only a very small fraction (0.001, their Fig. 18) of lines of
sight, toward local diffuse clouds.  This value, which is already a
high envelope, might have been even larger were it at the finer
resolution of starlight measurements.  Combined with the 3\,\% limit
from stars, this would apparently imply $\ratio > 6$, significantly
higher than our mean value $\ratio = \resr$.

However, for statistical reasons, these two estimates of $\ratio$
cannot be compared straightforwardly: \Planck\ statistics are based on
(almost) full-sky data, while those of \cite{SMF75} are based on less
than 300 stars.  A more consistent statistical comparison can be
sought.  Analyzing Fig.~9 of \cite{SMF75} and limiting our analysis to
stars satisfying $0.15 < \ebv < \ebvsubmax$ as in our selection
criterion (\Sect~\ref{diffusecriterion}), the upper envelope $\pv \le
9\,\%\,\ebv$ is approximately the $96\,\%$ percentile of $\pv/\ebv$
(in this interval of $\ebv$, 8 stars out of about 200 lie above this
envelope).
The corresponding 96\,\% percentile of $\PsI$ in our selected sample
is $\PsI=14.5\,\%$ (8 stars out of $\nmainQUcorrBPM$ above that line).
As a complement, we can obtain an estimate of the 96\,\% percentile of
the full \Planck\ map by smoothing the 353\GHz\ maps with a Gaussian
of 5\arcmin\ and selecting those pixels with $0.15 < \ebvsub <
\ebvsubmax$.  The 96\,\% percentile of $\PsI$ (after debiasing) in
this sample of over $10^7$ pixels is found to be 13.2\%.  Combining
these estimates based on consistent percentiles implies $\ratio$ in
the range 4.2--4.6, compatible with our direct, and more rigorous,
result.

\end{appendix}

% from the style guide
%A&A LaTEX puts double space between institution names at the end of
%Planck papers. To fix this, put \raggedright immediately
%before \end{document}.

\raggedright

\end{document}

%% file: Planck.tex
\def\setsymbol#1#2{\expandafter\def\csname #1\endcsname{#2}}
\def\getsymbol#1{\csname #1\endcsname}

\def\Planck{\textit{Planck}}

\def\all2013resultspapers{\nocite{planck2013-p01, planck2013-p02, planck2013-p02a, planck2013-p02d, planck2013-p02b, planck2013-p03, planck2013-p03c, planck2013-p03f, planck2013-p03d, planck2013-p03e, planck2013-p01a, planck2013-p06, planck2013-p03a, planck2013-pip88, planck2013-p08, planck2013-p11, planck2013-p12, planck2013-p13, planck2013-p14, planck2013-p15, planck2013-p05b, planck2013-p17, planck2013-p09, planck2013-p09a, planck2013-p20, planck2013-p19, planck2013-pipaberration, planck2013-p05, planck2013-p05a, planck2013-pip56, planck2013-p06b}}

\newbox\tablebox    \newdimen\tablewidth
\def\leaderfil{\leaders\hbox to 5pt{\hss.\hss}\hfil}
\def\endPlancktable{\tablewidth=\columnwidth 
    $$\hss\copy\tablebox\hss$$
    \vskip-\lastskip\vskip -2pt}

\def\tablenote#1 #2\par{\begingroup \parindent=0.8em
    \abovedisplayshortskip=0pt\belowdisplayshortskip=0pt
    \noindent
    $$\hss\vbox{\hsize\tablewidth \hangindent=\parindent \hangafter=1 \noindent
    \hbox to \parindent{$^#1$\hss}\strut#2\strut\par}\hss$$
    \endgroup}
\def\doubleline{\vskip 3pt\hrule \vskip 1.5pt \hrule \vskip 5pt}

\def\L2{\ifmmode L_2\else $L_2$\fi}

\def\DeltaT{\ifmmode \Delta T\else $\Delta T$\fi}
\def\deltat{\ifmmode \Delta t\else $\Delta t$\fi}
\def\fknee{\ifmmode f_{\rm knee}\else $f_{\rm knee}$\fi}
\def\Fmax{\ifmmode F_{\rm max}\else $F_{\rm max}$\fi}
\def\solar{\ifmmode{\rm M}_{\mathord\odot}\else${\rm M}_{\mathord\odot}$\fi}
\def\Msolar{\ifmmode{\rm M}_{\mathord\odot}\else${\rm M}_{\mathord\odot}$\fi}
\def\Lsolar{\ifmmode{\rm L}_{\mathord\odot}\else${\rm L}_{\mathord\odot}$\fi}

\def\inv{\ifmmode^{-1}\else$^{-1}$\fi}
\def\mo{\ifmmode^{-1}\else$^{-1}$\fi}
\def\sup#1{\ifmmode ^{\rm #1}\else $^{\rm #1}$\fi}
\def\expo#1{\ifmmode \times 10^{#1}\else $\times 10^{#1}$\fi}
\def\,{\thinspace}
\def\lsim{\mathrel{\raise .4ex\hbox{\rlap{$<$}\lower 1.2ex\hbox{$\sim$}}}}
\def\gsim{\mathrel{\raise .4ex\hbox{\rlap{$>$}\lower 1.2ex\hbox{$\sim$}}}}

\def\simprop{\mathrel{\raise .4ex\hbox{\rlap{$\propto$}\lower 1.2ex\hbox{$\sim$}}}}
\def\deg{\ifmmode^\circ\else$^\circ$\fi}
\def\pdeg{\ifmmode $\setbox0=\hbox{$^{\circ}$}\rlap{\hskip.11\wd0 .}$^{\circ}
          \else \setbox0=\hbox{$^{\circ}$}\rlap{\hskip.11\wd0 .}$^{\circ}$\fi}
\def\arcs{\ifmmode {^{\scriptstyle\prime\prime}}
          \else $^{\scriptstyle\prime\prime}$\fi}
\def\arcm{\ifmmode {^{\scriptstyle\prime}}
          \else $^{\scriptstyle\prime}$\fi}
\newdimen\sa  \newdimen\sb
\def\parcs{\sa=.07em \sb=.03em
     \ifmmode \hbox{\rlap{.}}^{\scriptstyle\prime\kern -\sb\prime}\hbox{\kern -\sa}
     \else \rlap{.}$^{\scriptstyle\prime\kern -\sb\prime}$\kern -\sa\fi}
\def\parcm{\sa=.08em \sb=.03em
     \ifmmode \hbox{\rlap{.}\kern\sa}^{\scriptstyle\prime}\hbox{\kern-\sb}
     \else \rlap{.}\kern\sa$^{\scriptstyle\prime}$\kern-\sb\fi}
\def\ra[#1 #2 #3.#4]{#1\sup{h}#2\sup{m}#3\sup{s}\llap.#4}
\def\dec[#1 #2 #3.#4]{#1\deg#2\arcm#3\arcs\llap.#4}
\def\deco[#1 #2 #3]{#1\deg#2\arcm#3\arcs}
\def\rra[#1 #2]{#1\sup{h}#2\sup{m}}

\def\dots{\relax\ifmmode \ldots\else $\ldots$\fi}
\def\WHzsr{\ifmmode $W\,Hz\mo\,sr\mo$\else W\,Hz\mo\,sr\mo\fi}
\def\mHz{\ifmmode $\,mHz$\else \,mHz\fi}
\def\GHz{\ifmmode $\,GHz$\else \,GHz\fi}
\def\mKs{\ifmmode $\,mK\,s$^{1/2}\else \,mK\,s$^{1/2}$\fi}
\def\muKs{\ifmmode \,\mu$K\,s$^{1/2}\else \,$\mu$K\,s$^{1/2}$\fi}
\def\muKRJs{\ifmmode \,\mu$K$_{\rm RJ}$\,s$^{1/2}\else \,$\mu$K$_{\rm RJ}$\,s$^{1/2}$\fi}
\def\muKHz{\ifmmode \,\mu$K\,Hz$^{-1/2}\else \,$\mu$K\,Hz$^{-1/2}$\fi}
\def\MJysr{\ifmmode \,$MJy\,sr\mo$\else \,MJy\,sr\mo\fi}
\def\MJysrmK{\ifmmode \,$MJy\,sr\mo$\,mK$_{\rm CMB}\mo\else \,MJy\,sr\mo\,mK$_{\rm CMB}\mo$\fi}
\def\microns{\ifmmode \,\mu$m$\else \,$\mu$m\fi}

\def\muK{\ifmmode \,\mu$K$\else \,$\mu$\hbox{K}\fi}
\def\microK{\ifmmode \,\mu$K$\else \,$\mu$\hbox{K}\fi}
\def\muW{\ifmmode \,\mu$W$\else \,$\mu$\hbox{W}\fi}
\def\kms{\ifmmode $\,km\,s$^{-1}\else \,km\,s$^{-1}$\fi}
\def\kmsMpc{\ifmmode $\,\kms\,Mpc\mo$\else \,\kms\,Mpc\mo\fi}
\setsymbol{LFI:center:frequency:70GHz:units}{70.3\,GHz}
\setsymbol{LFI:center:frequency:44GHz:units}{44.1\,GHz}
\setsymbol{LFI:center:frequency:30GHz:units}{28.5\,GHz}

\setsymbol{LFI:center:frequency:70GHz}{70.3}
\setsymbol{LFI:center:frequency:44GHz}{44.1}
\setsymbol{LFI:center:frequency:30GHz}{28.5}

\setsymbol{LFI:center:frequency:LFI18:Rad:M:units}{71.7\GHz}
\setsymbol{LFI:center:frequency:LFI19:Rad:M:units}{67.5\GHz}
\setsymbol{LFI:center:frequency:LFI20:Rad:M:units}{69.2\GHz}
\setsymbol{LFI:center:frequency:LFI21:Rad:M:units}{70.4\GHz}
\setsymbol{LFI:center:frequency:LFI22:Rad:M:units}{71.5\GHz}
\setsymbol{LFI:center:frequency:LFI23:Rad:M:units}{70.8\GHz}
\setsymbol{LFI:center:frequency:LFI24:Rad:M:units}{44.4\GHz}
\setsymbol{LFI:center:frequency:LFI25:Rad:M:units}{44.0\GHz}
\setsymbol{LFI:center:frequency:LFI26:Rad:M:units}{43.9\GHz}
\setsymbol{LFI:center:frequency:LFI27:Rad:M:units}{28.3\GHz}
\setsymbol{LFI:center:frequency:LFI28:Rad:M:units}{28.8\GHz}
\setsymbol{LFI:center:frequency:LFI18:Rad:S:units}{70.1\GHz}
\setsymbol{LFI:center:frequency:LFI19:Rad:S:units}{69.6\GHz}
\setsymbol{LFI:center:frequency:LFI20:Rad:S:units}{69.5\GHz}
\setsymbol{LFI:center:frequency:LFI21:Rad:S:units}{69.5\GHz}
\setsymbol{LFI:center:frequency:LFI22:Rad:S:units}{72.8\GHz}
\setsymbol{LFI:center:frequency:LFI23:Rad:S:units}{71.3\GHz}
\setsymbol{LFI:center:frequency:LFI24:Rad:S:units}{44.1\GHz}
\setsymbol{LFI:center:frequency:LFI25:Rad:S:units}{44.1\GHz}
\setsymbol{LFI:center:frequency:LFI26:Rad:S:units}{44.1\GHz}
\setsymbol{LFI:center:frequency:LFI27:Rad:S:units}{28.5\GHz}
\setsymbol{LFI:center:frequency:LFI28:Rad:S:units}{28.2\GHz}

\setsymbol{LFI:center:frequency:LFI18:Rad:M}{71.7}
\setsymbol{LFI:center:frequency:LFI19:Rad:M}{67.5}
\setsymbol{LFI:center:frequency:LFI20:Rad:M}{69.2}
\setsymbol{LFI:center:frequency:LFI21:Rad:M}{70.4}
\setsymbol{LFI:center:frequency:LFI22:Rad:M}{71.5}
\setsymbol{LFI:center:frequency:LFI23:Rad:M}{70.8}
\setsymbol{LFI:center:frequency:LFI24:Rad:M}{44.4}
\setsymbol{LFI:center:frequency:LFI25:Rad:M}{44.0}
\setsymbol{LFI:center:frequency:LFI26:Rad:M}{43.9}
\setsymbol{LFI:center:frequency:LFI27:Rad:M}{28.3}
\setsymbol{LFI:center:frequency:LFI28:Rad:M}{28.8}
\setsymbol{LFI:center:frequency:LFI18:Rad:S}{70.1}
\setsymbol{LFI:center:frequency:LFI19:Rad:S}{69.6}
\setsymbol{LFI:center:frequency:LFI20:Rad:S}{69.5}
\setsymbol{LFI:center:frequency:LFI21:Rad:S}{69.5}
\setsymbol{LFI:center:frequency:LFI22:Rad:S}{72.8}
\setsymbol{LFI:center:frequency:LFI23:Rad:S}{71.3}
\setsymbol{LFI:center:frequency:LFI24:Rad:S}{44.1}
\setsymbol{LFI:center:frequency:LFI25:Rad:S}{44.1}
\setsymbol{LFI:center:frequency:LFI26:Rad:S}{44.1}
\setsymbol{LFI:center:frequency:LFI27:Rad:S}{28.5}
\setsymbol{LFI:center:frequency:LFI28:Rad:S}{28.2}

\setsymbol{LFI:white:noise:sensitivity:70GHz:units}{134.7\muKs}
\setsymbol{LFI:white:noise:sensitivity:44GHz:units}{164.7\muKs}
\setsymbol{LFI:white:noise:sensitivity:30GHz:units}{143.4\muKs}

\setsymbol{LFI:white:noise:sensitivity:70GHz}{134.7}
\setsymbol{LFI:white:noise:sensitivity:44GHz}{164.7}
\setsymbol{LFI:white:noise:sensitivity:30GHz}{143.4}

\setsymbol{LFI:white:noise:sensitivity:LFI18:Rad:M:units}{512.0\muKs}
\setsymbol{LFI:white:noise:sensitivity:LFI19:Rad:M:units}{581.4\muKs}
\setsymbol{LFI:white:noise:sensitivity:LFI20:Rad:M:units}{590.8\muKs}
\setsymbol{LFI:white:noise:sensitivity:LFI21:Rad:M:units}{455.2\muKs}
\setsymbol{LFI:white:noise:sensitivity:LFI22:Rad:M:units}{492.0\muKs}
\setsymbol{LFI:white:noise:sensitivity:LFI23:Rad:M:units}{507.7\muKs}
\setsymbol{LFI:white:noise:sensitivity:LFI24:Rad:M:units}{462.2\muKs}
\setsymbol{LFI:white:noise:sensitivity:LFI25:Rad:M:units}{413.6\muKs}
\setsymbol{LFI:white:noise:sensitivity:LFI26:Rad:M:units}{478.6\muKs}
\setsymbol{LFI:white:noise:sensitivity:LFI27:Rad:M:units}{277.7\muKs}
\setsymbol{LFI:white:noise:sensitivity:LFI28:Rad:M:units}{312.3\muKs}
\setsymbol{LFI:white:noise:sensitivity:LFI18:Rad:S:units}{465.7\muKs}
\setsymbol{LFI:white:noise:sensitivity:LFI19:Rad:S:units}{555.6\muKs}
\setsymbol{LFI:white:noise:sensitivity:LFI20:Rad:S:units}{623.2\muKs}
\setsymbol{LFI:white:noise:sensitivity:LFI21:Rad:S:units}{564.1\muKs}
\setsymbol{LFI:white:noise:sensitivity:LFI22:Rad:S:units}{534.4\muKs}
\setsymbol{LFI:white:noise:sensitivity:LFI23:Rad:S:units}{542.4\muKs}
\setsymbol{LFI:white:noise:sensitivity:LFI24:Rad:S:units}{399.2\muKs}
\setsymbol{LFI:white:noise:sensitivity:LFI25:Rad:S:units}{392.6\muKs}
\setsymbol{LFI:white:noise:sensitivity:LFI26:Rad:S:units}{418.6\muKs}
\setsymbol{LFI:white:noise:sensitivity:LFI27:Rad:S:units}{302.9\muKs}
\setsymbol{LFI:white:noise:sensitivity:LFI28:Rad:S:units}{285.3\muKs}

\setsymbol{LFI:white:noise:sensitivity:LFI18:Rad:M}{512.0}
\setsymbol{LFI:white:noise:sensitivity:LFI19:Rad:M}{581.4}
\setsymbol{LFI:white:noise:sensitivity:LFI20:Rad:M}{590.8}
\setsymbol{LFI:white:noise:sensitivity:LFI21:Rad:M}{455.2}
\setsymbol{LFI:white:noise:sensitivity:LFI22:Rad:M}{492.0}
\setsymbol{LFI:white:noise:sensitivity:LFI23:Rad:M}{507.7}
\setsymbol{LFI:white:noise:sensitivity:LFI24:Rad:M}{462.2}
\setsymbol{LFI:white:noise:sensitivity:LFI25:Rad:M}{413.6}
\setsymbol{LFI:white:noise:sensitivity:LFI26:Rad:M}{478.6}
\setsymbol{LFI:white:noise:sensitivity:LFI27:Rad:M}{277.7}
\setsymbol{LFI:white:noise:sensitivity:LFI28:Rad:M}{312.3}
\setsymbol{LFI:white:noise:sensitivity:LFI18:Rad:S}{465.7}
\setsymbol{LFI:white:noise:sensitivity:LFI19:Rad:S}{555.6}
\setsymbol{LFI:white:noise:sensitivity:LFI20:Rad:S}{623.2}
\setsymbol{LFI:white:noise:sensitivity:LFI21:Rad:S}{564.1}
\setsymbol{LFI:white:noise:sensitivity:LFI22:Rad:S}{534.4}
\setsymbol{LFI:white:noise:sensitivity:LFI23:Rad:S}{542.4}
\setsymbol{LFI:white:noise:sensitivity:LFI24:Rad:S}{399.2}
\setsymbol{LFI:white:noise:sensitivity:LFI25:Rad:S}{392.6}
\setsymbol{LFI:white:noise:sensitivity:LFI26:Rad:S}{418.6}
\setsymbol{LFI:white:noise:sensitivity:LFI27:Rad:S}{302.9}
\setsymbol{LFI:white:noise:sensitivity:LFI28:Rad:S}{285.3}

\setsymbol{LFI:knee:frequency:70GHz:units}{29.5\mHz}
\setsymbol{LFI:knee:frequency:44GHz:units}{56.2\mHz}
\setsymbol{LFI:knee:frequency:30GHz:units}{113.7\mHz}

\setsymbol{LFI:knee:frequency:70GHz}{29.5}
\setsymbol{LFI:knee:frequency:44GHz}{56.2}
\setsymbol{LFI:knee:frequency:30GHz}{113.7}

\setsymbol{LFI:knee:frequency:LFI18:Rad:M:units}{16.3\mHz}
\setsymbol{LFI:knee:frequency:LFI19:Rad:M:units}{15.1\mHz}
\setsymbol{LFI:knee:frequency:LFI20:Rad:M:units}{18.7\mHz}
\setsymbol{LFI:knee:frequency:LFI21:Rad:M:units}{37.2\mHz}
\setsymbol{LFI:knee:frequency:LFI22:Rad:M:units}{12.7\mHz}
\setsymbol{LFI:knee:frequency:LFI23:Rad:M:units}{34.6\mHz}
\setsymbol{LFI:knee:frequency:LFI24:Rad:M:units}{46.2\mHz}
\setsymbol{LFI:knee:frequency:LFI25:Rad:M:units}{24.9\mHz}
\setsymbol{LFI:knee:frequency:LFI26:Rad:M:units}{67.6\mHz}
\setsymbol{LFI:knee:frequency:LFI27:Rad:M:units}{187.4\mHz}
\setsymbol{LFI:knee:frequency:LFI28:Rad:M:units}{122.2\mHz}
\setsymbol{LFI:knee:frequency:LFI18:Rad:S:units}{17.7\mHz}
\setsymbol{LFI:knee:frequency:LFI19:Rad:S:units}{22.0\mHz}
\setsymbol{LFI:knee:frequency:LFI20:Rad:S:units}{8.7\mHz}
\setsymbol{LFI:knee:frequency:LFI21:Rad:S:units}{25.9\mHz}
\setsymbol{LFI:knee:frequency:LFI22:Rad:S:units}{15.8\mHz}
\setsymbol{LFI:knee:frequency:LFI23:Rad:S:units}{129.8\mHz}
\setsymbol{LFI:knee:frequency:LFI24:Rad:S:units}{100.9\mHz}
\setsymbol{LFI:knee:frequency:LFI25:Rad:S:units}{38.9\mHz}
\setsymbol{LFI:knee:frequency:LFI26:Rad:S:units}{58.9\mHz}
\setsymbol{LFI:knee:frequency:LFI27:Rad:S:units}{104.4\mHz}
\setsymbol{LFI:knee:frequency:LFI28:Rad:S:units}{40.7\mHz}

\setsymbol{LFI:knee:frequency:LFI18:Rad:M}{16.3}
\setsymbol{LFI:knee:frequency:LFI19:Rad:M}{15.1}
\setsymbol{LFI:knee:frequency:LFI20:Rad:M}{18.7}
\setsymbol{LFI:knee:frequency:LFI21:Rad:M}{37.2}
\setsymbol{LFI:knee:frequency:LFI22:Rad:M}{12.7}
\setsymbol{LFI:knee:frequency:LFI23:Rad:M}{34.6}
\setsymbol{LFI:knee:frequency:LFI24:Rad:M}{46.2}
\setsymbol{LFI:knee:frequency:LFI25:Rad:M}{24.9}
\setsymbol{LFI:knee:frequency:LFI26:Rad:M}{67.6}
\setsymbol{LFI:knee:frequency:LFI27:Rad:M}{187.4}
\setsymbol{LFI:knee:frequency:LFI28:Rad:M}{122.2}
\setsymbol{LFI:knee:frequency:LFI18:Rad:S}{17.7}
\setsymbol{LFI:knee:frequency:LFI19:Rad:S}{22.0}
\setsymbol{LFI:knee:frequency:LFI20:Rad:S}{8.7}
\setsymbol{LFI:knee:frequency:LFI21:Rad:S}{25.9}
\setsymbol{LFI:knee:frequency:LFI22:Rad:S}{15.8}
\setsymbol{LFI:knee:frequency:LFI23:Rad:S}{129.8}
\setsymbol{LFI:knee:frequency:LFI24:Rad:S}{100.9}
\setsymbol{LFI:knee:frequency:LFI25:Rad:S}{38.9}
\setsymbol{LFI:knee:frequency:LFI26:Rad:S}{58.9}
\setsymbol{LFI:knee:frequency:LFI27:Rad:S}{104.4}
\setsymbol{LFI:knee:frequency:LFI28:Rad:S}{40.7}

\setsymbol{LFI:slope:70GHz:units}{$-1.03$\mHz}
\setsymbol{LFI:slope:44GHz:units}{$-0.89$\mHz}
\setsymbol{LFI:slope:30GHz:units}{$-0.87$\mHz}

\setsymbol{LFI:slope:70GHz}{$-1.03$}
\setsymbol{LFI:slope:44GHz}{$-0.89$}
\setsymbol{LFI:slope:30GHz}{$-0.87$}

\setsymbol{LFI:slope:LFI18:Rad:M:units}{$-1.04$\mHz}
\setsymbol{LFI:slope:LFI19:Rad:M:units}{$-1.09$\mHz}
\setsymbol{LFI:slope:LFI20:Rad:M:units}{$-0.69$\mHz}
\setsymbol{LFI:slope:LFI21:Rad:M:units}{$-1.56$\mHz}
\setsymbol{LFI:slope:LFI22:Rad:M:units}{$-1.01$\mHz}
\setsymbol{LFI:slope:LFI23:Rad:M:units}{$-0.96$\mHz}
\setsymbol{LFI:slope:LFI24:Rad:M:units}{$-0.83$\mHz}
\setsymbol{LFI:slope:LFI25:Rad:M:units}{$-0.91$\mHz}
\setsymbol{LFI:slope:LFI26:Rad:M:units}{$-0.95$\mHz}
\setsymbol{LFI:slope:LFI27:Rad:M:units}{$-0.87$\mHz}
\setsymbol{LFI:slope:LFI28:Rad:M:units}{$-0.88$\mHz}
\setsymbol{LFI:slope:LFI18:Rad:S:units}{$-1.15$\mHz}
\setsymbol{LFI:slope:LFI19:Rad:S:units}{$-1.00$\mHz}
\setsymbol{LFI:slope:LFI20:Rad:S:units}{$-0.95$\mHz}
\setsymbol{LFI:slope:LFI21:Rad:S:units}{$-0.92$\mHz}
\setsymbol{LFI:slope:LFI22:Rad:S:units}{$-1.01$\mHz}
\setsymbol{LFI:slope:LFI23:Rad:S:units}{$-0.95$\mHz}
\setsymbol{LFI:slope:LFI24:Rad:S:units}{$-0.73$\mHz}
\setsymbol{LFI:slope:LFI25:Rad:S:units}{$-1.16$\mHz}
\setsymbol{LFI:slope:LFI26:Rad:S:units}{$-0.79$\mHz}
\setsymbol{LFI:slope:LFI27:Rad:S:units}{$-0.82$\mHz}
\setsymbol{LFI:slope:LFI28:Rad:S:units}{$-0.91$\mHz}

\setsymbol{LFI:slope:LFI18:Rad:M}{$-1.04$}
\setsymbol{LFI:slope:LFI19:Rad:M}{$-1.09$}
\setsymbol{LFI:slope:LFI20:Rad:M}{$-0.69$}
\setsymbol{LFI:slope:LFI21:Rad:M}{$-1.56$}
\setsymbol{LFI:slope:LFI22:Rad:M}{$-1.01$}
\setsymbol{LFI:slope:LFI23:Rad:M}{$-0.96$}
\setsymbol{LFI:slope:LFI24:Rad:M}{$-0.83$}
\setsymbol{LFI:slope:LFI25:Rad:M}{$-0.91$}
\setsymbol{LFI:slope:LFI26:Rad:M}{$-0.95$}
\setsymbol{LFI:slope:LFI27:Rad:M}{$-0.87$}
\setsymbol{LFI:slope:LFI28:Rad:M}{$-0.88$}
\setsymbol{LFI:slope:LFI18:Rad:S}{$-1.15$}
\setsymbol{LFI:slope:LFI19:Rad:S}{$-1.00$}
\setsymbol{LFI:slope:LFI20:Rad:S}{$-0.95$}
\setsymbol{LFI:slope:LFI21:Rad:S}{$-0.92$}
\setsymbol{LFI:slope:LFI22:Rad:S}{$-1.01$}
\setsymbol{LFI:slope:LFI23:Rad:S}{$-0.95$}
\setsymbol{LFI:slope:LFI24:Rad:S}{$-0.73$}
\setsymbol{LFI:slope:LFI25:Rad:S}{$-1.16$}
\setsymbol{LFI:slope:LFI26:Rad:S}{$-0.79$}
\setsymbol{LFI:slope:LFI27:Rad:S}{$-0.82$}
\setsymbol{LFI:slope:LFI28:Rad:S}{$-0.91$}

\setsymbol{LFI:FWHM:70GHz:units}{13\parcm01}
\setsymbol{LFI:FWHM:44GHz:units}{27\parcm92}
\setsymbol{LFI:FWHM:30GHz:units}{32\parcm65}

\setsymbol{LFI:FWHM:70GHz}{13.01}
\setsymbol{LFI:FWHM:44GHz}{27.92}
\setsymbol{LFI:FWHM:30GHz}{32.65}

\setsymbol{LFI:FWHM:LFI18:units}{13\parcm39}
\setsymbol{LFI:FWHM:LFI19:units}{13\parcm01}
\setsymbol{LFI:FWHM:LFI20:units}{12\parcm75}
\setsymbol{LFI:FWHM:LFI21:units}{12\parcm74}
\setsymbol{LFI:FWHM:LFI22:units}{12\parcm87}
\setsymbol{LFI:FWHM:LFI23:units}{13\parcm27}
\setsymbol{LFI:FWHM:LFI24:units}{22\parcm98}
\setsymbol{LFI:FWHM:LFI25:units}{30\parcm46}
\setsymbol{LFI:FWHM:LFI26:units}{30\parcm31}
\setsymbol{LFI:FWHM:LFI27:units}{32\parcm65}
\setsymbol{LFI:FWHM:LFI28:units}{32\parcm66}

\setsymbol{LFI:FWHM:LFI18}{13.39}
\setsymbol{LFI:FWHM:LFI19}{13.01}
\setsymbol{LFI:FWHM:LFI20}{12.75}
\setsymbol{LFI:FWHM:LFI21}{12.74}
\setsymbol{LFI:FWHM:LFI22}{12.87}
\setsymbol{LFI:FWHM:LFI23}{13.27}
\setsymbol{LFI:FWHM:LFI24}{22.98}
\setsymbol{LFI:FWHM:LFI25}{30.46}
\setsymbol{LFI:FWHM:LFI26}{30.31}
\setsymbol{LFI:FWHM:LFI27}{32.65}
\setsymbol{LFI:FWHM:LFI28}{32.66}

\setsymbol{LFI:FWHM:uncertainty:LFI18:units}{0.170\arcm}
\setsymbol{LFI:FWHM:uncertainty:LFI19:units}{0.174\arcm}
\setsymbol{LFI:FWHM:uncertainty:LFI20:units}{0.170\arcm}
\setsymbol{LFI:FWHM:uncertainty:LFI21:units}{0.156\arcm}
\setsymbol{LFI:FWHM:uncertainty:LFI22:units}{0.164\arcm}
\setsymbol{LFI:FWHM:uncertainty:LFI23:units}{0.171\arcm}
\setsymbol{LFI:FWHM:uncertainty:LFI24:units}{0.652\arcm}
\setsymbol{LFI:FWHM:uncertainty:LFI25:units}{1.075\arcm}
\setsymbol{LFI:FWHM:uncertainty:LFI26:units}{1.131\arcm}
\setsymbol{LFI:FWHM:uncertainty:LFI27:units}{1.266\arcm}
\setsymbol{LFI:FWHM:uncertainty:LFI28:units}{1.287\arcm}

\setsymbol{LFI:FWHM:uncertainty:LFI18}{0.170}
\setsymbol{LFI:FWHM:uncertainty:LFI19}{0.174}
\setsymbol{LFI:FWHM:uncertainty:LFI20}{0.170}
\setsymbol{LFI:FWHM:uncertainty:LFI21}{0.156}
\setsymbol{LFI:FWHM:uncertainty:LFI22}{0.164}
\setsymbol{LFI:FWHM:uncertainty:LFI23}{0.171}
\setsymbol{LFI:FWHM:uncertainty:LFI24}{0.652}
\setsymbol{LFI:FWHM:uncertainty:LFI25}{1.075}
\setsymbol{LFI:FWHM:uncertainty:LFI26}{1.131}
\setsymbol{LFI:FWHM:uncertainty:LFI27}{1.266}
\setsymbol{LFI:FWHM:uncertainty:LFI28}{1.287}

\setsymbol{HFI:center:frequency:100GHz:units}{100\,GHz}
\setsymbol{HFI:center:frequency:143GHz:units}{143\,GHz}
\setsymbol{HFI:center:frequency:217GHz:units}{217\,GHz}
\setsymbol{HFI:center:frequency:353GHz:units}{353\,GHz}
\setsymbol{HFI:center:frequency:545GHz:units}{545\,GHz}
\setsymbol{HFI:center:frequency:857GHz:units}{857\,GHz}

\setsymbol{HFI:center:frequency:100GHz}{100}
\setsymbol{HFI:center:frequency:143GHz}{143}
\setsymbol{HFI:center:frequency:217GHz}{217}
\setsymbol{HFI:center:frequency:353GHz}{353}
\setsymbol{HFI:center:frequency:545GHz}{545}
\setsymbol{HFI:center:frequency:857GHz}{857}

\setsymbol{HFI:Ndetectors:100GHz}{8}
\setsymbol{HFI:Ndetectors:143GHz}{11}
\setsymbol{HFI:Ndetectors:217GHz}{12}
\setsymbol{HFI:Ndetectors:353GHz}{12}
\setsymbol{HFI:Ndetectors:545GHz}{3}
\setsymbol{HFI:Ndetectors:857GHz}{4}

\setsymbol{HFI:FWHM:Maps:100GHz:units}{9\parcm88}
\setsymbol{HFI:FWHM:Maps:143GHz:units}{7\parcm18}
\setsymbol{HFI:FWHM:Maps:217GHz:units}{4\parcm87}
\setsymbol{HFI:FWHM:Maps:353GHz:units}{4\parcm65}
\setsymbol{HFI:FWHM:Maps:545GHz:units}{4\parcm72}
\setsymbol{HFI:FWHM:Maps:857GHz:units}{4\parcm39}
\setsymbol{HFI:FWHM:Maps:100GHz}{9.88}
\setsymbol{HFI:FWHM:Maps:143GHz}{7.18}
\setsymbol{HFI:FWHM:Maps:217GHz}{4.87}
\setsymbol{HFI:FWHM:Maps:353GHz}{4.65}
\setsymbol{HFI:FWHM:Maps:545GHz}{4.72}
\setsymbol{HFI:FWHM:Maps:857GHz}{4.39}

\setsymbol{HFI:beam:ellipticity:Maps:100GHz}{1.15}
\setsymbol{HFI:beam:ellipticity:Maps:143GHz}{1.01}
\setsymbol{HFI:beam:ellipticity:Maps:217GHz}{1.06}
\setsymbol{HFI:beam:ellipticity:Maps:353GHz}{1.05}
\setsymbol{HFI:beam:ellipticity:Maps:545GHz}{1.14}
\setsymbol{HFI:beam:ellipticity:Maps:857GHz}{1.19}

\setsymbol{HFI:FWHM:Mars:100GHz:units}{9\parcm37}
\setsymbol{HFI:FWHM:Mars:143GHz:units}{7\parcm04}
\setsymbol{HFI:FWHM:Mars:217GHz:units}{4\parcm68}
\setsymbol{HFI:FWHM:Mars:353GHz:units}{4\parcm43}
\setsymbol{HFI:FWHM:Mars:545GHz:units}{3\parcm80}
\setsymbol{HFI:FWHM:Mars:857GHz:units}{3\parcm67}

\setsymbol{HFI:FWHM:Mars:100GHz}{9.37}
\setsymbol{HFI:FWHM:Mars:143GHz}{7.04}
\setsymbol{HFI:FWHM:Mars:217GHz}{4.68}
\setsymbol{HFI:FWHM:Mars:353GHz}{4.43}
\setsymbol{HFI:FWHM:Mars:545GHz}{3.80}
\setsymbol{HFI:FWHM:Mars:857GHz}{3.67}

\setsymbol{HFI:beam:ellipticity:Mars:100GHz}{1.18}
\setsymbol{HFI:beam:ellipticity:Mars:143GHz}{1.03}
\setsymbol{HFI:beam:ellipticity:Mars:217GHz}{1.14}
\setsymbol{HFI:beam:ellipticity:Mars:353GHz}{1.09}
\setsymbol{HFI:beam:ellipticity:Mars:545GHz}{1.25}
\setsymbol{HFI:beam:ellipticity:Mars:857GHz}{1.03}

\setsymbol{HFI:CMB:relative:calibration:100GHz}{$\lsim 1\%$}
\setsymbol{HFI:CMB:relative:calibration:143GHz}{$\lsim 1\%$}
\setsymbol{HFI:CMB:relative:calibration:217GHz}{$\lsim 1\%$}
\setsymbol{HFI:CMB:relative:calibration:353GHz}{$\lsim 1\%$}
\setsymbol{HFI:CMB:relative:calibration:545GHz}{}
\setsymbol{HFI:CMB:relative:calibration:857GHz}{}

\setsymbol{HFI:CMB:absolute:calibration:100GHz}{$\lsim 2\%$}
\setsymbol{HFI:CMB:absolute:calibration:143GHz}{$\lsim 2\%$}
\setsymbol{HFI:CMB:absolute:calibration:217GHz}{$\lsim 2\%$}
\setsymbol{HFI:CMB:absolute:calibration:353GHz}{$\lsim 2\%$}
\setsymbol{HFI:CMB:absolute:calibration:545GHz}{}
\setsymbol{HFI:CMB:absolute:calibration:857GHz}{}

\setsymbol{HFI:FIRAS:gain:calibration:accuracy:statistical:100GHz}{}
\setsymbol{HFI:FIRAS:gain:calibration:accuracy:statistical:143GHz}{}
\setsymbol{HFI:FIRAS:gain:calibration:accuracy:statistical:217GHz}{}
\setsymbol{HFI:FIRAS:gain:calibration:accuracy:statistical:353GHz}{2.5\%}
\setsymbol{HFI:FIRAS:gain:calibration:accuracy:statistical:545GHz}{1\%}
\setsymbol{HFI:FIRAS:gain:calibration:accuracy:statistical:857GHz}{0.5\%}

\setsymbol{HFI:FIRAS:gain:calibration:accuracy:systematic:100GHz}{}
\setsymbol{HFI:FIRAS:gain:calibration:accuracy:systematic:143GHz}{}
\setsymbol{HFI:FIRAS:gain:calibration:accuracy:systematic:217GHz}{}
\setsymbol{HFI:FIRAS:gain:calibration:accuracy:systematic:353GHz}{}
\setsymbol{HFI:FIRAS:gain:calibration:accuracy:systematic:545GHz}{7\%}
\setsymbol{HFI:FIRAS:gain:calibration:accuracy:systematic:857GHz}{7\%}

\setsymbol{HFI:FIRAS:zero:point:accuracy:100GHz:units}{0.8\MJysr}
\setsymbol{HFI:FIRAS:zero:point:accuracy:143GHz:units}{}
\setsymbol{HFI:FIRAS:zero:point:accuracy:217GHz:units}{}
\setsymbol{HFI:FIRAS:zero:point:accuracy:353GHz:units}{1.4\MJysr}
\setsymbol{HFI:FIRAS:zero:point:accuracy:545GHz:units}{2.2\MJysr}
\setsymbol{HFI:FIRAS:zero:point:accuracy:857GHz:units}{1.7\MJysr}

\setsymbol{HFI:FIRAS:zero:point:accuracy:100GHz}{0.8}
\setsymbol{HFI:FIRAS:zero:point:accuracy:143GHz}{}
\setsymbol{HFI:FIRAS:zero:point:accuracy:217GHz}{}
\setsymbol{HFI:FIRAS:zero:point:accuracy:353GHz}{1.4}
\setsymbol{HFI:FIRAS:zero:point:accuracy:545GHz}{2.2}
\setsymbol{HFI:FIRAS:zero:point:accuracy:857GHz}{1.7}

\setsymbol{HFI:unit:conversion:100GHz:units}{0.2415\MJysrmK}
\setsymbol{HFI:unit:conversion:143GHz:units}{0.3694\MJysrmK}
\setsymbol{HFI:unit:conversion:217GHz:units}{0.4811\MJysrmK}
\setsymbol{HFI:unit:conversion:353GHz:units}{0.2883\MJysrmK}
\setsymbol{HFI:unit:conversion:545GHz:units}{0.05826\MJysrmK}
\setsymbol{HFI:unit:conversion:857GHz:units}{0.002238\MJysrmK}

\setsymbol{HFI:unit:conversion:100GHz}{0.2415}
\setsymbol{HFI:unit:conversion:143GHz}{0.3694}
\setsymbol{HFI:unit:conversion:217GHz}{0.4811}
\setsymbol{HFI:unit:conversion:353GHz}{0.2883}
\setsymbol{HFI:unit:conversion:545GHz}{0.05826}
\setsymbol{HFI:unit:conversion:857GHz}{0.002238}

\setsymbol{HFI:colour:correction:alpha=-2:V1.01:100GHz}{0.9893}
\setsymbol{HFI:colour:correction:alpha=-2:V1.01:143GHz}{0.9759}
\setsymbol{HFI:colour:correction:alpha=-2:V1.01:217GHz}{1.0007}
\setsymbol{HFI:colour:correction:alpha=-2:V1.01:353GHz}{1.0028}
\setsymbol{HFI:colour:correction:alpha=-2:V1.01:545GHz}{1.0019}
\setsymbol{HFI:colour:correction:alpha=-2:V1.01:857GHz}{0.9889}

\setsymbol{HFI:colour:correction:alpha=0:V1.01:100GHz}{1.0008}
\setsymbol{HFI:colour:correction:alpha=0:V1.01:143GHz}{1.0148}
\setsymbol{HFI:colour:correction:alpha=0:V1.01:217GHz}{0.9909}
\setsymbol{HFI:colour:correction:alpha=0:V1.01:353GHz}{0.9888}
\setsymbol{HFI:colour:correction:alpha=0:V1.01:545GHz}{0.9878}
\setsymbol{HFI:colour:correction:alpha=0:V1.01:857GHz}{1.0014}

\providecommand{\sorthelp}[1]{}

%% file: polar_definitions.tex
\newcommand{\polang}{\rm \psi}                  %polarization angle (CMB convention)
\def\covar{\tens{C}}
\def\sigII{\covar_{II}}  % variance of I ( the dimension is dimension(I)^2, the same as for covariances)
\def\sigQQ{\covar_{QQ}} % variance of Q
\def\sigUU{\covar_{UU}} % variance of U
\def\sigIQ{\covar_{IQ}} % covariance between I and Q (same dimension as for sigII)
\def\sigIU{\covar_{IU}} % covariance between I and U
\def\sigQU{\covar_{QU}}  %covariance between U and Q
\def\sigQsIQsI{\covar_{Q/I,Q/I}} 
\def\sigQsIUsI{\covar_{Q/I,U/I}}
\def\sigUsIUsI{\covar_{U/I,U/I}}

\def\Scovar{\mathscr{C}}

\def\MatC{[\covar]}

%% file: PIP76_Bootstrap_Results_DX9_22sep2014.tex
\def\RmainQUcorrBPM{4.1}
\def\dRmainQUcorrBPM{0.2}
\def\nmainQUcorrBPM{206}
\def\RmainQUNOTcorrBPM{4.3}
\def\dRmainQUNOTcorrBPM{0.2}
\def\nmainQUNOTcorrBPM{196}

\def\RmainPdebiasedcorrBPM{4.2}
\def\dRmainPdebiasedcorrBPM{0.1}

\def\RmainPbiasedcorrBPM{4.2}
\def\dRmainPbiasedcorrBPM{0.1}

\def\RSNRAcorrBPM{4.1}
\def\dRSNRAcorrBPM{0.1}

\def\SNRC{10}
\def\RSNRCcorrBPM{4.3}
\def\dRSNRCcorrBPM{0.2}

\def\RDISMcorrBPM{4.0}
\def\dRDISMcorrBPM{0.2}

\def\ebvsubmaxA{0.3}
\def\RebvsubmaxAcorrBPM{4.7}
\def\dRebvsubmaxAcorrBPM{0.4}
\def\nebvsubmaxAcorrBPM{42}

\def\ebvsubmaxB{0.4}
\def\RebvsubmaxBcorrBPM{4.7}
\def\dRebvsubmaxBcorrBPM{0.2}
\def\nebvsubmaxBcorrBPM{82}

\def\ebvsubmaxC{0.6}
\def\RebvsubmaxCcorrBPM{4.2}
\def\dRebvsubmaxCcorrBPM{0.1}
\def\nebvsubmaxCcorrBPM{153}

\def\ROrthocorrBPM{4.2}
\def\dROrthocorrBPM{0.2}
\def\nOrthocorrBPM{112}

\def\NHratioA{1.2}
\def\RNHratioAcorrBPM{4.2}
\def\dRNHratioAcorrBPM{0.2}
\def\nNHratioAcorrBPM{121}

\def\NHratioD{1.8}
\def\RNHratioDcorrBPM{4.1}
\def\dRNHratioDcorrBPM{0.1}
\def\nNHratioDcorrBPM{231}

\def\NHratioE{2.0}
\def\RNHratioEcorrBPM{4.1}
\def\dRNHratioEcorrBPM{0.2}
\def\nNHratioEcorrBPM{251}

\def\NHratioF{3.0}

\def\nNHratioFcorrBPM{284}

\def\HeightA{100}
\def\RHeightAcorrBPM{4.1}
\def\dRHeightAcorrBPM{0.1}
\def\nHeightAcorrBPM{136}

\def\HeightB{150}
\def\RHeightBcorrBPM{4.1}
\def\dRHeightBcorrBPM{0.2}
\def\nHeightBcorrBPM{91}

\def\HeightC{200}
\def\RHeightCcorrBPM{4.2}
\def\dRHeightCcorrBPM{0.2}
\def\nHeightCcorrBPM{53}

\def\Heightalone{100}
\def\RHeightalonecorrBPM{4.0}
\def\dRHeightalonecorrBPM{0.2}
\def\nHeightalonecorrBPM{172}

\def\SmoothB{3}

\def\RSmoothCcorrBPM{4.1}
\def\dRSmoothCcorrBPM{0.2}

\def\SmoothD{8}
\def\RSmoothDcorrBPM{4.2}
\def\dRSmoothDcorrBPM{0.2}

\def\RSmoothEcorrBPM{4.0}
\def\dRSmoothEcorrBPM{0.2}

\def\RFMcorrBPM{3.9}
\def\dRFMcorrBPM{0.3}
\def\nFMcorrBPM{32}

\def\RVAcorrBPM{4.2}
\def\dRVAcorrBPM{0.2}
\def\nVAcorrBPM{48}

\def\RWEcorrBPM{4.1}
\def\dRWEcorrBPM{0.2}
\def\nWEcorrBPM{53}

\def\RSAcorrBPM{4.4}
\def\dRSAcorrBPM{0.2}
\def\nSAcorrBPM{97}

\def\RKRcorrBPM{4.1}
\def\dRKRcorrBPM{0.1}
\def\nKRcorrBPM{135}

\def\RNorthcorrBPM{4.0}
\def\dRNorthcorrBPM{0.3}

\def\RSouthcorrBPM{4.2}
\def\dRSouthcorrBPM{0.1}

\def\RHighLatcorrBPM{4.2}
\def\dRHighLatcorrBPM{0.3}
\def\nHighLatcorrBPM{95}
\def\RHighLatNOTcorrBPM{4.1}
\def\dRHighLatNOTcorrBPM{0.3}

\def\RLowLatcorrBPM{4.1}
\def\dRLowLatcorrBPM{0.1}
\def\nLowLatcorrBPM{111}
\def\RLowLatNOTcorrBPM{4.4}
\def\dRLowLatNOTcorrBPM{0.2}

\def\RFancorrBPM{4.2}
\def\dRFancorrBPM{0.2}
\def\nFancorrBPM{65}

\def\RAquilacorrBPM{4.9}
\def\dRAquilacorrBPM{0.3}
\def\nAquilacorrBPM{20}

\def\RAriacorrBPM{4.1}
\def\dRAriacorrBPM{0.4}
\def\nAriacorrBPM{22}

\def\RnoFancorrBPM{4.2}
\def\dRnoFancorrBPM{0.2}

\def\RIsAvHighLatcorrBPM{1.24}
\def\dRIsAvHighLatcorrBPM{0.07}

\def\RIsAvLowLatcorrBPM{1.25}
\def\dRIsAvLowLatcorrBPM{0.04}

\def\RIsAvFancorrBPM{1.20}
\def\dRIsAvFancorrBPM{0.04}

\def\RIsAvAquilacorrBPM{1.27}
\def\dRIsAvAquilacorrBPM{0.16}

\def\RIsAvAriacorrBPM{1.46}
\def\dRIsAvAriacorrBPM{0.07}

\def\RIsAvFMcorrBPM{1.29}
\def\dRIsAvFMcorrBPM{0.08}

\def\RIsAvVAcorrBPM{1.24}
\def\dRIsAvVAcorrBPM{0.06}

\def\RIsAvWEcorrBPM{1.26}
\def\dRIsAvWEcorrBPM{0.07}

\def\RIsAvSAcorrBPM{1.33}
\def\dRIsAvSAcorrBPM{0.04}

\def\RIsAvKRcorrBPM{1.33}
\def\dRIsAvKRcorrBPM{0.03}

\def\RRvcorrBPM{4.0}
\def\dRRvcorrBPM{0.2}
\def\nRvcorrBPM{69}

\def\RlmaxcorrBPM{4.0}
\def\dRlmaxcorrBPM{0.3}
\def\nlmaxcorrBPM{34}

\def\RrmvzodicorrBPM{4.1}
\def\dRrmvzodicorrBPM{0.2}

\def\RkeepzodicorrBPM{4.1}
\def\dRkeepzodicorrBPM{0.2}

\def\RCMBsubstractedcorrBPM{4.1}
\def\dRCMBsubstractedcorrBPM{0.2}

%% file: PIP76_PpBootstrap_Results_DX9_22sep2014.tex
\def\RPpmainQUNOTcorrBPM{5.5}
\def\dRPpmainQUNOTcorrBPM{0.2}

\def\RPpmainPdebiasedcorrBPM{5.4}
\def\dRPpmainPdebiasedcorrBPM{0.1}

\def\RPpmainPbiasedcorrBPM{5.4}
\def\dRPpmainPbiasedcorrBPM{0.1}

\def\PpSNRA{1}
\def\RPpSNRAcorrBPM{5.3}
\def\dRPpSNRAcorrBPM{0.2}
\def\nPpSNRAcorrBPM{268}

\def\RPpSNRCcorrBPM{5.4}
\def\dRPpSNRCcorrBPM{0.2}
\def\nPpSNRCcorrBPM{68}

\def\RPpDISMcorrBPM{5.7}
\def\dRPpDISMcorrBPM{0.2}
\def\nPpDISMcorrBPM{284}

\def\RPpebvsubmaxAcorrBPM{5.6}
\def\dRPpebvsubmaxAcorrBPM{0.4}

\def\RPpebvsubmaxBcorrBPM{5.7}
\def\dRPpebvsubmaxBcorrBPM{0.2}

\def\RPpebvsubmaxCcorrBPM{5.3}
\def\dRPpebvsubmaxCcorrBPM{0.2}

\def\RPpOrthocorrBPM{5.5}
\def\dRPpOrthocorrBPM{0.2}

\def\RPpHeightAcorrBPM{5.2}
\def\dRPpHeightAcorrBPM{0.1}

\def\RPpHeightBcorrBPM{5.3}
\def\dRPpHeightBcorrBPM{0.2}

\def\RPpHeightCcorrBPM{5.2}
\def\dRPpHeightCcorrBPM{0.2}

\def\RPpHeightalonecorrBPM{5.2}
\def\dRPpHeightalonecorrBPM{0.2}

\def\RPpSmoothCcorrBPM{5.3}
\def\dRPpSmoothCcorrBPM{0.2}

\def\RPpSmoothDcorrBPM{5.4}
\def\dRPpSmoothDcorrBPM{0.2}

\def\RPpSmoothEcorrBPM{5.2}
\def\dRPpSmoothEcorrBPM{0.3}

\def\RPpFMcorrBPM{5.1}
\def\dRPpFMcorrBPM{0.3}

\def\RPpVAcorrBPM{5.2}
\def\dRPpVAcorrBPM{0.3}

\def\RPpWEcorrBPM{5.2}
\def\dRPpWEcorrBPM{0.2}

\def\RPpSAcorrBPM{5.3}
\def\dRPpSAcorrBPM{0.3}

\def\RPpKRcorrBPM{5.3}
\def\dRPpKRcorrBPM{0.2}

\def\RPpNorthcorrBPM{5.1}
\def\dRPpNorthcorrBPM{0.3}

\def\RPpSouthcorrBPM{5.4}
\def\dRPpSouthcorrBPM{0.2}

\def\RPpHighLatcorrBPM{5.2}
\def\dRPpHighLatcorrBPM{0.3}

\def\RPpHighLatNOTcorrBPM{5.0}
\def\dRPpHighLatNOTcorrBPM{0.3}

\def\RPpLowLatcorrBPM{5.3}
\def\dRPpLowLatcorrBPM{0.2}

\def\RPpLowLatNOTcorrBPM{5.7}
\def\dRPpLowLatNOTcorrBPM{0.2}

\def\RPpFanQUcorrBPM{5.3}
\def\dRPpFanQUcorrBPM{0.2}

\def\RPpAquilaQUcorrBPM{5.6}
\def\dRPpAquilaQUcorrBPM{0.3}

\def\RPpAriaQUcorrBPM{6.0}
\def\dRPpAriaQUcorrBPM{0.5}

\def\RPpnoFanQUcorrBPM{5.4}
\def\dRPpnoFanQUcorrBPM{0.2}

\def\RPpRvcorrBPM{5.2}
\def\dRPpRvcorrBPM{0.2}

\def\RPplmaxcorrBPM{5.1}
\def\dRPplmaxcorrBPM{0.3}

\def\RPpQUcorrBPM{5.3}
\def\dRPpQUcorrBPM{0.2}

\def\RPpFMcorrBPM{5.1}
\def\dRPpFMcorrBPM{0.3}

\def\RPpVAcorrBPM{5.2}
\def\dRPpVAcorrBPM{0.3}

\def\RPpWEcorrBPM{5.2}
\def\dRPpWEcorrBPM{0.2}

\def\RPpSAcorrBPM{5.3}
\def\dRPpSAcorrBPM{0.3}

\def\RPpKRcorrBPM{5.3}
\def\dRPpKRcorrBPM{0.2}

\def\RPpCMBsubstractedcorrBPM{5.2}
\def\dRPpCMBsubstractedcorrBPM{0.2}
\def\nPpCMBsubstractedcorrBPM{203}

%% file: PIP76_Fit_Results_DX9_22sep2014.tex
\def\RfitQUcorrBPM{-4.13}
\def\dRfitQUcorrBPM{0.06}
\def\nfitQUcorrBPM{206}
\def\bfitQUcorrBPM{0.0006}
\def\dbfitQUcorrBPM{0.0007}
\def\PearsonfitQUcorrBPM{-0.93}
\def\chifitQUcorrBPM{1.64}

\def\RfitQcorrBPM{-3.90}
\def\dRfitQcorrBPM{0.09}

\def\bfitQcorrBPM{0.0060}
\def\dbfitQcorrBPM{0.0010}
\def\PearsonfitQcorrBPM{-0.92}

\def\RfitUcorrBPM{-4.03}
\def\dRfitUcorrBPM{0.14}

\def\bfitUcorrBPM{-0.0014}
\def\dbfitUcorrBPM{0.0009}
\def\PearsonfitUcorrBPM{-0.87}

\def\RfitPcorrBPM{4.18}
\def\dRfitPcorrBPM{0.04}

\def\PearsonfitPcorrBPM{0.74}
\def\chifitPcorrBPM{1.77}

\def\RfitPpQUcorrBPM{-5.32}
\def\dRfitPpQUcorrBPM{0.06}

\def\bfitPpQUcorrBPM{0.0020}
\def\dbfitPpQUcorrBPM{0.0009}
\def\PearsonfitPpQUcorrBPM{-0.95}
\def\chifitPpQUcorrBPM{2.29}

\def\RfitPpQcorrBPM{-5.17}
\def\dRfitPpQcorrBPM{0.09}

\def\bfitPpQcorrBPM{0.0077}
\def\dbfitPpQcorrBPM{0.0013}
\def\PearsonfitPpQcorrBPM{-0.94}

\def\RfitPpUcorrBPM{-5.07}
\def\dRfitPpUcorrBPM{0.13}

\def\bfitPpUcorrBPM{-0.0016}
\def\dbfitPpUcorrBPM{0.0013}
\def\PearsonfitPpUcorrBPM{-0.88}

\def\RfitPpPcorrBPM{5.41}
\def\dRfitPpPcorrBPM{0.04}

\def\PearsonfitPpPcorrBPM{0.87}
\def\chifitPpPcorrBPM{2.98}

\def\PearsonfitQUNOTcorrBPM{-0.94}
\def\chifitQUNOTcorrBPM{1.79}

\def\PearsonfitPpQUNOTcorrBPM{-0.95}
\def\chifitPpQUNOTcorrBPM{2.56}

%% file: PIP_76_Proj_7_9_Martin_authors_and_institutes.tex
%This author list corresponds to \title{Author list for SVN PIP\_76\_Proj\_7\_9\_Martin: Thermal dust emission at 353 GHz}
%Prepared by R. Leonardi (rleonardi@sciops.esa.int), ESAC/ESA
%This version is from Fri May 30 13:32:58 2014 CET
%\subtitle{There are 195 co-authors in this list}
\author{\small
Planck Collaboration:
P.~A.~R.~Ade\inst{78}
\and
N.~Aghanim\inst{54}
\and
D.~Alina\inst{83, 10}
\and
G.~Aniano\inst{54}
\and
C.~Armitage-Caplan\inst{81}
\and
M.~Arnaud\inst{67}
\and
M.~Ashdown\inst{64, 6}
\and
F.~Atrio-Barandela\inst{18}
\and
J.~Aumont\inst{54}
\and
C.~Baccigalupi\inst{77}
% ...
}
\institute{\small
APC, AstroParticule et Cosmologie, Universit\'{e} Paris Diderot, CNRS/IN2P3, CEA/lrfu, Observatoire de Paris, Sorbonne Paris Cit\'{e}, 10, rue Alice Domon et L\'{e}onie Duquet, 75205 Paris Cedex 13, France\\
\and
Aalto University Mets\"{a}hovi Radio Observatory and Dept of Radio Science and Engineering, P.O. Box 13000, FI-00076 AALTO, Finland\\
\and
African Institute for Mathematical Sciences, 6-8 Melrose Road, Muizenberg, Cape Town, South Africa\\
\and
Agenzia Spaziale Italiana Science Data Center, Via del Politecnico snc, 00133, Roma, Italy\\
\and
Agenzia Spaziale Italiana, Viale Liegi 26, Roma, Italy\\
\and
Astrophysics Group, Cavendish Laboratory, University of Cambridge, J J Thomson Avenue, Cambridge CB3 0HE, U.K.\\
\and
Astrophysics \& Cosmology Research Unit, School of Mathematics, Statistics \& Computer Science, University of KwaZulu-Natal, Westville Campus, Private Bag X54001, Durban 4000, South Africa\\
\and
Atacama Large Millimeter/submillimeter Array, ALMA Santiago Central Offices, Alonso de Cordova 3107, Vitacura, Casilla 763 0355, Santiago, Chile\\
\and
CITA, University of Toronto, 60 St. George St., Toronto, ON M5S 3H8, Canada\\
\and
CNRS, IRAP, 9 Av. colonel Roche, BP 44346, F-31028 Toulouse cedex 4, France\\
\and
California Institute of Technology, Pasadena, California, U.S.A.\\
\and
Centro de Estudios de F\'{i}sica del Cosmos de Arag\'{o}n (CEFCA), Plaza San Juan, 1, planta 2, E-44001, Teruel, Spain\\
\and
Computational Cosmology Center, Lawrence Berkeley National Laboratory, Berkeley, California, U.S.A.\\
\and
Consejo Superior de Investigaciones Cient\'{\i}ficas (CSIC), Madrid, Spain\\
\and
DSM/Irfu/SPP, CEA-Saclay, F-91191 Gif-sur-Yvette Cedex, France\\
\and
DTU Space, National Space Institute, Technical University of Denmark, Elektrovej 327, DK-2800 Kgs. Lyngby, Denmark\\
\and
D\'{e}partement de Physique Th\'{e}orique, Universit\'{e} de Gen\`{e}ve, 24, Quai E. Ansermet,1211 Gen\`{e}ve 4, Switzerland\\
\and
Departamento de F\'{\i}sica Fundamental, Facultad de Ciencias, Universidad de Salamanca, 37008 Salamanca, Spain\\
\and
Departamento de F\'{\i}sica, Universidad de Oviedo, Avda. Calvo Sotelo s/n, Oviedo, Spain\\
\and
Department of Astronomy and Astrophysics, University of Toronto, 50 Saint George Street, Toronto, Ontario, Canada\\
\and
Department of Astrophysics/IMAPP, Radboud University Nijmegen, P.O. Box 9010, 6500 GL Nijmegen, The Netherlands\\
\and
Department of Physics \& Astronomy, University of British Columbia, 6224 Agricultural Road, Vancouver, British Columbia, Canada\\
\and
Department of Physics and Astronomy, Dana and David Dornsife College of Letter, Arts and Sciences, University of Southern California, Los Angeles, CA 90089, U.S.A.\\
\and
Department of Physics and Astronomy, University College London, London WC1E 6BT, U.K.\\
\and
Department of Physics, Florida State University, Keen Physics Building, 77 Chieftan Way, Tallahassee, Florida, U.S.A.\\
\and
Department of Physics, Gustaf H\"{a}llstr\"{o}min katu 2a, University of Helsinki, Helsinki, Finland\\
\and
Department of Physics, Princeton University, Princeton, New Jersey, U.S.A.\\
\and
Department of Physics, University of California, Santa Barbara, California, U.S.A.\\
\and
Department of Physics, University of Illinois at Urbana-Champaign, 1110 West Green Street, Urbana, Illinois, U.S.A.\\
\and
Dipartimento di Fisica e Astronomia G. Galilei, Universit\`{a} degli Studi di Padova, via Marzolo 8, 35131 Padova, Italy\\
\and
Dipartimento di Fisica e Scienze della Terra, Universit\`{a} di Ferrara, Via Saragat 1, 44122 Ferrara, Italy\\
\and
Dipartimento di Fisica, Universit\`{a} La Sapienza, P. le A. Moro 2, Roma, Italy\\
\and
Dipartimento di Fisica, Universit\`{a} degli Studi di Milano, Via Celoria, 16, Milano, Italy\\
\and
Dipartimento di Fisica, Universit\`{a} degli Studi di Trieste, via A. Valerio 2, Trieste, Italy\\
\and
Dipartimento di Fisica, Universit\`{a} di Roma Tor Vergata, Via della Ricerca Scientifica, 1, Roma, Italy\\
\and
Discovery Center, Niels Bohr Institute, Blegdamsvej 17, Copenhagen, Denmark\\
\and
Dpto. Astrof\'{i}sica, Universidad de La Laguna (ULL), E-38206 La Laguna, Tenerife, Spain\\
\and
European Southern Observatory, ESO Vitacura, Alonso de Cordova 3107, Vitacura, Casilla 19001, Santiago, Chile\\
\and
European Space Agency, ESAC, Planck Science Office, Camino bajo del Castillo, s/n, Urbanizaci\'{o}n Villafranca del Castillo, Villanueva de la Ca\~{n}ada, Madrid, Spain\\
\and
European Space Agency, ESTEC, Keplerlaan 1, 2201 AZ Noordwijk, The Netherlands\\
\and
Helsinki Institute of Physics, Gustaf H\"{a}llstr\"{o}min katu 2, University of Helsinki, Helsinki, Finland\\
\and
INAF - Osservatorio Astrofisico di Catania, Via S. Sofia 78, Catania, Italy\\
\and
INAF - Osservatorio Astronomico di Padova, Vicolo dell'Osservatorio 5, Padova, Italy\\
\and
INAF - Osservatorio Astronomico di Roma, via di Frascati 33, Monte Porzio Catone, Italy\\
\and
INAF - Osservatorio Astronomico di Trieste, Via G.B. Tiepolo 11, Trieste, Italy\\
\and
INAF/IASF Bologna, Via Gobetti 101, Bologna, Italy\\
\and
INAF/IASF Milano, Via E. Bassini 15, Milano, Italy\\
\and
INFN, Sezione di Bologna, Via Irnerio 46, I-40126, Bologna, Italy\\
\and
INFN, Sezione di Roma 1, Universit\`{a} di Roma Sapienza, Piazzale Aldo Moro 2, 00185, Roma, Italy\\
\and
INFN/National Institute for Nuclear Physics, Via Valerio 2, I-34127 Trieste, Italy\\
\and
IPAG: Institut de Plan\'{e}tologie et d'Astrophysique de Grenoble, Universit\'{e} Grenoble Alpes, IPAG, F-38000 Grenoble, France, CNRS, IPAG, F-38000 Grenoble, France\\
\and
Imperial College London, Astrophysics group, Blackett Laboratory, Prince Consort Road, London, SW7 2AZ, U.K.\\
\and
Infrared Processing and Analysis Center, California Institute of Technology, Pasadena, CA 91125, U.S.A.\\
\and
Institut d'Astrophysique Spatiale, CNRS (UMR8617) Universit\'{e} Paris-Sud 11, B\^{a}timent 121, Orsay, France\\
\and
Institut d'Astrophysique de Paris, CNRS (UMR7095), 98 bis Boulevard Arago, F-75014, Paris, France\\
\and
Institute for Space Sciences, Bucharest-Magurale, Romania\\
\and
Institute of Astronomy, University of Cambridge, Madingley Road, Cambridge CB3 0HA, U.K.\\
\and
Institute of Theoretical Astrophysics, University of Oslo, Blindern, Oslo, Norway\\
\and
Instituto de Astrof\'{\i}sica de Canarias, C/V\'{\i}a L\'{a}ctea s/n, La Laguna, Tenerife, Spain\\
\and
Instituto de Astronomia, Geof\'{\i}sica e Ci\^{e}ncias Atmosf\'{e}ricas, Universidade de S\~{a}o Paulo, S\~{a}o Paulo, SP 05508-090, Brazil\\
\and
Instituto de F\'{\i}sica de Cantabria (CSIC-Universidad de Cantabria), Avda. de los Castros s/n, Santander, Spain\\
\and
Jet Propulsion Laboratory, California Institute of Technology, 4800 Oak Grove Drive, Pasadena, California, U.S.A.\\
\and
Jodrell Bank Centre for Astrophysics, Alan Turing Building, School of Physics and Astronomy, The University of Manchester, Oxford Road, Manchester, M13 9PL, U.K.\\
\and
Kavli Institute for Cosmology Cambridge, Madingley Road, Cambridge, CB3 0HA, U.K.\\
\and
LAL, Universit\'{e} Paris-Sud, CNRS/IN2P3, Orsay, France\\
\and
LERMA, CNRS, Observatoire de Paris, 61 Avenue de l'Observatoire, Paris, France\\
\and
Laboratoire AIM, IRFU/Service d'Astrophysique - CEA/DSM - CNRS - Universit\'{e} Paris Diderot, B\^{a}t. 709, CEA-Saclay, F-91191 Gif-sur-Yvette Cedex, France\\
\and
Laboratoire Traitement et Communication de l'Information, CNRS (UMR 5141) and T\'{e}l\'{e}com ParisTech, 46 rue Barrault F-75634 Paris Cedex 13, France\\
\and
Laboratoire de Physique Subatomique et de Cosmologie, Universit\'{e} Joseph Fourier Grenoble I, CNRS/IN2P3, Institut National Polytechnique de Grenoble, 53 rue des Martyrs, 38026 Grenoble cedex, France\\
\and
Laboratoire de Physique Th\'{e}orique, Universit\'{e} Paris-Sud 11 \& CNRS, B\^{a}timent 210, 91405 Orsay, France\\
\and
Lawrence Berkeley National Laboratory, Berkeley, California, U.S.A.\\
\and
Max-Planck-Institut f\"{u}r Astrophysik, Karl-Schwarzschild-Str. 1, 85741 Garching, Germany\\
\and
National University of Ireland, Department of Experimental Physics, Maynooth, Co. Kildare, Ireland\\
\and
Niels Bohr Institute, Blegdamsvej 17, Copenhagen, Denmark\\
\and
Observational Cosmology, Mail Stop 367-17, California Institute of Technology, Pasadena, CA, 91125, U.S.A.\\
\and
Optical Science Laboratory, University College London, Gower Street, London, U.K.\\
\and
SISSA, Astrophysics Sector, via Bonomea 265, 34136, Trieste, Italy\\
\and
School of Physics and Astronomy, Cardiff University, Queens Buildings, The Parade, Cardiff, CF24 3AA, U.K.\\
\and
Space Sciences Laboratory, University of California, Berkeley, California, U.S.A.\\
\and
Special Astrophysical Observatory, Russian Academy of Sciences, Nizhnij Arkhyz, Zelenchukskiy region, Karachai-Cherkessian Republic, 369167, Russia\\
\and
Sub-Department of Astrophysics, University of Oxford, Keble Road, Oxford OX1 3RH, U.K.\\
\and
UPMC Univ Paris 06, UMR7095, 98 bis Boulevard Arago, F-75014, Paris, France\\
\and
Universit\'{e} de Toulouse, UPS-OMP, IRAP, F-31028 Toulouse cedex 4, France\\
% ...
}